\documentclass[12pt]{article}

\usepackage[utf8]{inputenc}

\usepackage[margin=0.82in]{geometry}
\usepackage{setspace}

\usepackage{amsmath, amsthm, amssymb, amsfonts}
\usepackage{bbm}
\usepackage{bigints}

\usepackage{graphicx}
\graphicspath{{Pictures/}}
\usepackage{caption}
\usepackage{subcaption}
\usepackage{float}
\usepackage{algpseudocode}

\usepackage{booktabs}
\usepackage{multirow}

\usepackage{tikz}
\makeatletter
\IfFileExists{tikzlibrarybayesnet.code.tex}{\usetikzlibrary{bayesnet}}{}
\makeatother

\usepackage[toc,page]{appendix}

\usepackage{natbib}
\usepackage{url}

\usepackage[ruled]{algorithm2e}

\usepackage{xcolor}
\usepackage{enumitem}
\usepackage{enumerate}
\usepackage{authblk}
\usepackage{framed}
\usepackage[normalem]{ulem}
\usepackage{titlesec}
\usepackage{comment}
\usepackage{pdfpages}
\usepackage{epstopdf}
\usepackage{authblk}

\usepackage[colorlinks,citecolor=blue,urlcolor=blue]{hyperref}

\usepackage[english]{babel}
\usepackage{graphicx,color}
\usepackage{framed}
\usepackage[normalem]{ulem}
\usepackage{amsmath}
\usepackage{amsthm}
\usepackage{amssymb}
\usepackage{amsfonts}
\usepackage{enumerate}
\usepackage{algpseudocode}
\usepackage{comment}
\usepackage{setspace}
\usepackage{url} 

\usepackage{bm}
\usepackage{natbib}
\usepackage{multirow}
\usepackage{makecell}
\usepackage{xcolor}
\usepackage{enumitem}
\usepackage{hyperref}
\usepackage{lineno}

\theoremstyle{definition}
\newtheorem{theo}{\textbf{Theorem}}

\newtheorem{D}{Definition}

\newtheorem{asum}{Assumption}
\newtheorem{rema}{Remark}

\def\bm{\boldsymbol}

\newtheorem{Remark}{Remark}
\newtheorem{propositionA}{Proposition}

\newtheorem{theoremS}{Theorem}

\newtheorem{Le}{Lemma}

\title{\bf Functional-SVD for Heterogeneous Trajectories: Case Studies in Health \thanks{This research was supported in part by the NSF Grant CAREER-2203741 and NIH Grants R01HL169347 and R01HL168940. A. R. Zhang thanks Tailen Hsing for helpful discussions. }}
  \author{Jianbin Tan\\
    Department of Biostatistics \& Bioinformatics,\\ Duke University, Durham, NC, USA\\
    Pixu Shi\hspace{.2cm}\\
    Department of Biostatistics \& Bioinformatics,\\ Duke University, Durham, NC, USA\\
    and \\
    Anru R. Zhang\thanks{Email of correspondence: \texttt{anru.zhang@duke.edu}} \\
    Department of Biostatistics \& Bioinformatics \\  and Department of Computer Science, \\ Duke University, Durham, NC, USA}

\date{}

\begin{document}
\maketitle

\begin{abstract}
Trajectory data, including time series and longitudinal measurements, are increasingly common in health-related domains such as biomedical research and epidemiology. 
Real-world trajectory data frequently exhibit heterogeneity across subjects such as patients, sites, and subpopulations, yet many traditional methods are not designed to accommodate such heterogeneity in data analysis. 
To address this, we propose a unified framework, termed Functional Singular Value Decomposition (FSVD), for statistical learning with heterogeneous trajectories. 
We establish the theoretical foundations of FSVD and develop a corresponding estimation algorithm that accommodates noisy and irregular observations. 
We further adapt FSVD to a wide range of trajectory-learning tasks, including dimension reduction, factor modeling, regression, clustering, and data completion, while preserving its ability to account for heterogeneity, leverage inherent smoothness, and handle irregular sampling.  
Through extensive simulations, we demonstrate that FSVD-based methods consistently outperform existing approaches across these tasks. 
Finally, we apply FSVD to a COVID-19 case-count dataset and electronic health record datasets, showcasing its effective performance in global and subgroup pattern discovery and factor analysis.
\end{abstract}


\setstretch{1.5} 
\section{Introduction}

Trajectory data, including sequential or longitudinal measurements over time, are ubiquitous in applications such as biomedical research, health informatics, and epidemiology \citep{yao2005functional,huang2008functional,nie2022recovering,zhang2024individualized,tan2024graphical,luo2024functional}. 
Data measurements in these domains are usually collected at discrete and often irregular time points across subjects (e.g., patients, sites, or subpopulations), yielding trajectories that may exhibit substantial heterogeneity due to inherent subject-level variability.
Common learning tasks for these data include extracting shared temporal patterns (e.g., epidemic progression \citep{dong2020interactive}), identifying subgroups (e.g., patient trajectory clustering \citep{zhu2024functional}) and quantifying associations (e.g., linking health-related trajectories to disease risk \citep{luo2024functional}).
These goals naturally align with core tasks such as dimension reduction, clustering, and regression, motivating an urgent need for a versatile statistical framework that can accommodate heterogeneous trajectories.

In the statistical literature, trajectory data are also referred to as functional data \citep{wang2016functional}, which view each subject's trajectory as a functional sample for data analysis.
A central tool in this area is functional principal component analysis (FPCA), which underpins a broad range of learning tasks involving dimension reduction, including functional completion, linear regression, clustering, canonical correlation analysis, and additive modeling \citep{yao2005functional_reg,chiou2007functional,muller2008functional,hsing2015theoretical,morris2015functional,kraus2015components,scheipl2015functional,wang2016functional,reiss2017methods,imaizumi2018pca,kneip2020optimal}. 
Despite its success, standard FPCA-based methodology typically relies on a homogeneity assumption, namely that trajectories are realizations from a single underlying population. 
This assumption can be violated in many applications where trajectories arise from heterogeneous subpopulations or different sources.
Here, we focus on two real-world examples:
\begin{itemize}[itemsep=1pt, topsep=2pt, partopsep=2pt]
\item {\it Epidemic dynamic data}: Epidemic dynamic data \citep{dong2020interactive} consist of time-varying trajectories of reported cases across regions and are often used in public health to compare how outbreaks evolve globally over time. 
Although FPCA has been applied to summarize these regional curves, heterogeneity in shape across regions---due to differing interventions or population characteristics \citep{tian2021effects,tan2022transmission}---may violate the homogeneity assumption and render FPCA inappropriate.

\item {\it Electronic health records}: ICU electronic health records (EHRs) contain longitudinal trajectories of clinical features collected from patients admitted to intensive care units \citep{johnson2020mimic}, reflecting biologically meaningful temporal patterns that are crucial for monitoring patients' health status. Dimension reduction on patient-level EHR data can provide an interpretable summary for assessing and tracking the status of a patient.
While FPCA has been widely applied in longitudinal data analysis \citep{yao2005functional,yao2005functional_reg,chiou2007functional,wang2016functional}, it may not be suitable for EHR data due to the non-identically distributed nature of clinical measurements across different features.
\end{itemize}

Alternatively, factor models provide another dimension reduction approach that can account for heterogeneity in temporal trajectories. 
However, existing methods \citep{bai2003inferential,lam2011estimation,lam2012factor} often rely on regularly spaced temporal sampling and stationarity conditions to define valid factor loadings. 
These assumptions may not be suitable for irregular or nonstationary data, such as those arising from epidemic outbreaks or ICU patients.

To overcome the limitations of existing methods, we propose a new framework called functional singular value decomposition (FSVD), tailored for \sout{the} dimension reduction and feature extraction of heterogeneous trajectory data.
Specifically, the FSVD of $n$ trajectories $X_1,\ldots, X_n$ is defined as
\begin{equation}\label{Intro_FSVD}
        X_i(t) = \sum_{r \ge 1} \rho_r a_{ir} \phi_r(t),\ i =1,\ldots,n.
\end{equation}
Here, $\bm{a}_r:=(a_{1r},\ldots,a_{nr})^\top$, $r\geq 1$, are orthonormal $n$-dimensional singular vectors, $\phi_r$s are orthonormal singular functions, and $\rho_r$s are singular values. 
The first contribution of this paper is to validate the proposed framework by proving the existence of the FSVD in \eqref{Intro_FSVD}, without requiring homogeneity conditions of trajectory data. 
Accordingly, we establish an FSVD algorithm that accommodates practical settings where trajectories are irregularly observed.

\begin{figure}[h]
\begin{center}
\includegraphics[scale=0.3]{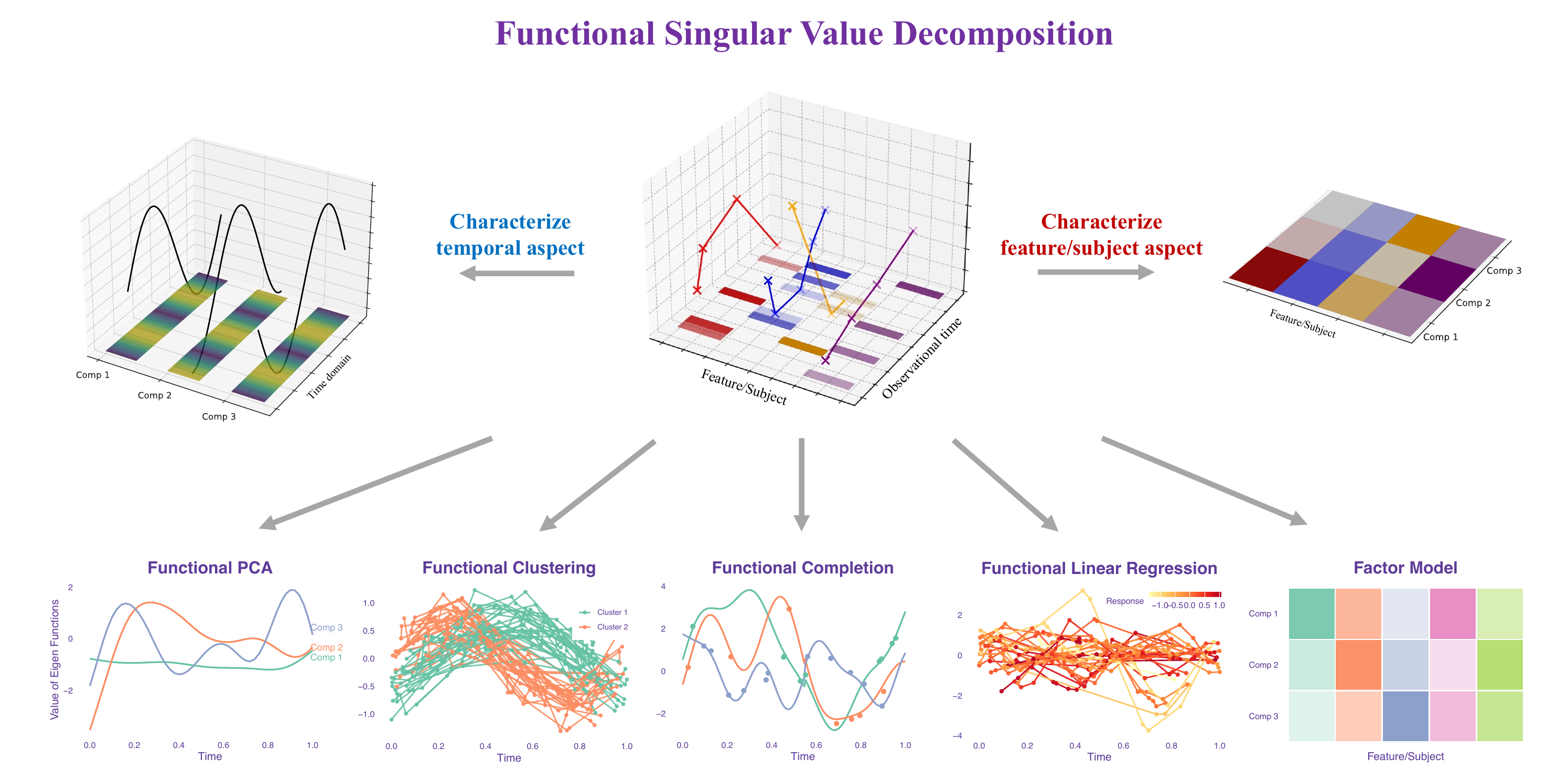}
\end{center}
\caption{An illustration of tasks associated with FSVD.} \label{FSVD_relation}
\end{figure}

Next, we introduce the concepts of intrinsic basis functions and intrinsic basis vectors for characterizing the heterogeneous structure, which unify several crucial dimension reduction methods for trajectory data within the FSVD framework.
Using these concepts, we connect the FSVD framework to several core learning tasks, including functional completion, clustering, and linear regression, where dimension reduction plays a central role. 
Unlike traditional methods, FSVD performs these tasks without relying on restrictive assumptions such as sample homogeneity or regularly spaced time grids. 
This makes it highly adaptable to real-world trajectory data that are heterogeneous, non-stationary, and irregularly sampled. 
See Figure~\ref{FSVD_relation} for an illustration of the supported tasks.

We apply FSVD to two real-world applications.
In a dataset recording the case counts of SARS-CoV-2 infections across 64 regions in 2020, FSVD effectively captures heterogeneous trajectory patterns that FPCA fails to identify, providing an effective evaluation of global and subgroup patterns for epidemic trajectory discovery.
In EHR datasets, FSVD leverages cross-feature structure for factor analysis of patient-specific patterns, offering improved imputation performance and providing guidance for understanding a patient’s health status.

\subsection{Related Work}\label{sec:related-work}
FSVD is connected to a broad range of work in functional data analysis, PCA, and SVD.

\paragraph*{PCA and SVD versus Functional-PCA and Functional-SVD.}
Principal Component Analysis (PCA) and Singular Value Decomposition (SVD) are closely related and widely used for dimensionality reduction and feature extraction in matrix data. 
PCA is a statistical method that models data as samples of random vectors and performs dimensionality reduction based on the covariance matrix, whereas SVD is a linear algebra technique that factorizes any deterministic or random data matrix into low-rank components. 
While PCA relies on the covariance matrix, it can also be computed using SVD on the centered data matrix, which effectively bypasses explicit covariance computation and is especially advantageous when the feature dimensionality exceeds the sample size. 
Beyond their interrelation, SVD has broader applications in many machine learning methods, such as sparse PCA \citep{witten2009penalized}, canonical correlation analysis \citep{witten2009penalized}, and matrix completion \citep{candes2012exact}, demonstrating its versatility.

A similar juxtaposition can be drawn between FPCA and FSVD as in the relationship between PCA and SVD. 
FPCA typically involves estimating covariance functions, a complex task that requires substantial data and smoothness assumptions on the covariance structure \citep{yao2005functional, hsing2015theoretical, zhang2022nonparametric}. 
This approach assumes that all trajectory samples share the same mean and covariance functions, which may be restrictive for real-world data. 
In contrast, FSVD performs dimension reduction directly on the observed trajectories, bypassing the need for covariance estimation. 
This makes FSVD a flexible alternative, particularly well-suited for heterogeneous datasets where shared mean and covariance structures may not hold.

\paragraph*{Comparison with Existing Functional PCA- and SVD-type methods.}

Most existing methods for trajectory-data dimension reduction share a similar philosophy as PCA by adopting linear combinations of random components as low-dimensional representations of the data. 
They can be broadly grouped into two frameworks: the first focuses on the functional aspect and projects the data onto deterministic basis functions, and the second focuses on the tabular (e.g., feature or subject) aspect and projects the data onto deterministic basis vectors. 

Methods under the first framework often project functions into deterministic eigenfunctions using Karhunen--Lo\`eve (KL) expansions and their extensions. 
For example, FPCA adopts the KL expansion for homogeneous functional data \citep{ramsay1997functional, hsing2015theoretical}; finite mixtures of KL expansions are used to account for clustering structures within heterogeneous functional data \citep{chiou2007functional,jacques2013funclust}; separable KL expansions handle separable covariance structures among dependent functional data \citep{zapata2022partial, liang2021modeling,tan2024graphical}; and other extensions of KL expansions serve different purposes \citep{chiou2014multivariate, chen2017modelling,happ2018multivariate}. 
Methods under the second framework focusing on the tabular aspect include factor models for multivariate time series \citep{lam2011estimation, lam2012factor}, which reduce the subject/features' dimensions via deterministic factor loadings.

Compared to the above methods, FSVD offers a unified framework for heterogeneous trajectory data and is capable of providing dimension reduction for both functional and tabular aspects. 
This allows FSVD to accomplish the tasks of both FPCA and factor models, making it suitable for a wider range of scenarios where various types of data structures need to be captured and interpreted.

Other SVD-type methods have also appeared in the literature on functional data analysis. 
\citet{yang2011functional} focused on the cross-covariance between functional datasets, building upon the SVD of compact operators in functional analysis. 
\citet{huang2008functional, huang2009analysis, zhang2013robust} implemented SVD-type methods to decompose regularly observed functional data by enforcing continuity on the singular vectors associated with time. 
\citet{nie2022recovering} applied SVD-type methods to decompose independent and identically distributed (i.i.d.) functional data. 
\citet{han2023guaranteed} extended SVD to regularly observed tensor functional data under fixed design.
However, the assumptions of regular observation and i.i.d.\ structure are often impractical for many real-world datasets, and a fixed design may also be unsuitable for learning tasks such as clustering and regression.
In contrast to prior works, FSVD accommodates irregular observations and heterogeneous random-function settings, providing foundational theoretical guarantees that were previously unavailable.

\paragraph*{Organization} 

The rest of this article is organized as follows. 
Section~\ref{sec:data_illu} provides a data illustration to highlight heterogeneity and motivate our scientific problem. 
To address this, we introduce the modeling framework of FSVD for fully observed trajectory data, and then develop an estimation procedure for noisy and irregularly observed trajectories in Section~\ref{sec:FSVD for DOFD}. 
In Section~\ref{sec: task}, we present the concepts of intrinsic basis functions and intrinsic basis vectors under the FSVD framework, showing how they encode different structural aspects of heterogeneous trajectories. 
Accordingly, we demonstrate the capability of FSVD in performing a range of learning tasks for heterogeneous trajectory data. 
Section~\ref{sec:dat_ana} showcases the application of FSVD to two real datasets. Section~\ref{sec: heter} discusses the theoretical foundations of FSVD. Section~\ref{sec:dis} concludes with a discussion. 
Additional theory and proofs, implementation details, simulation results, and supporting resources are provided in Parts~A--D of the Supplementary Materials, respectively. 
The codes and datasets are publicly available at 
\href{https://github.com/Jianbin-Tan/Functional-Singular-Value-Decompostion}{https://github.com/Jianbin-Tan/Functional-Singular-Value-Decompostion}. 
An \texttt{R} package named \texttt{FSVD} is also available at 
\href{https://github.com/Tan-jianbin/FSVD}{https://github.com/Tan-jianbin/FSVD}.

\section{Data and Exploratory Analysis}\label{sec:data_illu}

\begin{figure}[h]
    \centering
    \includegraphics[scale = 0.4, trim=0cm 1cm 0cm 0 cm, clip]{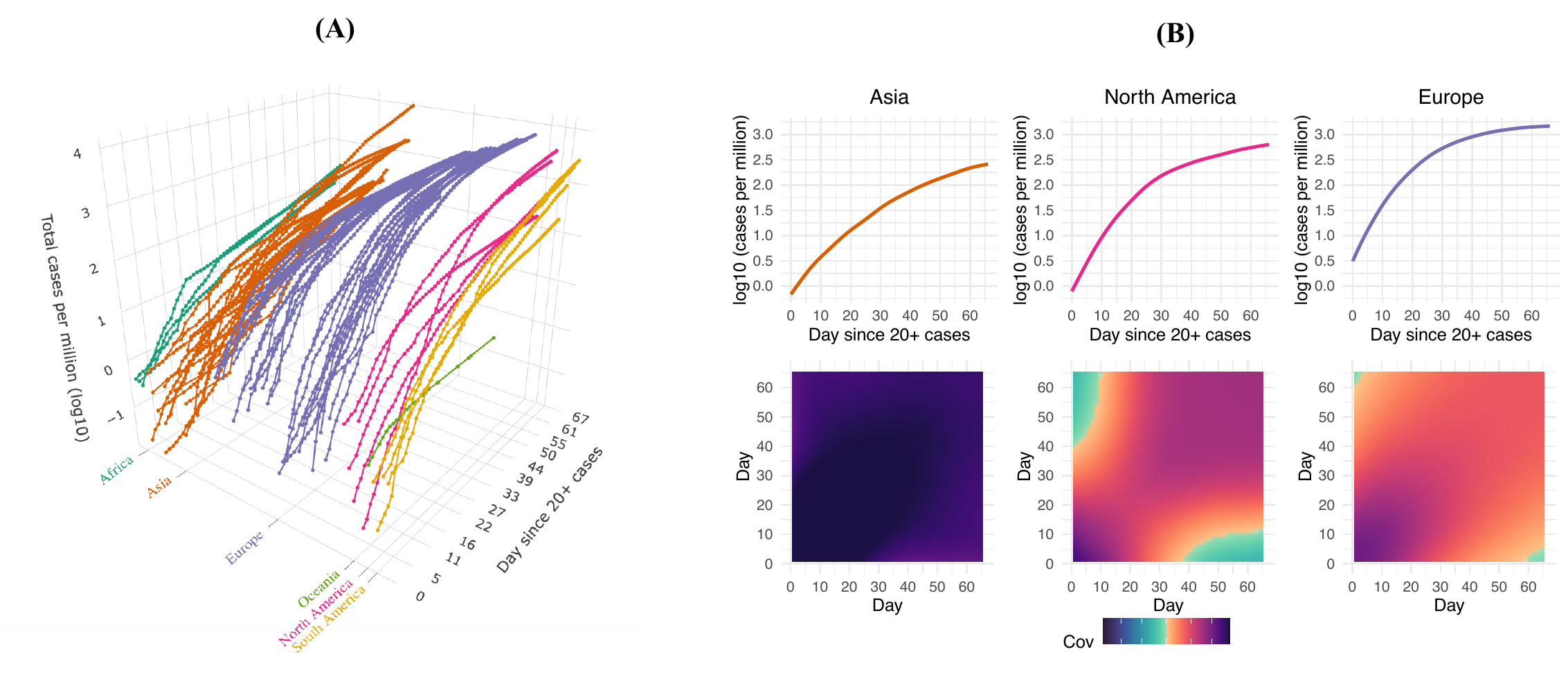}
    \caption{
        \textbf{(A)} Epidemic trajectory data across regions from different continents; the locations of these regions are shown in Figure~\ref{COVID19_tra} below. 
        \textbf{(B)} Estimated mean (first row) and covariance (second row) functions from trajectory data for three continents. Estimates are obtained using the \texttt{R} package \texttt{fdapace} \citep{yao2005functional}.
    }
    \label{fig:data_combine}
\end{figure}

We analyze two health trajectory datasets where heterogeneous sampling and progression can break standard methods, and show how our approach resolves these issues.
\paragraph*{COVID-19 Case Trajectories}  
Our first application analyzes cumulative COVID-19 case counts per million people (on the log scale) from 64 regions in 2020 \citep{dong2020interactive}. 
For each region, we extract case counts over 67 consecutive days, beginning with the first day that reported at least 20 confirmed cases. 
We focus on days when the cumulative counts changed, yielding 64 irregularly observed trajectories. Figure~\ref{fig:data_combine}(A) displays these trajectories, with most regions showing a similar monotonically increasing trend. 
Understanding global and subgroup dynamics in these data is crucial for revealing outbreak patterns and for describing how case trajectories change over time in relation to public health interventions \citep{carroll2020time,tian2021effects}. 
A common approach is to apply dimension reduction techniques, such as FPCA \citep{hsing2015theoretical}, to extract dominant modes of variation in trajectory data.  
However, these methods assume that the trajectories are i.i.d.\ samples drawn from a common population. 
This assumption may not hold, as suggested by Figure~\ref{fig:data_combine}(B), which shows clear differences in the mean and covariance functions across different regions.
To address this, we aim to develop new methodology to extract global and subgroup dynamics from these heterogeneous epidemic trajectories.

\paragraph*{MIMIC-IV Clinical Data}

Our second application focuses on the MIMIC-IV electronic health records dataset \citep{johnson2020mimic}, which contains de-identified longitudinal records from ICU stays at the Beth Israel Deaconess Medical Center between 2008 and 2019. 
This dataset includes a rich set of clinical measurements from different clinical aspects. 
The observation periods vary across patients, and for a given patient, the timing of measurements also differs across features. 
Accordingly, the patient-level longitudinal data form irregular trajectories across clinical features.

\begin{figure}[h]
    \centering
    \includegraphics[width=0.95\linewidth]{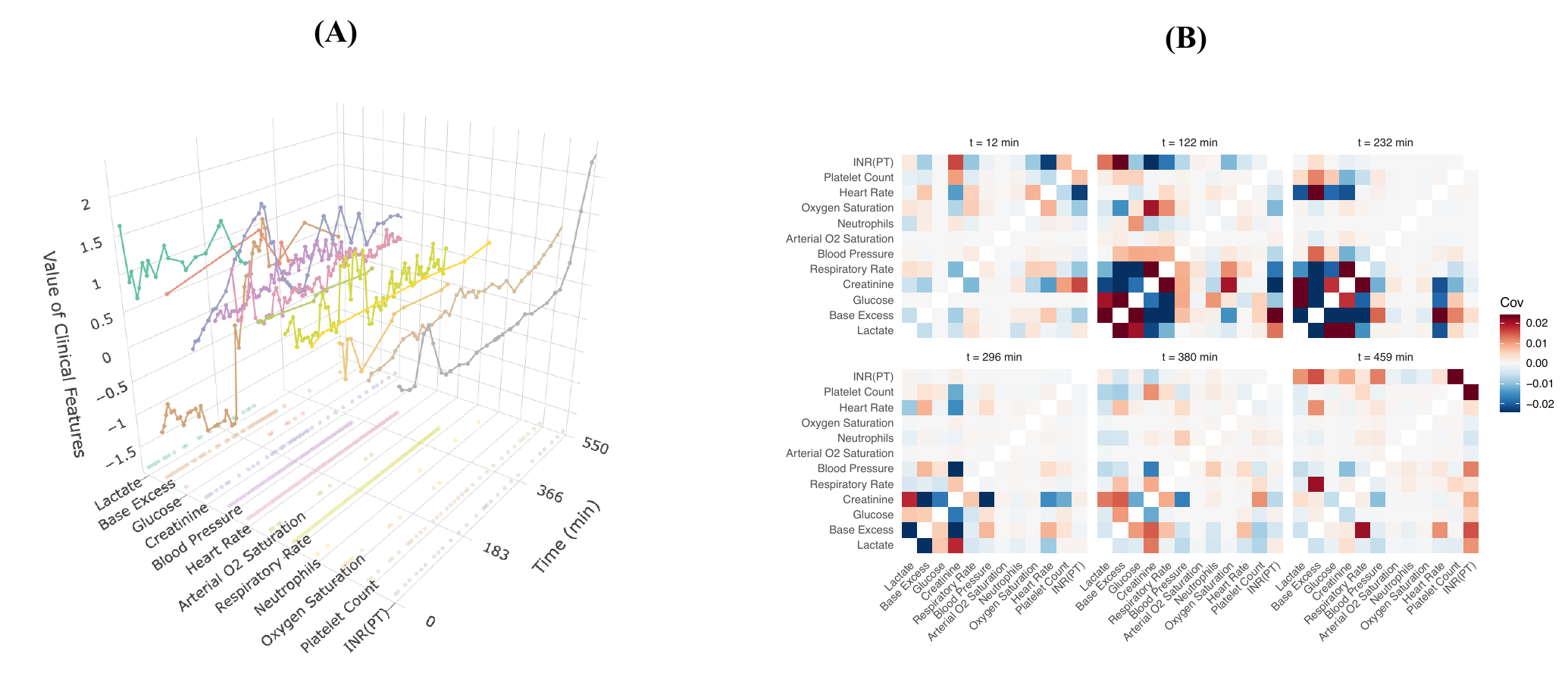}
    \caption{
       \textbf{(A)} Clinical longitudinal data across different features. Observed time points are indicated in the $x$–$y$ plane, and the zero point marks the time of admission, and all features are normalized to remove unit-specific effects;
\textbf{(B)} Local kernel-weighted cross-covariance between clinical features at six selected time points, which computes the cross-covariance matrix (with the diagonal set to zero) of clinical features locally over time. 
The definition of the local kernel-weighted cross-covariance for irregular trajectory data can be found in Part~D.1 of the Supplementary Materials.
    }
    \label{fig:data_combine_B}
\end{figure}

For illustration, we analyze a multivariate ICU time series from a single patient, consisting of 12 routinely monitored variables—Heart Rate, Respiratory Rate, Arterial O$_2$ Saturation, Blood Pressure, Oxygen Saturation, Base Excess, Glucose, Creatinine, INR (PT), Lactate, Platelet Count, and Neutrophils (see Table~S2 in the Supplementary Materials for definitions), recorded over a 580-minute window after admission (Figure~\ref{fig:data_combine_B}). We selected these variables because they are clinically relevant, interpretable, commonly measured in the ICU, and jointly characterize complementary physiological systems, including cardiopulmonary function and oxygenation, hemodynamics, metabolic/renal status, coagulation, and inflammatory response.

By Figure~\ref{fig:data_combine_B}(B), we observe clear time-varying co-movement among clinical features, reflected by cross-covariance patterns that evolve across time points. Understanding such joint dependence can help characterize a patient’s underlying physiological state and inform personalized treatment decisions.
To this end, factor-based approaches are widely used to capture such co-movement patterns in multivariate data \citep{bai2003inferential, lam2011estimation, lam2012factor}, typically relying on regular temporal sampling or stationary dynamics (e.g., assuming the mean or cross-covariance remains constant over time).
However, these conditions may not hold in clinical data, as illustrated in Figure~\ref{fig:data_combine_B}, where we observe highly irregular temporal sampling and distinct cross-covariance differences across time points.
Therefore, we aim to develop a new factor analysis framework for a patient’s clinical trajectories that can accommodate both irregular sampling and nonstationarity.

\section{FSVD for Heterogeneous Trajectories}\label{sec:FSVD for DOFD}

Let $\mathcal{T}$ be a bounded closed interval in $\mathbb{R}$. Without loss of generality, we assume $\mathcal{T} = [0,1]$ throughout this article.
Denote $\mathcal{L}^2(\mathcal{T})$ as the Hilbert space of square-integrable functions on $\mathcal{T}$, with the inner product $\langle \cdot,\cdot \rangle$ and norm $\|\cdot\|:=\sqrt{\langle \cdot,\cdot \rangle}$, where $\langle f, g\rangle=\int_{\mathcal{T}}f(t)g(t)\ \mathrm{d}t$ for $f,g\in \mathcal{L}^2(\mathcal{T})$. 
For any vector $\bm{a} = (a_1, \dots, a_n)^\top$, we also denote $\|\cdot\|$ the Euclidean norm of $\bm{a}$, 
that is,
\(
\|\bm{a}\|_2 := \sqrt{\sum_{i=1}^n a_i^2}.
\)
Define $\operatorname{span}(f_1,\dots,f_n)$ as the linear subspace spanned by $f_1,\dots,f_n\in \mathcal{L}^2(\mathcal{T})$.
Let $\mathbb{I}(\cdot)$ be the indicator function and $[Z]$ be the set of integers $\{1,\dots,Z\}$. Define $(a)_{+}:=\max(a,0)$.
We use $\operatorname{rank}(\cdot)$ to denote the rank of a matrix. 
For two sequences of non-negative real values $\{a_n\}$ and $\{b_n\}$, we say $a_n \lesssim b_n$ (or $b_n\gtrsim a_n$) if there exists a constant $C>0$ such that $a_n \leq Cb_n$ (or $a_n \geq Cb_n$) for all $n$. We denote $a_n \asymp b_n$ if $a_n\lesssim b_n$ and $a_n\gtrsim b_n$. $\bm{I}_m$ is the identity matrix of dimension $m \times m$.

In the following, we denote the latent trajectories from the datasets in Figures~\ref{fig:data_combine} and \ref{fig:data_combine_B} as \( X_1, \ldots, X_n \), where \( n \) is the total number of trajectories. We focus on the following model:
\begin{eqnarray}\label{mea_model}
Y_{ij}=X_{i}(T_{ij}) + \varepsilon_{ij},\quad j\in [J_i],\ i\in [n],
\end{eqnarray}
where $\big\{T_{ij}; j\in [J_i]\big\}$ is the observable time points for trajectory $X_{i}$, $\big\{\varepsilon_{ij};j\in [J_i]\big\}$ are mean-zero noise variables, and $\big\{Y_{ij}; j\in [J_i]\big\}$ are the noisy discrete observations of $X_i$ for each $i$. In this model, we allow the observation time points to be irregular, i.e., $\big\{T_{ij};j\in [J_i]\big\}$ may vary across different $i$. 

Before detailing the procedure of FSVD, we first introduce some preliminaries in the context of reproducing kernel Hilbert space (RKHS). Let $\mathcal{H}$ be a Hilbert space of functions on $\mathcal{T}$ with inner product $\langle \cdot, \cdot \rangle_{\mathcal{H}}$ and norm $\|\cdot\|_{\mathcal{H}}$. The space $\mathcal{H}$ is called an RKHS if there exists a kernel $\mathbb{K}$ on $\mathcal{T}\times\mathcal{T}$ such that $\mathbb{K}(t,\cdot)\in \mathcal{H}\ \text{and}\  f(t)=\langle f,\mathbb{K}(t,\cdot)\rangle_{\mathcal{H}}$, $\forall t\in \mathcal{T}$ and $f\in \mathcal{H}$. We denote $\mathcal{H}$ as $\mathcal{H}(\mathbb{K})$ because it can be shown that $\mathbb{K}$, the reproducing kernel of $\mathcal{H}$, is unique to $\mathcal{H}$. 

\subsection{Foundations of Functional Singular Value Decomposition}\label{sec:FSVD}
In this subsection, we propose functional singular value decomposition (FSVD) for modeling a collection of trajectories $X_i$.
Suppose that \( X_1, \ldots, X_n \in \mathcal{H} \), where \( \mathcal{H} \subseteq \mathcal{L}^2(\mathcal{T}) \) is a Hilbert space. The FSVD for \(X_i\)s is defined as:
\begin{equation}\label{eq:X-a-xi_1}
    X_i(t) = \sum_{r \ge 1} \rho_r a_{ir} \phi_r(t),\quad i \in [n],
\end{equation}
where \( \rho_1 \geq \dots \geq \rho_R > 0 \) are the singular values,
\(
\bm{a}_r := (a_{1r}, \ldots, a_{nr})^\top \in \mathbb{R}^n, \quad r \in [R],
\)
are the singular vectors, and \( \phi_1, \ldots, \phi_R \in \mathcal{H} \) are the singular functions. Here, \( R \leq n \) denotes the rank of the expansion.
The vectors \( \bm{a}_1, \ldots, \bm{a}_R \) and functions \( \phi_1, \ldots, \phi_R \) are orthonormal in the sense that
\(
\bm{a}_r^{\top} \bm{a}_{r^{\prime}} = \langle \phi_r, \phi_{r^{\prime}} \rangle = \mathbb{I}(r = r^{\prime})
\quad \text{for } r, r^{\prime} \in [R].
\)
The uniqueness of FSVD \eqref{eq:X-a-xi_1} is characterized by Proposition~S1 of the Supplementary Materials,
which shows that when \(\rho_r\)s are distinct, the singular functions/vectors are unique up to a sign flip.

In Section~\ref{sec: heter}, we show that \eqref{eq:X-a-xi_1} exists as long as \( X_i \)s lie in the same Hilbert space \(\mathcal{H}\).
This allows \( X_i \)s to be neither independent nor identically distributed across functions or time, a weaker condition than the i.i.d.\ assumption imposed on functional data \citep{ramsay1997functional, yao2005functional, li2010uniform, hsing2015theoretical, wang2016functional}
or the stationarity assumptions in time series analysis \citep{lam2011estimation, lam2012factor}.

Similar to matrix SVD, the $r$th singular component of $X_1,\ldots, X_n$ provides the optimal rank-one approximation for these trajectories after subtracting the first $(r-1)$ singular components.
Specifically, consider $g_{i0}$, $i \in [n]$, as zero functions, and let $g_{ir}$, $i \in [n]$, be defined as the minimizers $f_i$ obtained from
\begin{equation}\label{eun: FSVD_s}
    \min_{f\in \mathcal{H}}\min_{f_1,\dots,f_n \in \operatorname{span}(f)} \sum_{i=1}^n \bigg\|X_i - \sum_{l=0}^{r-1} g_{il} - f_i\bigg\|^2.
\end{equation}
Define
\(
\rho_r^0 := \sqrt{\sum_{i=1}^n \|g_{ir}\|^2},\
\phi_r^0:= g_{ir} / \|g_{ir}\|,
\)
and
\(
\bm{a}_r^0 := \big(\langle g_{1r}, \phi_r^0 \rangle, \dots, \langle g_{nr}, \phi_r^0 \rangle\big)^\top / \rho_r^0.
\)
Then, $\{\rho_r^0, \bm{a}_r^0, \phi_r^0;$ $\ r \in [R]\}$ forms the FSVD of $X_1,\dots, X_n$. See Proposition~S2 in the Supplementary Materials for details and proof.

\subsection{Rank-One Kernel Ridge Regression}\label{sec:model-fsvd}

In this subsection, we focus on the FSVD based on the discrete observations \(Y_{ij}\)\sout{s}. 
To this end, we assume that \(X_1, \ldots, X_n\) are contained in an RKHS \(\mathcal{H}(\mathbb{K})\subset \mathcal{L}^2(\mathcal{T})\), which ensures that the singular components of \(X_i\)\sout{s} are also contained in \(\mathcal{H}(\mathbb{K})\).
Based on \eqref{eun: FSVD_s}, we propose to estimate the first singular component by computing the following kernel ridge regression:
\begin{equation}\label{opt_null}
      \min_{f\in \mathcal{H}(\mathbb{K})}\min_{f_1,\dots,f_n \in \operatorname{span}(f)}\sum_{i=1}^n\left(\frac{1}{J_i}\sum_{j=1}^{J_i}\big\{Y_{ij}-f_i(T_{ij})\big\}^2 + \nu \left\|\mathcal{P}f_i\right\|^2_{\mathcal{H}}\right).
\end{equation}
Here, \(\mathcal{P}\) is an operator from \(\mathcal{H}(\mathbb{K})\) onto a subspace to prevent overfitting,
and \(\nu\) is a tuning parameter.
We set \(f_i=a_{i1}\phi_1\) and \(\bm{a}_1=(a_{11},\dots,a_{n1})^\top\); then \eqref{opt_null} is equivalent to
\begin{eqnarray}\label{opt_2}
\min_{\bm{a}_1\in \mathbb{R}^n,\phi_1\in \mathcal{H}(\mathbb{K})}\sum_{i=1}^n\frac{1}{J_i}\sum_{j=1}^{J_i}\big\{Y_{ij}-a_{i1} \phi_1(T_{ij})\big\}^2+\nu\|\bm{a}_1\|^2\cdot\|\mathcal{P}\phi_1\|_{\mathcal{H}}^2.
\end{eqnarray}

First note that \eqref{opt_null} reduces to the standard kernel ridge regression \citep{gu2013smoothing} when \(n=1\).
For \(n>1\), the constraint \(f_1,\ldots,f_n \in \operatorname{span}(f)\) enables borrowing information across functions in the kernel ridge regressions.
Meanwhile, \eqref{opt_2}, being equivalent to \eqref{opt_null},
can be viewed as a penalized decomposition of \(Y_{ij}\)\sout{s},
analogous to existing SVD-type methods developed for matrices \citep{witten2009penalized} and time series \citep{zhang2013robust,yu2016temporal}.
Such SVD-type methods have also been applied to regularly observed functional data \citep{huang2008functional, huang2009analysis}
and to i.i.d.\ functional data \citep{nie2022recovering}.
In contrast, here we consider a more general setting of irregularly observed heterogeneous functional data.

Because the regularization of \(f_i\)\sout{s} in \eqref{opt_null} is transferred to \(\phi_1\) and \(\bm{a}_1\) in \eqref{opt_2}, the minimization over the function \(\phi_1\) can then be reformulated into a finite-dimensional optimization problem as demonstrated by the representer theorem. Details of this result can be found in Proposition~S3 of the Supplementary Materials.

\begin{rema}[Interpretation of Rank-One Regressions]
Note that \eqref{opt_null} does not assume \(Y_{ij}\) have zero mean, nor does it require subtracting an estimated mean function prior to analysis.
Instead, the regression directly seeks a rank-one approximation of the dominant structure in the data.
If the mean signal dominates, the rank-one component primarily estimates the dominant mean trend.
Otherwise, it may reflect a combination of mean, covariance-driven variation, and data randomness.
In these cases, \(\phi_1\) captures the dominant functional pattern across subjects, while the coefficients \(a_{i1}\) represent feature/subject-specific amplitudes independent of the functional structure.
Both objects are data-dependent rather than population-level quantities,
but they may reflect population-level information of \(X_i\)\sout{s} under suitable conditions (see Section~\ref{sec: heter} for a detailed demonstration).
\end{rema}

The choice of \(\mathbb{K}\) is crucial in \eqref{opt_null}.
For instance, the Sobolev kernel of order \(q\),
\(
\mathbb{K}(t,s)
= \sum_{m=0}^{q-1}\frac{t^m}{m!}\frac{s^m}{m!}
+ \int_0^1 \frac{(t-u)_{+}^{\,q-1}}{(q-1)!}\,\frac{(s-u)_{+}^{\,q-1}}{(q-1)!}\,\mathrm{d}u,
\)
induces an RKHS equivalent to the Sobolev space and thus encourages global smoothness.
Other kernels, such as the Mat\'ern kernel, can also be adopted in our framework to encourage desirable properties.
In practice, the kernel can be chosen based on the expected properties of the data,
allowing \eqref{opt_null} to adapt flexibly to a wide range of settings.

\subsection{Alternating Minimization for FSVD}\label{sec:alternative_minimization}
In this subsection, we adopt the Sobolev kernel to implement the FSVD,
which is a common RKHS choice for reflecting the smoothness of functional data \citep{yuan2010reproducing, hsing2015theoretical}.
The associated RKHS is the Sobolev space, defined as
\(
\mathcal{W}^q_2(\mathcal{T})
:=
\bigl\{f:\mathcal{T}\rightarrow \mathbb{R}\,;\ D^0f,\ldots,D^{q-1}f\ \text{are continuous and } D^qf\in \mathcal{L}^2(\mathcal{T}) \bigr\}
\subseteq \mathcal{L}^2(\mathcal{T}),
\)
where \(D^q\) denotes the \(q\)th-order differential operator.
Under this setting, the operator \(\mathcal{P}\) in \eqref{opt_2} can be taken as
\(
\|\mathcal{P}f\|_{\mathcal{H}}=\|D^q f\|,
\)
measuring the global smoothness of \(f\) via its \(q\)th derivative.

When \(\mathcal{H}(\mathbb{K})=\mathcal{W}^q_2(\mathcal{T})\) and \(\|\mathcal{P}\phi_1\|_{\mathcal{H}}=\|D^q\phi_1\|\), we have a simpler representer theorem for the optimization \eqref{opt_2}.
Specifically, suppose that \(J_i>q\), \(i\in [n]\).
Then there exist $w_{ij}\in \mathbb{R}$, $i\in [n]$ and $j\in [J_i]$ such that the minimizer \(\phi_1\) in \eqref{opt_2} can be represented
as \(\phi_1(t) = \sum_{m=1}^{M}{w}_{m}N_{m}(t)\), and \eqref{opt_2} can be transformed into
\begin{eqnarray}\label{opt_3}
\min_{\bm{a}_1\in \mathbb{R}^n,\bm{w}\in \mathbb{R}^M}\sum_{i=1}^n\frac{1}{J_i}\sum_{j=1}^{J_i}\bigg\{Y_{ij}-a_{i1}   \sum_{m=1}^Mw_{m}N_{m}(T_{ij})\bigg\}^2+\nu\|\bm{a}_1\|^2\cdot\bm{w}^\top \bm{H}\bm{w},
\end{eqnarray}
where \(\big\{N_{m};m\in [M]\big\}\) are the natural spline functions on knots \(\cup_{i\in [n]}\big\{T_{ij};j\in[J_i]\big\}\), with \(M\) being the number of unique time points across subjects;
\(\bm{w}=\big(w_{m};m\in [M]\big)^\top\in \mathbb{R}^M\); the matrix \(\bm{H}\in \mathbb{R}^{M\times M}\) has entries
\(\langle D^q N_{m'}, D^q N_{m''} \rangle\) for all \(m', m'' \in [M]\).
The above transformation can be proven by the theory of splines, e.g., Theorem~6.6.9 in \citet{hsing2015theoretical}.

We employ an alternating minimization approach to obtain the minimizers of \(\bm{a}_1\) and \(\bm{w}\) from \eqref{opt_3}.
Note that \(\bm{a}_1\) and \(\bm{w}\) are identifiable only up to a scalar multiplication; therefore, we always scale \(\bm{a}_1\) such that \(\|\bm{a}_1\| = 1\) during the alternating minimization.
This procedure is summarized in Algorithm~\ref{algo: FSVD}, with \(\tau = 10^{-5}\) and \(H = 500\) in our implementation.

\begin{algorithm}[h]
\caption{Alternating Minimization for Estimating the First Component}\label{algo: FSVD}
\footnotesize
\begin{algorithmic}[1]
\State \textbf{Input} initialization $\hat{\bm{a}}_1^{(0)}$, $\big\{Y_{ij};j\in [J_i],i\in [n]\big\}$, tuning parameter $\nu$, threshold value $\tau$, and maximum iteration number $H$.
\State $h=0$ and $\hat{\bm{a}}^{(0)}=\hat{\bm{a}}_1^{(0)}$.
\State \textbf{Repeat}
\State \quad Solve
\[
\hat{\bm{w}}
=
\arg\min_{\bm{w}\in \mathbb{R}^M}
\sum_{i=1}^n\frac{1}{J_i}\sum_{j=1}^{J_i}(\hat{a}_{i}^{(h)})^2
\bigg\{Y_{ij}/\hat{a}_{i}^{(h)}- \sum_{m=1}^M w_{m}N_{m}(T_{ij})\bigg\}^2
+\nu\bm{w}^\top \bm{H}\bm{w}.
\]
\State\quad $\widehat{\rho\phi}^{(h)}(T_{ij})=\sum_{m=1}^M\hat{w}_{m}N_{m}(T_{ij})$ for $i\in [n]$ and $j\in [J_i]$.
\State \quad  $\tilde{a}_{i}^{(h+1)}= \big\{\frac{1}{J_i}\sum_{j=1}^{J_i}Y_{ij}\widehat{\rho\phi}^{(h)}(T_{ij})\big\}/ \big\{\frac{1}{J_i}\sum_{j=1}^{J_i}\big(\widehat{\rho\phi}^{(h)}(T_{ij})\big)^2+\nu\hat{\bm{w}}^\top \bm{H}\hat{\bm{w}}\big\}$, $i=1,\ldots,n$.
\State \quad Update $\hat{\bm{a}}^{(h+1)}:= \left(\tilde{a}_{1}^{(h+1)},\ldots,\tilde{a}^{(h+1)}_{n}\right)^\top/\sqrt{\left(\tilde{a}_{1}^{(h+1)}\right)^2+\cdots+\left(\tilde{a}^{(h+1)}_{n}\right)^2}$.
\State \quad $h=h+1$.
\State \textbf{Until} $h\geq H$ or $\|\widehat{\rho\phi}^{(h-1)} - \widehat{\rho\phi}^{(h)}\|/\|\widehat{\rho\phi}^{(h-1)}\|\leq \tau$.
\State Set $\hat{\bm{a}}_1$, $\hat{\phi}_1$, and $\hat{\rho}_1$ as $\hat{\bm{a}}^{(h)}$,  $\widehat{\rho\phi}^{(h)}/\|\widehat{\rho\phi}^{(h)}\|$, and $\|\widehat{\rho\phi}^{(h)}\|$, respectively.
\State \textbf{Output} $\hat{\bm{a}}_1$, $\hat{\phi}_1$, and $\hat{\rho}_1$.
\end{algorithmic}
\end{algorithm}

Note that \eqref{opt_3} is a non-convex optimization problem requiring a suitable initialization \( \hat{\bm{a}}^{(0)} \).
To obtain \( \hat{\bm{a}}^{(0)} \), we first construct a time grid,
\(
\mathcal{T}_{\text{obs}} = \{T_m; m \in [M]\} := \bigcup_{i=1}^n \{T_{ij}; j \in [J_i]\},
\)
and form an incomplete data matrix
\(
\bm{Y}_{\text{inc}} = (Y_{iq}^{\text{inc}})_{i\in [n],\, q\in [M]} \in \mathbb{R}^{n\times M},
\)
where \( Y_{im}^{\text{inc}} = Y_{ij} \) if \( T_m \in \{T_{ij}; j \in [J_i]\} \), and is missing otherwise.
We then apply matrix completion \citep{candes2012exact} to impute missing values, perform SVD on the completed matrix, and use its first left singular vector as the FSVD initialization \( \hat{\bm{a}}^{(0)}_1 \).
In Part~C.1 of the Supplementary Materials, we show that this approach yields a minimizer comparable to that obtained from exhaustive random initializations, in terms of the proximity to the true singular components.

\begin{rema}[Tuning]
We propose a cross-validation (CV) criterion for selecting \( \nu \) in Algorithm~\ref{algo: FSVD}.
For each \( i \), \( \{T_{ij}, Y_{ij}; j \in [J_i]\} \) are randomly split into five folds:
\(
\{T_{ij}, Y_{ij}; j \in [J_i]\} = \bigcup_{m=1}^5 \{T_{ij,m}, Y_{ij,m}; j \in [J_{i,m}]\}, \quad \forall i \in [n],
\)
where each fold \( \{T_{ij,m}, Y_{ij,m}; j \in [J_{i,m}]\} \) is a proper subset of the data.
Let \( \hat{\rho}_1^{(-m)} \), \( \hat{\phi}_1^{(-m)} \), and \( \hat{\bm{a}}_1^{(-m)} \) denote the outputs of Algorithm~\ref{algo: FSVD} trained without the \( m \)th fold.
The CV error is defined as:
\(
\operatorname{CV}(\nu)
=
\frac{1}{5} \sum_{m=1}^5 \sum_{i=1}^n \frac{1}{J_i} \sum_{j=1}^{J_{i,m}}
\left\{ Y_{ij,m} - \hat{\rho}_1^{(-m)} \hat{a}_{i1}^{(-m)} \hat{\phi}_1^{(-m)}(T_{ij,m}) \right\}^2,
\)
where the summand is set to \(0\) if fold \(m\) is empty for subject \(i\).
The optimal \( \nu \) is chosen to minimize \( \operatorname{CV}(\nu) \).
\end{rema}

\begin{algorithm}[h]
\caption{General Procedure of FSVD}\label{algo: FSVD_R}
\footnotesize
\begin{algorithmic}[1]
\State \textbf{Input}
 observed data $\big\{Y_{ij};j\in [J_i],i\in [n]\big\}$ and $K>1$.
 \State Set $\tau$ and $H$.
 \State \textbf{Input} $\hat{\bm{a}}_1^{(0)}$, tuning parameter $\nu_1$.
 \State \textbf{Output} $\hat{\bm{a}}_1$, $\hat{\phi}_1$, and $\hat{\rho}_1$ from Algorithm~\ref{algo: FSVD}.
\State \textbf{For} $r=2,\dots,K$ \textbf{do}
\State \quad\textbf{Input} $\hat{\bm{a}}_r^{(0)}$, tuning parameter $\nu_r$.
\State \quad Calculate ${Y}_{ij}^{(r)}=Y_{ij}-\sum_{l=1}^{r-1}\hat{\rho}_{l}\hat{{a}}_{i{l}}\hat{\phi}_{l}(T_{ij})$, \ $j\in [J_i],i\in [n]$.
\State \quad Implement Algorithm~\ref{algo: FSVD} with $\hat{\bm{a}}_r^{(0)}$, $\big\{{Y}^{(r)}_{ij};j\in [J_i],i\in [n]\big\}$, $\nu_{r}$, $\tau$, and $H$.
\State \quad\textbf{Output} $\hat{\bm{a}}_r$, $\hat{\phi}_r$, $\hat{\rho}_r$.
\State \textbf{End for}
\end{algorithmic}
\end{algorithm}

Based on \eqref{eun: FSVD_s}, the $r$th singular component
can be estimated by sequentially applying Algorithm~\ref{algo: FSVD}, subtracting the previously
estimated $r-1$ components. This procedure is summarized in Algorithm~\ref{algo: FSVD_R}.
The maximum computational complexity of the algorithm is of order
\(
\Big(\sum_{i=1}^n J_i\Big)\cdot M^2 \cdot H K.
\)

\section{FSVD for Learning Tasks}\label{sec: task}

This section discusses the application of FSVD to tasks for trajectory/functional data that potentially connect with our data analysis. A simulation study evaluating FSVD for these tasks is provided in Part~\ref{sec: sim_SM} of the Supplementary Materials.

\subsection{Heterogeneous Dimension Reduction}\label{sec: HDR}
This subsection explores FSVD for dimension reduction in heterogeneous trajectories \(X_i\).
Here, we treat \(X_i\) as random objects, so their singular values, singular functions/vectors, and ranks are also random.
Moreover, heterogeneity means that \(X_i\), \(i \in [n]\), are not i.i.d., or that \(\bm{X}(t),\ t \in \mathcal{T}\), does not form a stationary multivariate time series over \(\mathcal{T}\), where $\bm{X}(t)=(X_1(t),\dots,X_n(t))^\top$.

In the following, we focus on two concepts: intrinsic basis functions (IBFs)
and intrinsic basis vectors (IBVs), which connect the FSVD of \(X_i\)s to various dimension reduction methods.

\begin{D}[Intrinsic Basis Functions]\label{def: intrinsic}
Suppose $X_1,\dots,X_n \in \mathcal{L}^2(\mathcal{T})$ are a sequence of random functions.
For a fixed $K$, the orthonormal basis functions $\{\varphi_k;k\leq K\}$ in $\mathcal{L}^2(\mathcal{T})$ are the intrinsic basis functions of $X_i$s if for any deterministic orthonormal functions $\{\tilde{\varphi}_k;k\leq K\}$ and any random variables $\tilde{\xi}_{ik}$s,
\begin{equation}\label{ineq:intrinsic-basis-function}
\sum_{i=1}^n\mathbb{E}\bigg\|X_i-\sum_{k=1}^K\xi_{ik}\varphi_k\bigg\|^2
\leq
\sum_{i=1}^n\mathbb{E}\bigg\|X_i-\sum_{k=1}^K\tilde{\xi}_{ik}\tilde{\varphi}_k\bigg\|^2,
\end{equation}
where $\xi_{ik}:=\langle X_i,\varphi_k\rangle$, $i\in [n]$ and $k\in [K]$.
\end{D}

\begin{D}[Intrinsic Basis Vectors]\label{def: intrinsic_v}
For random functions $\bm{X}(t),\ t \in \mathcal{T}$ and any fixed $K$, let $\bm{L} = (\bm{l}_1, \dots, \bm{l}_K) \in \mathbb{R}^{n \times K}$ be deterministic orthonormal vectors. These vectors are the intrinsic basis vectors of $\bm{X}$ if
\begin{equation*}
    \int_{\mathcal{T}}\mathbb{E}\big\|\bm{X}(t) - \bm{L} \bm{F}(t)\big\|^2\ \mathrm{d}t
    \leq
    \int_{\mathcal{T}}\mathbb{E}\big\|\bm{X}(t) - \tilde{\bm{L}} \tilde{\bm{F}}(t)\big\|^2\ \mathrm{d}t,
\end{equation*}
where $\bm{F}(t) = \bm{L}^\top \bm{X}(t)$ for $t \in \mathcal{T}$, and $\tilde{\bm{L}} \in \mathbb{R}^{n \times K}$ and $\tilde{\bm{F}}(t) \in \mathbb{R}^K$ consist of any
$K$ deterministic orthonormal vectors in $\mathbb{R}^n$ and any $K$ random functions in $\mathcal{L}^2(\mathcal{T})$, respectively.
\end{D}

The IBF concepts essentially extract the dominant functional patterns in the data, achieving a low-dimensional and parsimonious representation similar to the mean function and eigenfunctions for i.i.d.\ functional data.
Specifically, the IBFs are orthonormal deterministic functions such that the projection of $X_i$s onto these functions achieves the optimal rank-$K$ approximation:
\begin{eqnarray}\label{RKHS_model}
X_i(t)\approx\sum_{k=1}^{K} \xi_{ik} \varphi_k(t),\ t\in \mathcal{T},\ i\in [n].
\end{eqnarray}
The above generalizes the Karhunen--Lo\`eve formula for i.i.d.\ sampled $X_i$s (Theorem~7.3.5 in \citet{hsing2015theoretical}).
A key difference from the i.i.d.\ setting is that the deterministic functions \( \varphi_k \) may vary with \( n \), and \( \{\xi_{ik}; i \in [n]\} \) are not identically distributed and may be correlated across different subjects \( i \) and components \( k \).

Similarly, the IBVs of $X_i$s are deterministic vectors such that the projection of $\bm{X}$ onto these vectors achieves the optimal rank-$K$ dimension reduction, i.e.,
\begin{equation}\label{factor_model}
    \bm{X}(t) \approx \bm{L}\bm{F}(t),\qquad t \in \mathcal{T}.
\end{equation}
This model corresponds to the factor model of time series in the literature \citep{lam2011estimation, lam2012factor}: here, $\bm{X}(t)$ is viewed as a multivariate time series indexed by $t$, $\bm{F}(t)\in \mathbb{R}^{K}$ is the factor process, $K$ is the number of factors, and $\bm{L}\in \mathbb{R}^{n\times K}$ is a factor loading matrix.
A main difference between those in \eqref{factor_model} and \citet{lam2011estimation, lam2012factor} is that $\bm{X}(t)$, $t \in \mathcal{T}$, can be a nonstationary time series; that is, $\mathbb{E}\{\bm{X}(t)\}$ and $\operatorname{Cov}\{\bm{X}(t), \bm{X}(t+s)\}$ can vary with $t$.

Since for any invertible (or orthogonal) matrix $\bm{B} \in \mathbb{R}^{K \times K}$, we have
\(
\bm{L}\bm{F}(t) = (\bm{L}\bm{B}^{-1}) \big\{\bm{B}\bm{F}(t)\big\},\allowbreak \ t \in \mathcal{T},
\)
the factor series and factor loading matrix (or IBVs) of $\bm{X}$ are identifiable only up to an invertible (or orthogonal) transformation. The IBFs of $X_i$s also have a similar identifiability issue.

\begin{rema}[Distinctions between IBFs and IBVs]
    An essential distinction between \eqref{RKHS_model} and \eqref{factor_model} lies in which part of the low-rank structure is treated as deterministic. 
In \eqref{RKHS_model}, the basis functions $\varphi_k$s are deterministic and shared across subjects, so the heterogeneity is absorbed into the subject-specific random scores $\xi_{ik}$; this perspective emphasizes functional-mode dimension reduction, i.e., extracting a common set of temporal patterns that can parsimoniously represent each trajectory. 
In contrast, in \eqref{factor_model}, the loading matrix $\bm{L}$ is deterministic and shared over time, while the factor series $\bm{F}(t)$ is random; this perspective emphasizes subject/feature-mode dimension reduction, i.e., identifying a low-dimensional subspace in the tabular direction that explains the joint evolution of $\bm{X}(t)$ over $t$. 
\end{rema}

Regardless of the framework of IBFs or IBVs, the FSVD \eqref{eq:X-a-xi_1} provides a single decomposition that contains both ingredients: the singular functions $\phi_r$ span the dominant functional subspace, and the singular vectors $\bm{a}_r$ span the dominant tabular subspace. 
In particular, when $K$ is chosen so that the leading $K$ components give a good approximation of $X_i$s in \eqref{RKHS_model} and \eqref{factor_model}, FSVD implies that 
\(\operatorname{span}(\phi_1,\ldots,\phi_K)\approx\operatorname{span}(\varphi_1,\ldots,\varphi_K)\), and 
\(\operatorname{span}(\bm{a}_1,\ldots,\bm{a}_K)\approx\operatorname{span}(\bm{l}_1,\ldots,\bm{l}_K)\).
This means FSVD can be used to extract both IBFs and IBVs from $X_i$'s.

For the case where \(X_i\) is not observed directly but only through its noisy and discrete measurements \(Y_{ij}\), we consider the measurement model \eqref{mea_model} and assume the smoothness condition \(X_i \in \mathcal{W}_2^q(\mathcal{T})\) for \(i \in [n]\). Under these assumptions, we implement FSVD using Algorithm~\ref{algo: FSVD_R} to estimate the IBFs and IBVs, as summarized in Algorithm~\ref{algo: FM}.

\begin{algorithm}
\caption{Dimension Reduction / Factor Model Estimation by FSVD}\label{algo: FM}
\footnotesize
\begin{algorithmic}[1]
\State \textbf{Input}: Observed data $\big\{Y_{ij};\ i \in [n],\ j \in [J_i]\big\}$, rank $K$, and an invertible matrix $\bm{B} = (\bm{b}_1, \dots, \bm{b}_K)\in \mathbb{R}^{K\times K}$.
\State Obtain $\hat{\phi}_k$, $\bm{\hat{a}}_k$, and $\hat{\rho}_k$ for $k \in [K]$ using Algorithm~\ref{algo: FSVD_R} based on the observed data $Y_{ij}$.
\State \textbf{If} estimating IBFs:
    \State \quad Compute $(\hat{\varphi}_1, \ldots, \hat{\varphi}_K) = \bm{B}(\hat{\phi}_1, \ldots, \hat{\phi}_K)^\top$.
    \State \quad \textbf{Output}: $\hat{\varphi}_1, \ldots, \hat{\varphi}_K$.
\State \textbf{Else} for IBVs and factor analysis:
    \State \quad Compute $\hat{\bm{L}} := (\bm{\hat{a}}_1, \ldots, \bm{\hat{a}}_K)\bm{B}^{-1}$ and $\hat{\bm{F}} = \sum_{k=1}^K \hat{\rho}_k \bm{b}_k \hat{\phi}_k$.
    \State \quad \textbf{Output}: $\hat{\bm{L}}$ and $\hat{\bm{F}}$.
\end{algorithmic}
\end{algorithm}

In Section~\ref{sec: heter}, we establish mild conditions that ensure the existence of IBFs and IBVs for random trajectories \(X_i\). We show that the IBF framework is preferable when \(X_i\)s exhibit weak dependencies across trajectories, whereas IBVs are more suitable when \(X_i\)s exhibit strong inter-subject dependencies. The former setting is analogous to FPCA in functional data analysis, while the latter resembles factor models in time series analysis. In both cases, FSVD provides an effective approach for estimating the IBFs or IBVs for the purpose of dimension reduction, as detailed in Section~\ref{sec: heter}.

\subsection{Functional Completion}
FSVD can be directly applied to recover the entire trajectories of \( X_i \) from discrete data \(Y_{ij}\)s:
\begin{equation*}
    \hat{X}_i(t) = \sum_{r=1}^K \hat{\rho}_r \hat{a}_{ir} \hat{\phi}_r(t), 
\quad i \in [n],
\end{equation*}
where \(\hat{\rho}_r\), \(\hat{\phi}_r\), and \(\hat{a}_{ir}\) are obtained from Algorithm~\ref{algo: FSVD_R}. 
This procedure is referred to as functional completion, a common task in the analysis of incompletely observed functional data \citep{yao2005functional, kraus2015components, delaigle2016approximating, kneip2020optimal, nie2022recovering}. 
However, most existing functional completion methods assume that \( X_i \) are i.i.d.\ samples, which limits their applicability in heterogeneous settings. 
In contrast, FSVD is applicable to both homogeneous and heterogeneous cases, as well as to dependent trajectories, through its connections to IBFs and IBVs discussed in Section~\ref{sec: task}.

Functional completion via FSVD requires selecting the number of components \( K \).  
Under the IBF framework, if the noise terms \( \{\varepsilon_{ij}; j \in [J_i]\} \) in \eqref{mea_model} are i.i.d.\ mean-zero Gaussian with variance \( \sigma_i^2 \), we can choose \( K \) by minimizing the Akaike information criterion (AIC):
\begin{equation}\label{AIC}
   \text{AIC}(K) := \sum_{i=1}^n J_i \log(\hat{\sigma}^2_{i,K}) + 2nK,
\end{equation}
where
\(
\hat{\sigma}^2_{i,K} = \frac{1}{J_i} \sum_{j=1}^{J_i} \left\{ Y_{ij} - \sum_{k=1}^K \hat{\rho}_k \hat{a}_{ik} \hat{\phi}_{k}(T_{ij}) \right\}^2.
\)
This AIC is obtained by treating the model \eqref{mea_model} as a linear regression of \( Y_{ij} \) on \( (\hat{\phi}_1(T_{ij}), \dots, \hat{\phi}_K(T_{ij})) \), similar to the approach in \citet{li2013selecting}.  
Under the IBV framework, one may adopt the information criteria from factor models \citep{bai2002determining} to select \( K \).  
Both approaches are commonly used for component selection in functional data analysis, and we refer to \citet{li2013selecting} for a detailed discussion.

\subsection{Functional Clustering}

Next, we connect FSVD with the clustering of heterogeneous functional data, aiming to group the functional objects $X_i$ into distinct clusters \citep{wang2016functional}. A classic approach in the literature involves projecting $X_i$s onto a collection of basis functions \citep{james2003clustering,giacofci2013wavelet}, transforming the functions into vectors that enable the application of various clustering procedures.
Since these procedures require a prior selection of basis functions for the projection, \citet{chiou2007functional,jacques2013funclust} adopted data-driven methods to determine basis functions using eigenfunctions derived from FPCA. 
Here, we develop a new method for functional clustering using FSVD.

We assume that $X_i$s are independent but non-identically distributed random functions valued in $\mathcal{W}_2^q(\mathcal{T})$.
Following the settings in \citet{james2003clustering, giacofci2013wavelet}, 
we assume that 
\begin{equation}\label{model_clu}
Y_{ij}=X_i(T_{ij})+\varepsilon_{ij}=\sum_{k=1}^K\xi_{ik}\varphi_{k}(T_{ij})+\varepsilon_{ij},
\end{equation}
where $\varphi_k$s are deterministic basis functions, $\xi_{ik}$s are random scores, $\varepsilon_{ij}$s are mean-zero white noise independent of $X_{i}$s, and $T_{ij}$ can vary across $i$. Here, $\{\bm{\xi}_i := (\xi_{i1}, \dots, \xi_{iK})^\top; i \in [n]\}$ can be grouped into $H$ distinct clusters, with $Z_{i}$ denoting the cluster membership for the $i$th function. 

To obtain $Z_{i}$, we assume $Z_{1},\ldots,Z_n$ are i.i.d.\ latent variables following a multinomial distribution on $\{1,\ldots,H\}$ with 
$\mathbb{P}(Z_{i}=h)=\pi_h$, consistent with the model settings of \citet{james2003clustering, giacofci2013wavelet}.
For $Z_i=h$ in the $h$th cluster, we assume $\bm{\xi}_i\sim \operatorname{N}(\bm{\mu}_h,\bm{\Sigma}_h)$ and $\varepsilon_{ij}\sim \operatorname{N}(0,\sigma_h^2)$,
with $\bm{\mu}_h\in \mathbb{R}^{K}$ and $\bm{\Sigma}_h\in \mathbb{R}^{K\times K}$ as the mean and covariance matrix for $\bm{\xi}_i$s, and $\sigma_h^2$ as the variance of white noise.
Accordingly, $X_i$s in the $h$th cluster share the mean function $(\bm{\varphi}(t))^\top\bm{\mu}_h$ and the covariance function $(\bm{\varphi}(t))^\top\bm{\Sigma}_h\bm{\varphi}(s)$, and
\begin{eqnarray}\label{EM_Y}
      \bm{Y}_{i}\sim \operatorname{N}(\bm{\varphi}_i^\top\bm{\mu}_h,\bm{\varphi}_i^\top\bm{\Sigma}_h\bm{\varphi}_i+\sigma_h^2\bm{I}_{J_i})\ \text{if}\ Z_i=h,
\end{eqnarray}
where $\bm{\varphi}(t)=(\varphi_1(t),\dots,\varphi_K(t))^\top$, $\bm{Y}_i=(Y_{i1},\dots,Y_{iJ_i})^\top$, $\bm{\varphi}_i=(\bm{\varphi}(T_{i1}),\dots,\bm{\varphi}(T_{iJ_i}))\in \mathbb{R}^{J_i\times K}$. Under this setting, we employ an EM algorithm to estimate \( \mathbb{P}\{Z_i = h \mid \bm{Y}_i\} \) for \( h \in [H] \) and \( i \in [n] \) based on model~\eqref{EM_Y}. The detailed algorithm can be found in \citet{james2003clustering, giacofci2013wavelet}.

Unlike the existing literature \citep{james2003clustering, giacofci2013wavelet}, 
we do not pre-specify $\varphi_k$s in model~\eqref{model_clu}. 
Instead, we take $\varphi_k$s as the IBFs of $X_i$ and estimate them directly from the observations $Y_{ij}$ using Algorithm~\ref{algo: FM}.  
By the definition of IBFs \eqref{ineq:intrinsic-basis-function}, 
the number of basis functions we employ is minimal, 
thus avoiding the additional conditions required in \citet{james2003clustering, giacofci2013wavelet} 
to mitigate the effects of using an excessively large number of basis functions.  
We refer to this method as the FSVD-EM clustering algorithm; the detailed procedure is provided in Algorithm~\ref{algo: FC} of the Supplementary Materials.

The above method is designed for $X_i$s that share a common functional pattern. 
For more general settings where cluster-specific patterns are present, 
we may embed the FSVD algorithm within a functional clustering procedure, similar to FPCA-based clustering \citep{chiou2007functional}. 
Specifically, we first obtain the clustering membership using the FSVD-EM clustering algorithm. 
Next, we update the IBFs for each cluster separately and then re-cluster the functional data based on the updated IBFs. 
This procedure strengthens the shared patterns of functional data within each cluster, which may lead to more accurate estimation of IBFs, as suggested by Theorem~\ref{Theorem_FSVD} below.

\subsection{Functional Linear Regression}

The goal of functional linear regression is to model and capture the linear relationship between functional predictors and responses \citep{yao2005functional_reg,yuan2010reproducing,morris2015functional,reiss2017methods,imaizumi2018pca}. In particular, let \(\{X_i; i\in [n]\} \subseteq \mathcal{L}^2(\mathcal{T})\) denote the independent functional predictors defined on a domain \(\mathcal{T}\), and consider the following model:
\begin{eqnarray}\label{FLR_model}
    Z_i=\alpha+\langle \beta,X_i\rangle+\vartheta_i,\ i\in [n],
\end{eqnarray}
where \(Z_i \in \mathbb{R}\) is a scalar response, \(\alpha \in \mathbb{R}\) is an intercept, \(\beta \in \mathcal{L}^2(\mathcal{T})\) is the unknown coefficient function, and \(\vartheta_i\) is a noise term with finite variance. Our objective is to estimate \(\beta\) based on the responses \(\{Z_i;i\in [n]\}\) and discrete, noisy observations of the functional predictors \(\{X_i;i\in [n]\}\).

A variety of methods have been proposed for functional linear regression. One popular class, known as penalized functional regression (PFR), employs basis expansions or RKHS representations, coupled with regularization \citep{yuan2010reproducing,goldsmith2011penalized,goldsmith2012longitudinal,zhao2012wavelet,luo2024functional}.  
Although effective for densely sampled data, PFR methods can be less suitable for irregularly observed 
functional data in longitudinal settings \citep{reiss2017methods}. Another line of work applies FPCA to $X_i$s to extract basis functions, which are then substituted into \eqref{FLR_model} to estimate the coefficient \(\beta\) \citep{yao2005functional_reg,cai2006prediction}. However, these FPCA-based methods often assume i.i.d.\ functional data, which may not hold in practice.

The limitations above can be overcome by using FSVD for functional regression.
We first apply the FSVD to the predictors \(\{X_i;i\in [n]\}\) in model~\eqref{FLR_model}. This yields
\begin{eqnarray}\label{FLR_estimate}
    Z_i = \alpha + \sum_{r=1}^R \rho_r\, a_{ir}\,\langle \beta, \phi_r \rangle + \vartheta_i := \alpha + \sum_{r=1}^R \xi_{ir}\,\beta_r + \vartheta_i,
    \quad i \in [n],
\end{eqnarray}
where \(\xi_{ir} := \rho_r\, a_{ir}\) and \(\beta_r = \langle \beta, \phi_r\rangle\). Here, \(\{\phi_r;r\in [R]\}\) are the singular functions of \(\{X_i;i\in [n]\}\), and \(\beta_r\) is the projection of \(\beta\) onto \(\phi_r\). Suppose we only observe discrete and noisy samples $\{Y_{ij};j\in [J_i]\}$ from each \(X_i\). To estimate \(\beta\), we first apply Algorithm~\ref{algo: FSVD_R} to estimate \(\hat{\xi}_{ir} := \hat{\rho}_r\,\hat{a}_{ir}\) and \(\hat{\phi}_r\), for \(i \in [n]\) and \(r \in [K]\), and then substitute these into model~\eqref{FLR_estimate}. Subsequently, we can perform a least squares regression of $Z_i$ on \((\hat{\xi}_{i1}, \dots, \hat{\xi}_{iK})^\top\), $i\in [n]$, to estimate \(\hat{\alpha}\) and \(\{\hat{\beta}_r;r\in [K]\}\) and reconstruct \(\hat{\beta}\) as $\sum_{r=1}^K \hat{\beta}_r\, \hat{\phi}_r$. 
This process is summarized in Algorithm~\ref{algo: FLR}.

\begin{algorithm}[h]
\caption{Functional Linear Regression by FSVD}\label{algo: FLR}
\footnotesize
\begin{algorithmic}[1]
\State \textbf{Input:} Discrete and noisy observations \(\{Y_{ij} : j \in [J_i]\}\) of each \(X_i\), corresponding responses \(Z_i\), and the number of components \(K\).
\State Apply Algorithm~\ref{algo: FSVD_R} to $Y_{ij}$s to obtain \(\hat{\rho}_r\), \(\hat{a}_{ir}\), and \(\hat{\phi}_r\) for \(i \in [n]\) and \(r \in [K]\). Set \(\hat{\xi}_{ir} := \hat{\rho}_r\, \hat{a}_{ir}\).
\State Perform a least squares regression of \(Z_i\) on \(\{\hat{\xi}_{i1},\dots,\hat{\xi}_{iK}\}\) to obtain the estimates \(\hat{\alpha}\) and \(\{\hat{\beta}_r;r\in [K]\}\).
\State \textbf{Output:} \(\hat{\alpha}\) and \(\hat{\beta} = \sum_{r=1}^K \hat{\beta}_r\,\hat{\phi}_r\).
\end{algorithmic}
\end{algorithm}

\setlength{\topsep}{0pt}
\begin{rema}[Identifiability in Functional Linear Regression]
Unlike classical linear regression with finite-dimensional predictors, the functional coefficient \(\beta\) lies in an infinite-dimensional space. Suppose \(\beta\) is decomposed as 
\(
\beta = \sum_{r=1}^K \beta_r \phi_r + \beta_\perp,
\) 
where $\phi_r$s are the singular functions of $X_i$s, and \(\beta_\perp\) is the remainder term orthogonal to \(\operatorname{span}\{\phi_1, \dots, \phi_K\}\). When $X_i$s are assumed to be fixed, we have \(\langle \beta_\perp, X_i \rangle = 0\) for all \(i \in [n]\), so \(\beta_\perp\) has no influence on the functional regression model. Therefore, only the projection of \(\beta\) onto \(\operatorname{span}\{X_1, \dots, X_n\} = \operatorname{span}\{\phi_1, \dots, \phi_R\}\) is identifiable. To address this, we assume that \(\beta \in \operatorname{span}\{\phi_1, \dots, \phi_R\}\), a subspace where \(\beta\) is fully represented.
The bases of this subspace can be estimated via Step~2 of Algorithm~\ref{algo: FLR}.

For the random design, the identifiability issue can be addressed by assuming \( \beta \in \operatorname{span}\{\psi_k; k \geq 1\} \), where \( \psi_k \)s are the IBFs of \( X_i \)s. This assumption is reasonable since the identifiable subspace \( \operatorname{span}\{X_1, \dots, X_n\} \) can be optimally approximated by the IBFs in the sense of~\eqref{ineq:intrinsic-basis-function}. These functions, as in the fixed design, can be estimated via FSVD.
In contrast, FPCA-based methods \citep{cai2006prediction,hall2007methodology} impose stronger assumptions that \( \beta \) lies in the space spanned by the eigenfunctions of \( X_i \)s, which may not hold when \( X_i \)s are non-i.i.d., as the eigenfunctions are not well-defined in such cases.
\end{rema}

\section{Real Data Analysis}\label{sec:dat_ana}

We illustrate the practical utility of FSVD on two health trajectory datasets described in Section~\ref{sec:data_illu}. 
Our goals are to (i) summarize heterogeneous trajectories with a small number of interpretable phases, (ii) compare trajectories based on these phases, and (iii) improve trajectory reconstruction and prediction when observations are irregular or partially missing.

\paragraph*{Pattern Discovery of Epidemic Trajectory Data}

\begin{figure}[h]
\begin{center}
\includegraphics[scale = 0.45, trim=0cm 0.3cm 0cm 0cm, clip]{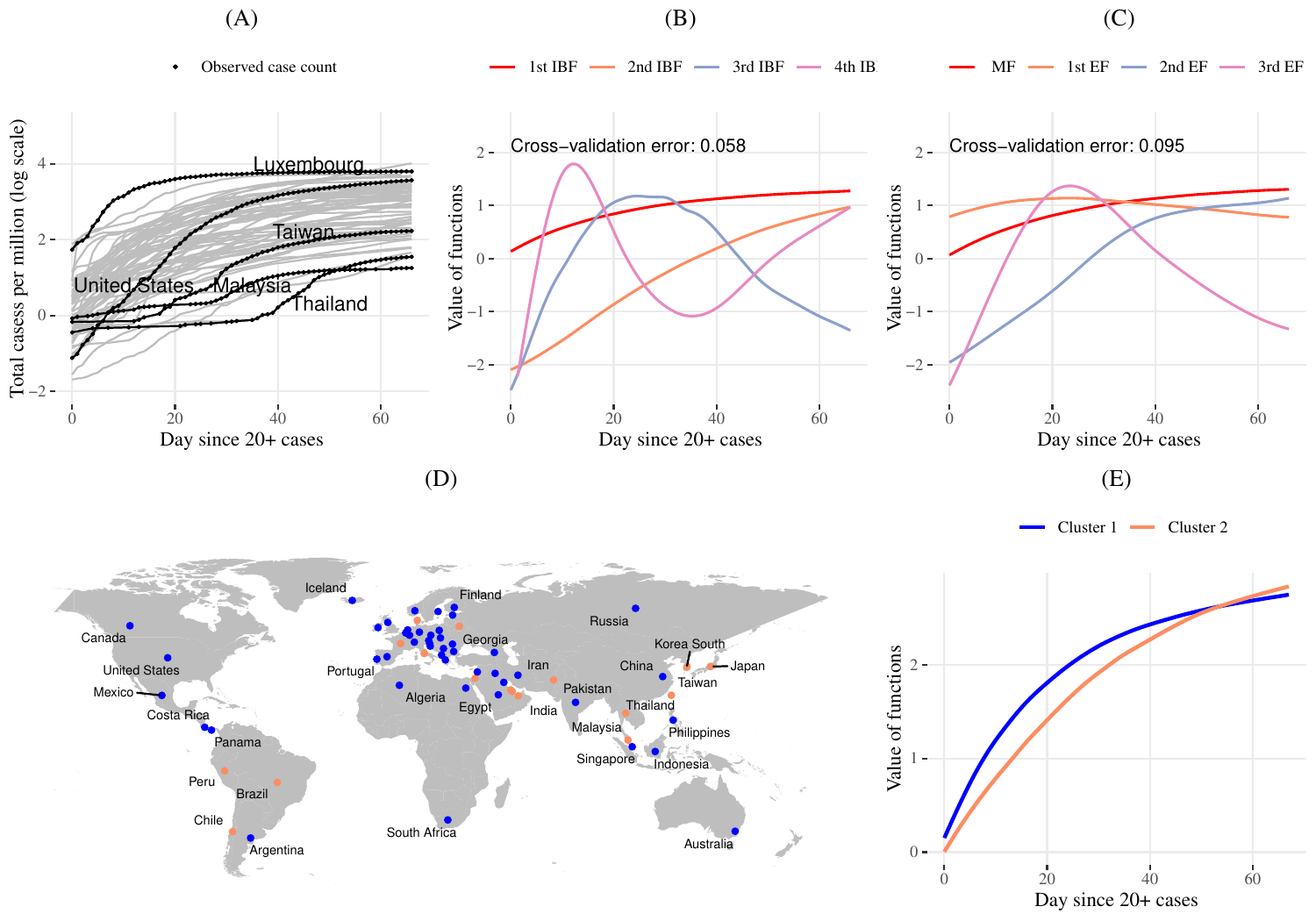}
\end{center}
\caption{\textbf{(A)} Irregularly observed data across different regions; \textbf{(B)} estimated intrinsic basis functions (IBFs) using Algorithm~\ref{algo: FM}, where $\bm{B}$ for the IBFs is chosen as the identity matrix, consistent with the theory in Section~\ref{sec: intrinsic}; \textbf{(C)} estimated mean function after normalization (MF) and estimated eigenfunctions (EFs) from FPCA; \textbf{(D)} clustering map for the dynamics from different regions; \textbf{(E)} estimated mean functions of two clusters.} \label{COVID19_tra}
\end{figure}

We analyze the 64 COVID-19 epidemic trajectories from \citet{carroll2020time} shown in Figure~\ref{COVID19_tra}(A). 
Such region-level case-count curves are a core input for comparative surveillance, but they can be heterogeneous due to differences in reporting practices, population structure, and the timing and intensity of public health measures \citep{tian2021effects,tan2023age}. 
Consistent with this, while many regions exhibit broadly similar growth trends, others (e.g., Luxembourg and Thailand) display distinct rise-and-slowdown patterns. 
Whereas \citet{carroll2020time} applied FPCA under a common-population assumption, we use FSVD to explicitly accommodate between-region heterogeneity and extract shared, recurring epidemic progression phases.

Figures~\ref{COVID19_tra}(B)--(C) compare the leading temporal patterns extracted by FSVD (IBFs) and FPCA (mean function and eigenfunctions). 
The IBFs provide an interpretable summary of epidemic progression phases shared across regions: (i) an overall growth baseline, together with (ii) early-stage acceleration, (iii) mid-course slowing/plateauing, and (iv) regime-switch behavior characterized by trend changes. 
Compared with FPCA, FSVD captures additional changes in trend (e.g., around days 15 and 35 in the 4th IBF, beyond the change around day 20 detected by both methods), which helps describe regions such as Thailand, Taiwan, and Luxembourg where the timing of exponential growth and plateau phases differs. 
In practice, these phases provide a compact way to compare regions---differences across regions are reflected in the strength and timing of a small number of shared patterns.

To quantify the practical benefits of these representations, we compare FSVD and FPCA on two tasks relevant to epidemic monitoring: trajectory reconstruction under partial observation and prediction of later outcomes from early dynamics.  
For trajectory reconstruction, we order each region's observations and form five cyclic folds to ensure coverage across the time frame; four folds are used for training and the remaining fold for evaluation. 
For prediction, we use each region's first 60 days as a functional predictor and the post-60-day mean as the response, and compare FSVD-based regression (Algorithm~\ref{algo: FLR}) with standard FPCA-based regression \citep{yao2005functional_reg,cai2006prediction,hall2007methodology} using leave-one-region-out cross-validation.

\begin{table}[h]
\centering
\caption{Average cross-validation mean squared errors for FSVD and FPCA across tasks.}
\label{CV_error_task}
\renewcommand{\arraystretch}{1.2}
\setlength\tabcolsep{10pt}
\footnotesize
\begin{tabular}{lcc}  
  \hline
  \textbf{Task} & \textbf{FPCA} & \textbf{FSVD} \\
  \hline
  Trajectory reconstruction & 0.0949 & 0.0662 \\
  Early-to-late prediction & 0.0175 & 0.0063 \\
  \hline
\end{tabular}
\end{table}

Table~\ref{CV_error_task} shows that FSVD improves performance in both tasks: it reduces reconstruction error by 30.24\% relative to FPCA and reduces prediction error by 64.00\%. 
These gains indicate that explicitly modeling heterogeneity yields a representation that is more effective for downstream epidemic monitoring tasks.

Finally, we use FSVD--EM clustering to group regions by their epidemic dynamics (Figure~\ref{COVID19_tra}(D)--(E)), yielding two clusters with clear geographic tendencies. 
Cluster~1, more common in Europe and North America and present in parts of Africa and Oceania, shows faster early growth followed by an earlier slowdown. 
Cluster~2, more prevalent in South America, South Asia, Southeast Asia, and parts of East Asia, shows slower early progression with a more prolonged increase before slowing. 
These results suggest that the timing of the growth-to-stabilization transition can be broadly shared within geographic regions yet heterogeneous across regions, plausibly reflecting differences in reporting practices and the timing/intensity of control measures and recurrent waves \citep{tian2021effects}.

In Part~D.3 of the Supplementary Materials, we re-estimate IBFs within each cluster and find that they are nearly identical to those in Figure~\ref{COVID19_tra}(B), indicating that regions largely share common progression phases and differ primarily in how strongly and when these phases manifest.

\paragraph*{Factor Analysis of Longitudinal Electronic Health Records}

We analyze patient-level longitudinal ICU electronic health record (EHR) data from \citet{johnson2020mimic}, where labs and vitals are measured irregularly and with substantial missingness. Such incomplete trajectories can hinder downstream tasks such as monitoring, risk modeling, and phenotyping, making accurate reconstruction and low-dimensional summaries practically important. For each patient, we retain clinical features with more than four nonzero measurements, and include patients with at least two such features, yielding 3705 patients. We apply FSVD with a factor-model representation to each patient to obtain patient-specific latent factor series and feature loadings, and assess performance on trajectory reconstruction.

\begin{figure}[h]
    \centering
    \includegraphics[width=0.5\linewidth]{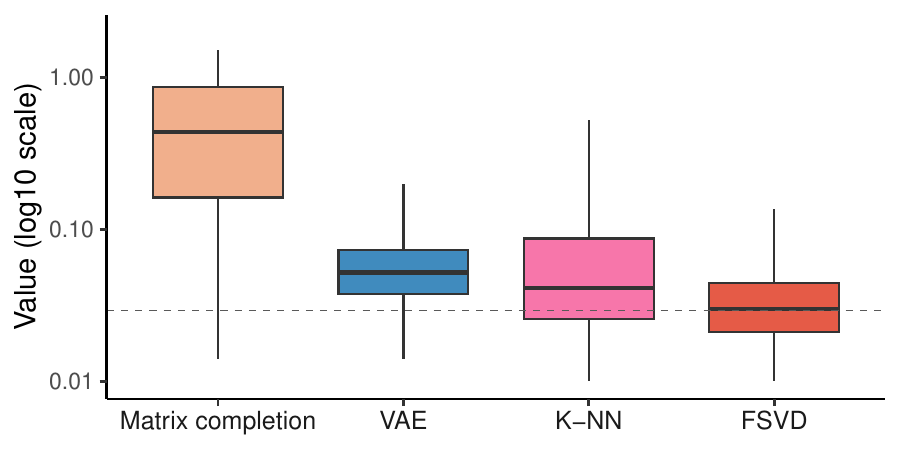}
    \caption{Boxplot of the imputation errors from 3705 patients for the four methods. 
    The horizontal dashed line indicates the median error from FSVD.}
    \label{fig:imputation}
\end{figure}

We compare FSVD with commonly used approaches for imputing heterogeneous clinical data, including matrix completion \citep{candes2012exact}, variational autoencoders (VAE; \citealp{nazabal2020handling}), and the $k$-nearest neighbor (K-NN) method \citep{bertsimas2018predictive}. For matrix completion, VAE, and K-NN, imputation is restricted to the grid of observed time points $\bigcup_{j \in [J_i]} \{T_{ij}\}$, whereas FSVD yields smooth reconstructions over the entire observation window, which is useful for summarizing trajectories at clinically meaningful times.

We evaluate each method using reconstruction error. For each patient, we split observations evenly into two folds to ensure balanced coverage across the time frame, fit each method on one fold, and compute mean squared error on the held-out fold. Figure~\ref{fig:imputation} shows the distribution of errors across 3705 patients; FSVD achieves lower reconstruction errors overall, indicating improved recovery of irregular EHR trajectories.

\begin{figure}[ht!]
\centering
\includegraphics[scale = 0.6]{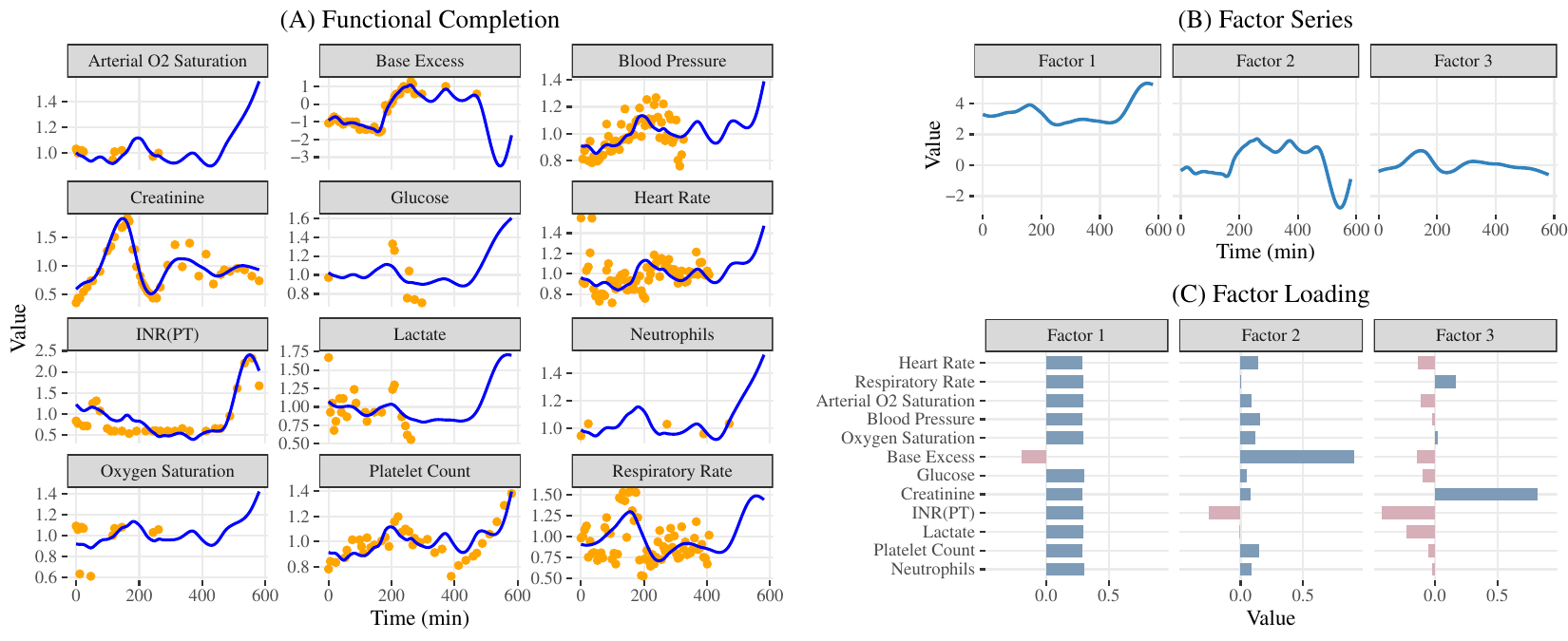}
\caption{\textbf{(A)} Longitudinal trajectories for 12 clinical features from a patient and their functional completion by FSVD. \textbf{(B)} The estimated factor series and 
 \textbf{(C)} the corresponding factor loadings for the electronic health record from a patient.} \label{Factor_mod_fig}
\end{figure}

\begin{figure}[h]
\begin{center}
\includegraphics[scale = 0.45,trim=0cm 0cm 0cm 0cm, clip]{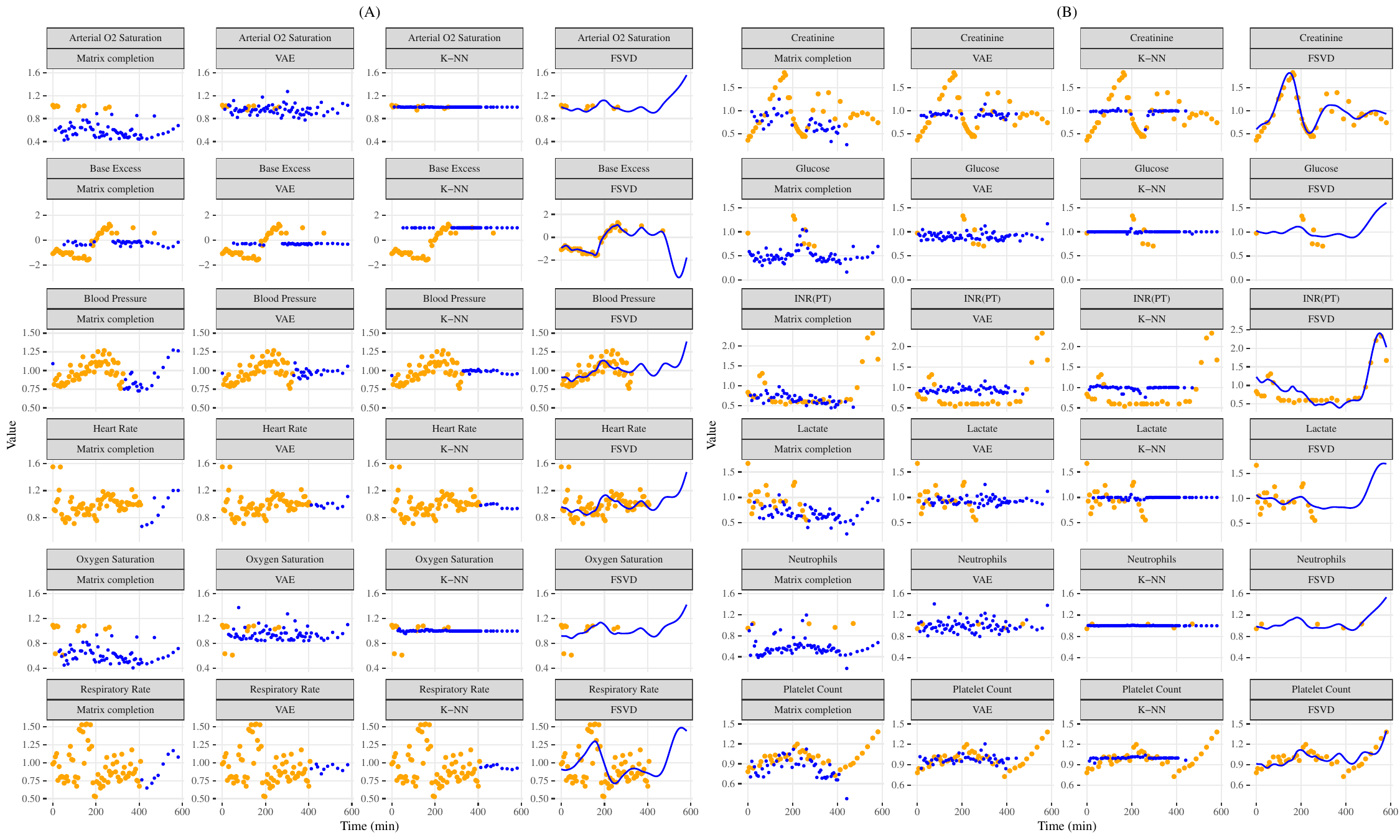}
\end{center}
\caption{Data imputation/functional completion for 12 clinical features by matrix completion, VAE, K-NN, and FSVD.}\label{Com_func}
\end{figure}

To illustrate interpretability at the patient level, we examine the example in Figure~\ref{fig:data_combine_B}(A). Using the information criterion of \citet{bai2002determining}, we select three latent factors from the 12 clinical features and estimate the factor loading matrix via Algorithm~\ref{algo: FM} (with $\bm{B}$ set to the identity matrix). Figure~\ref{Factor_mod_fig}(A) shows that FSVD provides smooth and clinically plausible reconstructions; Figure~\ref{Com_func} compares with competing methods.

Figures~\ref{Factor_mod_fig}(B)--(C) display the first three latent factor series and their corresponding feature loadings. The first factor loads broadly across multiple features and increases around 400 minutes after ICU admission, capturing shared co-movement among physiologic measures (e.g., Platelet Count, Heart Rate, and Respiratory Rate; Figure~\ref{Factor_mod_fig}(A)). The second factor reflects changes in Base Excess around 200 minutes and INR (PT) around 550 minutes; since these markers relate to acid--base balance and coagulation, the factor may represent a latent axis of blood chemistry and clotting dynamics.

Overall, FSVD highlights coordinated temporal changes across multiple clinical measures, with a notable transition during approximately 400--550 minutes after ICU admission in this example. This illustrates how FSVD can provide interpretable low-dimensional summaries of multivariate EHR trajectories and flag time windows of joint change for further clinical review; establishing clinical significance would require validation across patients and outcomes.

\section{Theory of FSVD}\label{sec: heter}
In this section, we establish the fundamental theory for dimension reduction of heterogeneous trajectory data using FSVD. We first propose the following theorem.

\begin{theo}[Existence of Functional Singular Value Decomposition]\label{th:functional-SVD}
Let \( \mathcal{H} \subseteq \mathcal{L}^2(\mathcal{T}) \) be a Hilbert space. Suppose $X_1,\ldots, X_n \in \mathcal{H}$. Then there exists an FSVD of $X_1,\ldots, X_n$:
\begin{equation}\label{eq:X-a-xi}
        X_i(t) = \sum_{r \ge 1} \rho_r a_{ir} \phi_r(t),\ i \in [n].
    \end{equation}
Here, $\phi_r$ and $\bm{a}_r$ are the $r$th eigenfunction of the kernel $\frac{1}{n}\sum_{i=1}^nX_i(t)X_i(s)$ and the $r$th eigenvector of the matrix $\int_{\mathcal{T}}\bm{X}(t)\bm{X}^\top(t)\ \mathrm{d}t$, respectively, corresponding to the eigenvalue $\rho_r^2$.
\end{theo}

Theorem~\ref{th:functional-SVD} can be proven by viewing functional data as an operator
$\mathcal{X}_n: \mathcal{H}\to\mathbb{R}^n$,
\(
\mathcal{X}_n: f \mapsto  \big(\langle X_1,f\rangle,\dots,\langle X_n,f\rangle\big)^\top,\ f\in \mathcal{H}.
\)
The compactness of $\mathcal{X}_n$ then leads to the FSVD of $\{X_i\}_{i=1}^n$ (see Part~A.1.1 of the Supplementary Materials for the derivations).

It can be seen that Theorem~\ref{th:functional-SVD} is applicable to both fixed and random designs for trajectory data $X_i$, and that the FSVD exists under mild conditions; importantly, this existence does not depend on population-level information about $X_i$s, making FSVD applicable to a wide range of settings.

In the remaining, we connect FSVD for estimating IBFs and IBVs of $X_i$s under random designs, exploring how FSVD can reveal the population-level structure of the trajectories.
The following focuses on the theory for fully observed trajectory data $X_i$, while the case of discretely observed data is discussed in Part~A.3 of the Supplementary Materials.

\subsection{Intrinsic Basis Functions}\label{sec: intrinsic}

This subsection explores the connection between IBFs and FSVD. For identifiability, we focus on the IBFs such that Definition~\ref{def: intrinsic} holds for any \(K\), rather than for a fixed \(K\). This leads to an ordered sequence of IBFs \(\{\varphi_k : k \geq 1\}\) where \(\{\varphi_k : k \leq K\}\) is always the first \(K\) IBFs of $X_i$s. The following equivalent conditions confirm the existence of the IBFs.

\begin{theo}\label{the: eui_in}
Assume $\{X_i(t);t\in \mathcal{T}\}$, $i\in[n]$, are mean-square continuous processes (i.e., the mean functions and covariance functions are continuous).
Then the following conditions are equivalent:
\begin{itemize}[itemsep=1pt, topsep=2pt, partopsep=2pt]
    \item[a.] The deterministic orthonormal basis functions $\{\varphi_k;k\geq 1\}$ are the IBFs of $X_i$s.
    \item[b.]  $\{\varphi_k;k\geq 1\}$ are eigenfunctions of the kernel $H_n(t,s):=\frac{1}{n}\,\mathbb{E}\sum_{i=1}^nX_i(t)X_i(s)$. 
    \item[c.] The orthonormal basis functions $\{\varphi_k;k\geq 1\}$ satisfy $\sum_{i=1}^n\mathbb{E}\,\xi_{ik_1}\xi_{ik_2}=0$ whenever $k_1\neq k_2$, where $\xi_{ik}:=\langle X_i,\varphi_k\rangle$, $i\in [n]$ and $k\geq 1$.
\end{itemize}   
\end{theo}

By b. in Theorem~\ref{the: eui_in}, the IBFs are unique up to a sign flip if the eigenvalues of $H_n$ are distinct. 

Define the empirical kernel
\(
\hat{H}_n(t,s):=\frac{1}{n}\sum_{i=1}^n X_i(t)X_i(s),
\)
which can be viewed as a noisy approximation to the kernel $H_n(t,s)$. Theorem~\ref{th:functional-SVD}
shows that the singular functions of $X_i$s, denoted by $\phi_k$s, are
the eigenfunctions of $\hat{H}_n(t,s)$. Therefore, by the equivalence of a. and b. in
Theorem~\ref{the: eui_in}, we can use the singular functions $\phi_k$ of $X_i$s to estimate their IBFs
$\varphi_k$.

To measure errors for IBFs, we define the sine distance between pairs of functions: 
\(
\operatorname{dist}(f, g) 
= \sqrt{1 - \left(\frac{\langle f, g \rangle}{\|f\| \cdot \|g\|}\right)^2},
\quad \forall f, g \in \mathcal{L}^2(\mathcal{T}).
\)
The distance between the IBFs and the singular functions of $X_i$s is quantified as follows.

\begin{theo}[Upper Bound on the Distance between Singular Functions and Intrinsic Basis Functions]\label{Theorem_FSVD}
Assume the conditions in Theorem~\ref{the: eui_in} hold. 
Let $X_i$, $i \in [n]$, be functional data (not necessarily independent or identically distributed), satisfying
\(
\sup_{i \in [n]} \mathbb{E}\|X_i\|^4 \leq C,
\)
for some constant $C$ independent of $n$.  
Suppose that for $k_1, k_2 \leq K$ with finite $K>1$,
\begin{equation}\label{condi_iden}
    \inf_{1 \leq k_1 \neq k_2 \leq K+1} 
    \left| \sum_{i=1}^n \mathbb{E}\,\xi_{ik_1}^2 
    - \sum_{i=1}^n \mathbb{E}\,\xi_{ik_2}^2 \right| 
    \gtrsim n^{1-2\delta}
\end{equation}
holds for some $0 \leq \delta < 1/4$.  
Then, for $1 \leq k \leq K$, with high probability,
\begin{equation*}
    \operatorname{dist}(\phi_k, \varphi_k) 
    \;\lesssim\; n^{2\delta -1/2}+n^{2\delta}\varrho,
\end{equation*}
where  
\(
\varrho 
:= \sup_{1 \leq i_1 < i_2 \leq n}
\sqrt{\frac{|\mathbb{E}Y_{i_1,i_2}|}
     {\sqrt{\mathbb{E}{Y_{i_1,i_1}}\cdot {\mathbb{E}Y_{i_2,i_2}}}}} \in [0,1],
\)
with
\(
Y_{i_1,i_2}
:= \int_0^1 \int_0^1 
\chi_{i_1}(t,s)\,\chi_{i_2}(t,s) \ \mathrm{d}t \,\mathrm{d}s, 
\) and 
\(
\chi_i(t,s) := X_i(t)X_i(s) - \mathbb{E}\big[X_i(t)X_i(s)\big].
\)
\end{theo}

The assumption \eqref{condi_iden} generalizes the eigen-gap condition commonly used in the functional data literature \citep{yao2005functional, hsing2015theoretical}, introducing a separation structure for the first $K$ IBFs shared across the \( n \) functions. 
Here, $\delta$ quantifies the strength of the shared structure, while $\varrho$ measures the dependencies across $X_i$s. 
Smaller values of $\delta$ and $\varrho$ correspond to stronger common patterns and weaker inter-function dependencies, leading to more accurate approximation of IBFs using $\phi_k$s.
In particular, when $\delta = \varrho = 0$, $\phi_k$s achieve root-$n$ consistency. 

\begin{rema}[Comparison of IBF and FSVD with Existing Base Concepts in Functional Data]

Functional data are inherently infinite-dimensional random variables, usually observed with noise and often sampled at irregular time points. Data-adaptive deterministic basis functions, such as the proposed IBFs, can efficiently capture important shared structures across these functional observations. They provide a low-dimensional representation that does not depend on the randomness of the samples, making them essential for tasks like functional regression and clustering.

Many basis function concepts in functional data analysis are closely related to the proposed IBFs. 
When \( X_1, \dots, X_n \) are i.i.d.\ centered random functions, the IBFs \( \varphi_k \) reduce to the eigenfunctions of the covariance function \( \operatorname{Cov}\{X_i(t), X_i(s)\} \), and \eqref{RKHS_model} becomes the FPCA of i.i.d.\ functional data \citep{hsing2015theoretical}.
Unlike FPCA, which requires estimating the covariance function of \( X_i \)s, FSVD avoids this step via Algorithm~\ref{algo: FSVD}.
This makes FSVD particularly useful when the covariance is difficult to estimate, such as in cases where the observed time points are relatively sparse. 

For the non-i.i.d.\ case, separability is a common condition used to capture shared patterns across random functions \( X_i \), assuming that their cross-covariance 
\(\operatorname{Cov}\{X_{i_1}(t), X_{i_2}(s)\}\) exhibits a separable structure \citep{zapata2022partial, liang2021modeling}. When \( X_1, \dots, X_n \) are centered, \citet{zapata2022partial} introduced a weak form of separability by assuming the existence of deterministic orthonormal functions \( \{\varphi_k; k \geq 1\} \) such that \( \mathbb{E}[\xi_{i_1 k_1} \xi_{i_2 k_2}] = 0 \) for all \( i_1, i_2 \in [n] \) when \( k_1 \neq k_2 \). These are called partially separable KL basis functions, capturing dominant functional patterns in the data \citep{zapata2022partial}. By the equivalence of a. and c. in Theorem~\ref{the: eui_in}, these partial bases are the IBFs.

In essence, IBFs subsume classical bases: for i.i.d.\ data, they coincide with FPCA eigenfunctions; and if the functional data have a separable covariance structure, they coincide with the separable KL basis (see Table~\ref{basis_comparison} below for a summary of special cases). 
These connections demonstrate that IBFs unify basis representations and provide an efficient framework for estimating bases across a wide range of functional data settings.

\end{rema}

\begin{table}[h]
\centering
\caption{Summary of forms of intrinsic basis functions under different settings.}
\label{basis_comparison}
\renewcommand{\arraystretch}{1.2}
\setlength\tabcolsep{6pt}
\footnotesize
\begin{tabular}{lll}
  \hline
  \textbf{Setting} & \textbf{Reference} & \textbf{Intrinsic basis functions becomes} \\ 
  \hline
  i.i.d.\ centered functional data & \citet{hsing2015theoretical} & Eigenfunctions \\
  i.i.d.\ non-centered functional data & \citet{nie2022recovering} & Optimal empirical basis functions \\
  Separable centered functional data & \citet{zapata2022partial} & Partial separable KL basis functions \\
  \hline
\end{tabular}
\end{table}

\subsection{Intrinsic Basis Vectors}
\label{sec:factor}

Note that the estimation of IBFs requires weak dependence among \( X_i \)s. 
This assumption may be violated when \( X_i \)s are collected from highly correlated features, leading to inaccurately estimated IBFs. 
For example, consider the extreme case where \( X_i = \xi_{1}\varphi_1 + \xi_{2}\varphi_2 \), \( i \in [n] \), 
with random coefficients \( \xi_1 \) and \( \xi_2 \), and IBFs \( \varphi_1 \) and \( \varphi_2 \). 
In this scenario, \( X_i \)s have arbitrary singular functions that depend on data randomness and fail to recover the true IBFs. 
As a result, the dependencies among functions undermine the interpretability of IBFs estimated via FSVD.

To address the above issue, we introduce the intrinsic basis vectors (IBVs), which instead capture the underlying structures in the tabular mode of functional data rather than in the functional mode.

The following theorem demonstrates that IBVs generally exist and can be derived from FSVD.

\begin{theo}\label{the: fac}
Assume $\sup_{i\in [n]}\mathbb{E}\|X_{i}\|^2\leq C$.
For any $K\leq n$, $\bm{L}\in \mathbb{R}^{n\times K}$ are the IBVs of $\{X_i(t); i\in [n],t\in \mathcal{T}\}$ if and only if there exists an orthogonal matrix $\bm{B}\in \mathbb{R}^{K\times K}$ such that $\bm{L}\bm{B}$ are the top-$K$ eigenvectors of $\mathbb{E}\int_{\mathcal{T}}\bm{X}(t)\bm{X}^\top(t)\, \mathrm{d}t$. 

Let $K = \operatorname{rank}\!\left(\mathbb{E}\int_{\mathcal{T}} \bm{X}(t)\bm{X}^\top(t)\, \mathrm{d}t\right) \geq 1$.  
The following conditions are equivalent:
\begin{itemize}[itemsep=1pt, topsep=2pt, partopsep=2pt]
    \item[a.] The deterministic orthogonal vectors $(\bm{l}_1,\dots,\bm{l}_K):=\bm{L} \in \mathbb{R}^{n \times K}$ are the IBVs of $X_i$s.
    \item[b.] \(
\mathbb{P}\!\left\{\, 
   \bm{X}(t) = \bm{L}\bm{F}(t) \ \text{for almost every } t \in \mathcal{T} 
   \,\right\} = 1,
   \ \text{where } \bm{F}(t) = \bm{L}^\top \bm{X}(t).
\)
    \item[c.] There exists a random matrix $\bm{B} \in \mathbb{R}^{K \times R}$, with $\bm{B}^\top \bm{B} = \bm{I}_R$, such that $\bm{L}\bm{B}$ are the singular vectors of
 ${X}_i$s, almost surely,
where $R$ is the FSVD rank of ${X}_i$s and less than or equal to $K$.
\end{itemize}
\end{theo}

Theorem~\ref{the: fac} shows that there always exists
\(
K = \operatorname{rank}\left(\mathbb{E}\int_{\mathcal{T}} \bm{X}(t)\bm{X}^\top(t)\, \mathrm{d}t\right)
\)
such that the first \(K\) IBVs make \eqref{factor_model} hold exactly.

The estimation of factor loadings reduces to the FSVD on $X_i$s as indicated by c. in Theorem~\ref{the: fac}. We formalize this result in the following theorem.

\begin{theo}[Relation between Singular Vectors and IBVs]\label{Theorem_FSVD_vector}
Assume $\sup_{i\in [n]}\mathbb{E}\|X_{i}\|^2\leq C$.  
Let 
\(
K = \operatorname{rank}\!\left(\mathbb{E}\int_{\mathcal{T}} \bm{X}(t)\bm{X}^\top(t)\, \mathrm{d}t\right),
\)
and let $\bm{L} = (\bm{l}_1,\dots,\bm{l}_K)$ denote the $K$ IBVs of $X_i$s with factor series $\bm{F}(t)$, $t \in \mathcal{T}$. 
Suppose that 
\begin{equation}\label{idenfi_fac}
    \int_{\mathcal{T}} \bm{F}(t)\bm{F}^\top(t)\, \mathrm{d}t \in \mathbb{R}^{K \times K}\ \text{is non-singular},
\end{equation}
or $\operatorname{rank}\!\left(\int_{\mathcal{T}} \bm{X}(t)\bm{X}^\top(t)\, \mathrm{d}t\right)=K$.
Then, almost surely, $R=K$, and 
$(\bm{a}_1,\cdots,\bm{a}_K)\bm{B}^{-1}$ are the factor loadings of $X_i$s,
with the factor series given by
\(
\bm{F}(t) = \sum_{r=1}^K \rho_r \bm{b}_r \phi_r(t), \ t \in \mathcal{T}.
\)
Here, $\rho_r$, $\bm{a}_r$, $\phi_r$, and $R$ are obtained from the FSVD of $X_i$s, 
and $\bm{B}=(\bm{b}_1,\dots,\bm{b}_K)\in \mathbb{R}^{K\times K}$ is any invertible matrix. 
\end{theo}

Condition~\eqref{idenfi_fac} implies that \( R = K \), ensuring that all factors in \( X_i \) are realized without degeneration.  
When this condition holds, the singular vectors obtained from FSVD can fully capture the signals associated with the IBVs, even when \( X_i \)s exhibit strong inter-trajectory dependencies.

\begin{rema}[Connection between FSVD and Existing Works on Factor Models of Time Series]\label{con_factor} 
In Table~\ref{fac_comparison}, we summarize two classical frameworks for factor models of time series and compare them with FSVD. We find that FSVD can be viewed as a continuous analogue of the factor model in \citet{bai2002determining}, incorporating temporal smoothness of the factor series through the RKHS framework. While \citet{lam2011estimation} also considers temporal structure in factor models, their method requires the time series to be stationary and observed at equally spaced time points. In contrast, our method accommodates non-stationary and irregularly observed time series. Meanwhile, FSVD allows the incorporation of prior knowledge about the factor series through the choice of different kernels \( \mathbb{K} \) for the RKHS \( \mathcal{H}(\mathbb{K}) \), thereby providing additional flexibility in factor modeling.
\end{rema}

\begin{table}[h]
\centering
\caption{Summary of assumptions in factor models for time series.}
\label{fac_comparison}
\renewcommand{\arraystretch}{1}
\setlength\tabcolsep{6pt}
\footnotesize
\begin{tabular}{lp{4cm}p{9cm}}  
  \hline
  \textbf{Reference} & \textbf{Observation Scheme} & \textbf{Assumption} \\
  \hline
  \citet{bai2002determining} & Regularly observed data & 
  $\displaystyle\lim_{J\rightarrow\infty}\frac{1}{J}\sum_{j=1}^J \bm{F}(t_j) \bm{F}^\top(t_j)$ is non-singular, where $\{t_1,\ldots,t_J\}$ denotes a time grid. \\

  \citet{lam2011estimation} & Equally spaced data & 
  $\{\bm{F}(t): t \in \mathbb{R}\}$ is a stationary sequence with non-singular autocovariance matrices. \\

  FSVD  & Irregularly observed data & 
  $\displaystyle\int_{\mathcal{T}} \bm{F}(t) \bm{F}^\top(t) \,\mathrm{d}t$ is non-singular and \( \{\bm{F}(t) : t \in \mathcal{T} \} \) consists of functions in an RKHS \( \mathcal{H}(\mathbb{K}) \). \\
  \hline
\end{tabular}
\end{table}

\section{Discussions}\label{sec:dis}

In this article, we establish the mathematical framework, implementation procedure, and statistical theory of FSVD for trajectory data with potential heterogeneity. By introducing intrinsic basis functions/vectors, FSVD provides a unified perspective on common learning tasks for trajectory data, capturing different structural aspects of the data. We demonstrate the advantages of FSVD through extensive simulations and two data analyses, showcasing its superior performance compared to existing methods.

Beyond the setting of this article, trajectory data with two-way heterogeneity have emerged in various real-world applications.
In these scenarios, the mean and covariance functions of random trajectories \( X_{ij}(t) \) may vary across subject \( i \) and/or feature \( j \), often involving complex subject--feature--trajectory tensor structures and varying time grids across $i$ or $j$ \citep{shi2024tempted, zhang2024individualized}. These complexities often require effective dimension reduction, which was historically achieved through techniques such as KL expansions \citep{chiou2014multivariate, zapata2022partial}, factor models \citep{zhang2024individualized}, and tensor SVD decompositions \citep{shi2024tempted, han2023guaranteed}. It would be interesting to establish their connections to our framework.

\bibliographystyle{apalike}

\bibliography{refbib}
\newpage

\appendix

\begin{sloppypar}

\section{Technical Proof}\label{SM: proof}
\paragraph*{Preliminary}
We first recall some notations.
Let $\mathcal{T}$ be a bounded closed interval in $\mathbb{R}$. Without loss of generality, we set $\mathcal{T}$ to be $[0,1]$ throughout this article.
Denote $\mathcal{L}^2(\mathcal{T})$ as the Hilbert space of square-integrable functions on $\mathcal{T}$ with the inner product $\langle \cdot,\cdot \rangle$ and norm $\|\cdot\|:=\sqrt{\langle \cdot,\cdot \rangle}$, where 
$$
\langle f, g\rangle=\int_{\mathcal{T}}f(t)g(t)\ \mathrm{d}t,\ \forall f,g\in \mathcal{L}^2(\mathcal{T}).
$$
We also use $\|\cdot\|$ to denote both the Euclidean norm of a vector and the Frobenius norm of a matrix in the following proof.
Denote $\overline{\mathcal{H}}$ as the closure of a set $\mathcal{H}$ from a Hilbert space in terms of its norm, and define $\operatorname{span}(f_1,\dots,f_n)$ as the functional space spanned by $f_1,\dots,f_n\in \mathcal{L}^2(\mathcal{T})$. 
Let $\mathbb{I}(\cdot)$ be the indicator function and $[Z]$ be the set of integers $\{1,\dots,Z\}$. Moreover, we denote that $f=\lim_{n\rightarrow\infty}f_n$ if $\lim_{n\rightarrow\infty}\|f-f_n\|= 0$.
    
Consider an operator $\mathcal{K}$ between two Hilbert spaces $\mathcal{H}_1$ and $\mathcal{H}_2$, each with inner product $\langle \cdot, \cdot\rangle_i$ and norms $\|\cdot\|_i$ for $i=1,2$.
Define $\operatorname{Dom}(\mathcal{K})$ as the domain of $\mathcal{K}$. Denote $\operatorname{Im}(\mathcal{K}):=\{\mathcal{K}x; x\in \operatorname{Dom}(\mathcal{K})\}$ and $\operatorname{Null}(\mathcal{K}):=\{x\in \operatorname{Dom}(\mathcal{K}); \mathcal{K}x=\bm{0}\}$ as the image and null spaces of $\mathcal{K}$, where $\bm{0}$ is the zero element in $\mathcal{H}_2$.
Define the multiplication of two operators $\mathcal{K}_1$ and $\mathcal{K}_2$ as $\mathcal{K}_1\mathcal{K}_2$ if $\operatorname{Im}(\mathcal{K}_2)\subset \operatorname{Dom}(\mathcal{K}_1)$. 
Besides, define the operator norm of $\mathcal{K}$ as $\|\mathcal{\mathcal{K}}\|_{\infty}=\sup\{ \|\mathcal{K}x\|_2;\|x\|_1\leq 1\}$, and denote $\mathcal{K}^*$ as the adjoint operator of an operator $\mathcal{K}$ if $$\langle \mathcal{K}f,g\rangle_2=\langle f,\mathcal{K}^*g\rangle_1 \quad \forall f\in \mathcal{H}_1 \text{ and } g\in \mathcal{H}_2.$$
Given an operator $\mathcal{K}$ from $\mathcal{H}_1$ to $\mathcal{H}_1$ such that $\|\mathcal{K}\|_{\infty}<\infty$, if there exist $e \neq 0\in \mathcal{H}_1$ and $\lambda\in \mathbb{R}$ obtaining
$
\mathcal{K}e=\lambda e,
$
we refer $\lambda$ and $e$ to as the eigenvalue and eigenfunction of $\mathcal{K}$, respectively. 
    
An operator $\mathcal{K}$ is compact if for any bounded sequence $\{x_N;N\geq 1\}$ in $\mathcal{H}_1$, $\{\mathcal{K}x_N;N\geq 1\}$ has a convergent subsequence in $\mathcal{H}_2$. For a compact operator $\mathcal{K}$, it has the following singular value decomposition
\begin{eqnarray*}
    \mathcal{K}f=\sum_{r=1}^\infty \rho_r\langle f,\phi_r\rangle_1 \psi_r,\ \forall f\in \mathcal{H}_1,
\end{eqnarray*}
where $\rho_r^2$ are the eigenvalues of both $\mathcal{K}^*\mathcal{K}$ and $\mathcal{K}\mathcal{K}^*$, 
$\{\phi_r \in \overline{\operatorname{Im}(\mathcal{K}^*\mathcal{K})} ; r \geq 1\}$ are the eigenfunctions of $\mathcal{K}^*\mathcal{K}$, 
and $\{\psi_r \in \overline{\operatorname{Im}(\mathcal{K}\mathcal{K}^*)} ; r \geq 1\}$ are the eigenfunctions of $\mathcal{K}\mathcal{K}^*$. 
See Theorem~4.3.1 in \citet{hsing2015theoretical} for more details.
    
Denote $\mathcal{H}$ as a Hilbert space of functions on $\mathcal{T}$ with inner product $\langle \cdot, \cdot \rangle_{\mathcal{H}}$ and norm $\|\cdot\|_{\mathcal{H}}$. The functional space $\mathcal{H}$ is called a reproducing kernel Hilbert space (RKHS) $\mathcal{H}(\mathbb{K})$ if there exists a kernel $\mathbb{K}$ on $\mathcal{T}\times\mathcal{T}$ such that $\mathbb{K}(t,\cdot)\in \mathcal{H}$ and
\begin{align*}
    f(t)=\langle f,\mathbb{K}(t,\cdot)\rangle_{\mathcal{H}},
\end{align*}
$\forall t\in \mathcal{T}$ and $f\in \mathcal{H}$.

For any semi-positive definite kernel $K(t, s)$ such that $\int_0^1\int_0^1(K(t,s))^2\ \mathrm{d}t\,\mathrm{d}s<\infty$, we call $\mathcal{K}$ an integral operator associated with $K(t,s)$ if 
$$
\mathcal{K}f=\int_{0}^1 K(t,s)f(s)\ \mathrm{d}s,
$$
$\forall f\in \mathcal{L}^2(\mathcal{T})$.
It can be shown that $\mathcal{K}$ is a compact self-adjoint operator, and the SVD of $\mathcal{K}$ leads to a spectral decomposition of $K(t,s)$:
\begin{eqnarray*}
    K(t,s)=\sum_{k=1}^{\infty} \lambda_k\psi_k(t)\psi_k(s),
\end{eqnarray*}
where $\lambda_k$ and $\psi_k$ are the eigenvalues and eigenvectors of $\mathcal{K}$, respectively. See Section 4.6 of \citet{hsing2015theoretical} for more details.

\
\subsection{Mathematical Foundation of FSVD}

\subsubsection{Proof of Theorem \ref{th:functional-SVD}}\label{proof_FSVD_SM}
\begin{proof}
Define the operator $\mathcal{X}_n:\mathcal{H}\to\mathbb{R}^n$ by
\begin{eqnarray}\label{X_ope}
    \mathcal{X}_n:f \mapsto \big(\langle X_1,f\rangle,\dots,\langle X_n,f\rangle\big)^\top,\ \forall f\in \mathcal{H}.
\end{eqnarray} 
Notice that for all $f\in \mathcal{L}^2(\mathcal{T})$ with $\|f\|\le 1$,
\begin{eqnarray*}
    \|\mathcal{X}_n f\|^2
    =\sum_{i=1}^n\langle X_i,f\rangle^2
    \leq \sum_{i=1}^n\|X_i\|^2\,\|f\|^2
    \leq \sum_{i=1}^n\|X_i\|^2.
\end{eqnarray*}
Since $\mathbb{E}\sum_{i=1}^n \|X_i\|^2$ is finite for any fixed $n$, it follows that
$\sum_{i=1}^n \|X_i\|^2<\infty$ almost surely. Therefore, for any realization of $\{X_i\}_{i=1}^n$,
$\mathcal{X}_n$ is a bounded operator. For any bounded sequence $\{f_N\}_{N\ge 1}$ in
$\mathcal{L}^2(\mathcal{T})$, the boundedness of $\mathcal{X}_n$ implies that
$\{\mathcal{X}_n f_N\}_{N\ge 1}$ is bounded in $\mathbb{R}^n$. By the Bolzano--Weierstrass theorem,
$\{\mathcal{X}_n f_N\}_{N\ge 1}$ admits a convergent subsequence in $\mathbb{R}^n$. Hence,
$\mathcal{X}_n$ is compact.

The compactness of $\mathcal{X}_n$ yields the following singular value decomposition:
\begin{eqnarray}\label{SVD_operator}
    \mathcal{X}_n f=\sum_{r=1}^\infty \rho_r\langle f,\phi_r\rangle \bm{a}_r,\ \forall f\in \mathcal{L}^2(\mathcal{T}),
\end{eqnarray}
where $\rho_r^2$ are the eigenvalues of both $\mathcal{X}_n^*\mathcal{X}_n$ and $\mathcal{X}_n\mathcal{X}_n^*$,
$\phi_r\in \overline{\operatorname{Im}(\mathcal{X}_n^*\mathcal{X}_n)}$, $r\ge 1$, are the eigenfunctions of
$\mathcal{X}_n^*\mathcal{X}_n$, and $\{\bm{a}_r\}$ are the eigenvectors of $\mathcal{X}_n\mathcal{X}_n^*$.

Next, we show that $\mathcal{X}_n^*\mathcal{X}_n$ is an integral operator with kernel
$\sum_{i=1}^n X_i(t)X_i(s)$, and that $\mathcal{X}_n\mathcal{X}_n^*$ is the linear transformation associated with the matrix
$\int_0^1 \bm{X}(t)\bm{X}^\top(t)\,\mathrm{d}t$. Consequently, $\{\phi_r\}$ and $\{\bm{a}_r\}$ are the
eigenfunctions/eigenvectors of $\sum_{i=1}^n X_i(t)X_i(s)$ and $\int_0^1 \bm{X}(t)\bm{X}^\top(t)\,\mathrm{d}t$,
respectively. To see this, note that
\begin{eqnarray*}
   (\mathcal{X}_n f)^\top \bm{c} 
   = \sum_{i=1}^n c_i \langle X_i, f \rangle
   = \bigg\langle f, \sum_{i=1}^n c_i X_i \bigg\rangle,
\end{eqnarray*}
for all $f\in \mathcal{L}^2(\mathcal{T})$ and $\bm{c}=(c_1,\dots,c_n)^\top\in\mathbb{R}^n$.
Thus, $\mathcal{X}_n^*$ maps $\bm{c}$ to $\sum_{i=1}^n c_i X_i$, i.e.,
\begin{eqnarray}\label{equa_*}
    \mathcal{X}_n^*\bm{c}=\sum_{i=1}^n c_i X_i.
\end{eqnarray}
Therefore,
\begin{eqnarray*}
   \mathcal{X}_n^*\mathcal{X}_n f
   =\sum_{i=1}^n \langle X_i,f\rangle X_i
   =\int_0^1 \bigg\{\sum_{i=1}^n X_i(t)X_i(s)\bigg\} f(s)\,\mathrm{d}s,
\end{eqnarray*}
for all $f\in \mathcal{L}^2(\mathcal{T})$, so $\mathcal{X}_n^*\mathcal{X}_n$ is indeed an integral operator with kernel
$\sum_{i=1}^n X_i(t)X_i(s)$. Similarly,
\begin{eqnarray*}
\mathcal{X}_n\mathcal{X}_n^*\bm{c}
=\left(\int_0^1 \bm{X}(t)\bm{X}^\top(t)\,\mathrm{d}t\right)\bm{c},
\qquad \forall \bm{c}\in\mathbb{R}^n,
\end{eqnarray*}
where $\bm{X}(t)=(X_1(t),\dots,X_n(t))^\top$. Hence, $\mathcal{X}_n\mathcal{X}_n^*$ is the linear transformation
associated with the matrix $\int_0^1 \bm{X}(t)\bm{X}^\top(t)\,\mathrm{d}t$.

Since $\mathcal{X}_n\mathcal{X}_n^*$ is an $n\times n$ matrix, we have $\rho_r=0$ for all $r>n$. Then \eqref{SVD_operator} leads to
\begin{eqnarray*}
   \langle X_i,f\rangle=\sum_{r=1}^R \rho_r\langle f,\phi_r\rangle a_{ir},\ \forall f\in \mathcal{L}^2(\mathcal{T})\ \text{and}\ i\in[n],
\end{eqnarray*}
where $R\le n$ is the rank of $\mathcal{X}_n$ and $a_{ir}$ denotes the $i$th entry of $\bm{a}_r$.

Let $\{f_N;N\ge 1\}$ be any orthonormal basis of $\mathcal{L}^2(\mathcal{T})$. Then, for each $i\in[n]$,
\begin{eqnarray*}
    X_i
    =\sum_{N=1}^\infty \langle X_i,f_N\rangle f_N
    =\sum_{N=1}^\infty \sum_{r=1}^R \rho_r\langle f_N,\phi_r\rangle a_{ir} f_N
    =\sum_{r=1}^R \rho_r \phi_r\, a_{ir},
\end{eqnarray*}
which gives the FSVD of $\{X_i\}_{i=1}^n$.

It remains to show that $\overline{\operatorname{Im}(\mathcal{X}_n^*\mathcal{X}_n)}\subset \mathcal{H}$ when
$X_i\in\mathcal{H}$ for all $i\in[n]$. This implies $\phi_r\in\mathcal{H}$ for all $r\ge 1$, since
$\phi_r\in\overline{\operatorname{Im}(\mathcal{X}_n^*\mathcal{X}_n)}$.
By projection theory, $\mathcal{L}^2(\mathcal{T})$ admits the orthogonal decomposition
\[
\mathcal{L}^2(\mathcal{T})=\mathcal{H}\oplus \mathcal{H}^\perp,
\]
where $\mathcal{H}^\perp$ is the orthogonal complement of $\mathcal{H}$ in the $L^2$ inner product. Moreover,
$\mathcal{H}^\perp\subset \operatorname{Null}(\mathcal{X}_n)$ since $X_1,\dots,X_n\in\mathcal{H}$. Therefore,
\begin{eqnarray*}
    \operatorname{Null}(\mathcal{X}_n)^\perp \subset (\mathcal{H}^\perp)^\perp=\mathcal{H}.
\end{eqnarray*}
By Theorem 3.3.7 in \citet{hsing2015theoretical},
\[
\operatorname{Null}(\mathcal{X}_n)^\perp
=\overline{\operatorname{Im}(\mathcal{X}_n^*)}
=\overline{\operatorname{Im}(\mathcal{X}_n^*\mathcal{X}_n)},
\]
which implies $\overline{\operatorname{Im}(\mathcal{X}_n^*\mathcal{X}_n)}\subset \mathcal{H}$. The proof is complete.
\end{proof}

\

\subsubsection{Additional Propositions}

The following proposition characterizes the uniqueness of FSVD.

\begin{propositionA}\label{th:uniqueness}
    If there exist two FSVDs of $X_1,\ldots, X_n$: $\big\{\rho_r, \bm{a}_r, \phi_r; r=1,\dots, R\big\}$, $\big\{\tilde{\rho}_r,\tilde{\bm{a}}_r,\tilde{\phi}_r; r=1,\dots,\tilde{R}\big\}$ such that $\rho_1\geq \cdots \geq \rho_{R}>0$, $\tilde{\rho}_1 \geq \cdots \geq \tilde{\rho}_{\tilde{R}}>0$, $\bm{a}_r^\top \bm{a}_{r^{\prime}} = \langle\phi_r,\phi_{r^{\prime}}\rangle = \tilde{\bm{a}}_r^\top \tilde{\bm{a}}_{r^{\prime}} = \langle\tilde{\phi}_r,\tilde{\phi}_{r^{\prime}}\rangle = \mathbb{I}(r=r^{\prime})$, and satisfying $\sum_{r=1}^R\rho_r \bm{a}_r \phi_r = \sum_{r=1}^{\tilde{R}}\tilde{\rho}_r \tilde{\bm{a}}_r \tilde{\phi}_r$,
    then $R=\tilde{R}$ and $\rho_r = \tilde{\rho}_r$ for all $r\in [R]$.
    
    If $\rho_1 > \cdots> \rho_R>0$ are distinct, then $(\tilde{\bm{a}}_r, \tilde{\phi}_r) = \pm (\bm{a}_r, \phi_r)$. If we have identical singular values: $\rho_{r_1-1} > \rho_{r_1}=\cdots= \rho_{r_2}>\rho_{r_2+1}$, then there exists an orthogonal matrix $\bm{B} \in \mathbb{R}^{(r_2-r_1+1)\times (r_2-r_1+1)}$ such that $(\tilde{\bm{a}}_{r_1},\dots, \tilde{\bm{a}}_{r_2}) =({\bm{a}}_{r_1},\dots, {\bm{a}}_{r_2}) \bm{B}$ and $(\tilde{\phi}_{r_1},\dots, \tilde{\phi}_{r_2}) =({\phi}_{r_1},\dots, {\phi}_{r_2})\bm{B}$.
\end{propositionA} 

\begin{proof}
If there exist two FSVDs of $X_1,\ldots, X_n$: $\big\{\rho_r, \bm{a}_r, \phi_r; r=1,\dots, R\big\}$, $\big\{\tilde{\rho}_r,\tilde{\bm{a}}_r,\tilde{\phi}_r; r=1,\dots,\tilde{R}\big\}$ such that $\rho_1\geq \cdots \geq \rho_{R}>0$, $\tilde{\rho}_1 \geq \cdots \geq \tilde{\rho}_{\tilde{R}}>0$,
$\bm{a}_r^\top \bm{a}_{r^{\prime}} = \langle\phi_r,\phi_{r^{\prime}}\rangle = \tilde{\bm{a}}_r^\top \tilde{\bm{a}}_{r^{\prime}} = \langle\tilde{\phi}_r,\tilde{\phi}_{r^{\prime}}\rangle = \mathbb{I}(r=r^{\prime})$, and satisfying 
$$
\sum_{r=1}^R\rho_r \bm{a}_r \phi_r = \sum_{r=1}^{\tilde{R}}\tilde{\rho}_r \tilde{\bm{a}}_r \tilde{\phi}_r.
$$
By Theorem~\ref{th:functional-SVD}, $\{\rho_r^2; r\in [R]\}$ and $\{{\tilde{\rho}}_r^2; r\in [\tilde{R}]\}$ are both the positive eigenvalues of $\mathcal{X}_n\mathcal{X}_n^*$. Therefore, $R=\tilde{R}$ and $\rho_r = \tilde{\rho}_r$ for all $r\in [R]$.

If $\rho_1>\cdots>\rho_R>0$ are distinct, then each eigenspace of $\mathcal{X}_n\mathcal{X}_n^*$ is one-dimensional. Hence $\tilde{\bm{a}}_r=\pm \bm{a}_r$ for all $r\in[R]$. Consequently,
\[
\tilde{\phi}_r=\frac{1}{\rho_r}(X_1,\dots,X_n)\tilde{\bm{a}}_r
=\pm \frac{1}{\rho_r}(X_1,\dots,X_n)\bm{a}_r
=\pm \phi_r,
\]
and thus $(\tilde{\bm{a}}_r,\tilde{\phi}_r)=\pm(\bm{a}_r,\phi_r)$.

If there exists a block of identical singular values, say $\rho_{r_1-1} > \rho_{r_1}=\cdots= \rho_{r_2}>\rho_{r_2+1}$. Then $(\tilde{\bm{a}}_{r_1},\dots, \tilde{\bm{a}}_{r_2})$ and $({\bm{a}}_{r_1},\dots, {\bm{a}}_{r_2})$ are both the eigenvectors of the matrix $\mathcal{X}_n\mathcal{X}_n^*$ corresponding to eigenvalue $\rho_{r_1}^2$. Consequently, there exists an orthogonal matrix $\bm{B} \in \mathbb{R}^{(r_2-r_1+1)\times (r_2-r_1+1)}$ such that 
$$
(\tilde{\bm{a}}_{r_1},\dots, \tilde{\bm{a}}_{r_2}) =({\bm{a}}_{r_1},\dots, {\bm{a}}_{r_2}) \bm{B}.
$$
This leads to
\begin{eqnarray*}
    (\tilde{\phi}_{r_1},\dots, \tilde{\phi}_{r_2})
    &=&\frac{1}{\rho_{r_1}}\big(X_1,\dots,X_n\big)(\tilde{\bm{a}}_{r_1},\dots, \tilde{\bm{a}}_{r_2})\\
    &=&\frac{1}{\rho_{r_1}}\big(X_1,\dots,X_n\big)({\bm{a}}_{r_1},\dots, {\bm{a}}_{r_2})\bm{B}\\
    &=&({\phi}_{r_1},\dots, {\phi}_{r_2})\bm{B}.
\end{eqnarray*}
The proof is complete.
\end{proof}

The following property shows that the $r$th singular component of $X_1,\ldots, X_n$ provides the optimal rank-one approximation for these functions after subtracting the first $(r-1)$ singular components. 
\begin{propositionA}\label{lemma: FSVD_s}
    Consider $g_{i0}$, $i\in [n]$, as zero functions, and let $g_{ir}$, $i\in [n]$, be defined by the minimizers of $f_i$s obtained from
    \begin{equation*}
      \min_{f\in \mathcal{H}}\min_{f_1,\dots,f_n \in \operatorname{span}(f)} \sum_{i=1}^n \bigg\|X_i - \sum_{l=0 }^{r-1}g_{il} - f_i\bigg\|^2.
    \end{equation*}
    Define $\rho_r^0:=\sqrt{\sum_{i=1}^n \|g_{ir}\|^2}$, $\phi_r^0=g_{ir}/\|g_{ir}\|$ and $\bm{a}_r^0:=(\langle g_{1r},\phi_r^0\rangle,\dots,\langle g_{nr},\phi_r^0\rangle)^\top/\rho_r^0$.
    Then, $\{\rho_r^0, \bm{a}_r^0, \phi_r^0;$
    $r \in [R]\}$ forms the FSVD of $X_1,\dots, X_n$. 
\end{propositionA}

\begin{proof}
Let $f_i=b_i g$, $i\in[n]$, for any $\bm{b}=(b_1,\cdots,b_n)^\top\in\mathbb{R}^n$ and $g\in\mathcal{H}$ satisfying $\|g\|=1$.
Denote
\begin{eqnarray*}
    X_i=\sum_{r=1}^R \rho_r^0 a_{ir}^0 \phi_r^0
\end{eqnarray*}
as the FSVD of $X_i$s, where $\rho_1^0\ge \rho_2^0\ge \cdots \ge \rho_R^0$.
Note that
\begin{eqnarray*}
   L(\bm{b},g)
   &=&\sum_{i=1}^n\|X_i-f_i\|^2
    =\sum_{i=1}^n\|X_i\|^2-2\sum_{i=1}^n b_i\langle X_i,g\rangle+\sum_{i=1}^n b_i^2\\
    &=&\sum_{i=1}^n\|X_i\|^2-2\sum_{i=1}^n\sum_{r=1}^R\rho_r^0 a_{ir}^0 b_i\langle \phi_r^0,g\rangle+\sum_{i=1}^n b_i^2\\
    &=&\sum_{i=1}^n\|X_i\|^2-2\sum_{r=1}^R\rho_r^0\langle \bm{a}_r^0,\bm{b}\rangle\langle \phi_r^0,g\rangle+\sum_{i=1}^n b_i^2,
\end{eqnarray*}
where $\bm{a}_r^0=(a_{1r}^0,\cdots,a_{nr}^0)^\top$.
Since $\sum_{r=1}^R\langle \bm{a}_r^0,\bm{b}\rangle^2\le \|\bm{b}\|^2$ and $\sum_{r=1}^R\langle \phi_r^0,g\rangle^2\le 1$, by the Cauchy--Schwarz inequality,
\begin{eqnarray*}
\sum_{r=1}^R\big|\langle \bm{a}_r^0,\bm{b}\rangle\langle \phi_r^0,g\rangle\big|
\le
\Big(\sum_{r=1}^R\langle \bm{a}_r^0,\bm{b}\rangle^2\Big)^{1/2}
\Big(\sum_{r=1}^R\langle \phi_r^0,g\rangle^2\Big)^{1/2}
\le \|\bm{b}\|.
\end{eqnarray*}
Therefore,
\begin{eqnarray*}
\sum_{r=1}^R\rho_r^0\langle \bm{a}_r^0,\bm{b}\rangle\langle \phi_r^0,g\rangle
\le
\sup_{r\in[R]}\{\rho_r^0\}\sum_{r=1}^R\big|\langle \bm{a}_r^0,\bm{b}\rangle\langle \phi_r^0,g\rangle\big|
\le \rho_1^0\|\bm{b}\|.
\end{eqnarray*}
Hence, for any $\bm{b}$ and $g$,
\begin{eqnarray*}
    L(\bm{b},g)
    \ge
    \sum_{i=1}^n\|X_i\|^2-2\rho_1^0\|\bm{b}\|+\|\bm{b}\|^2
    =L(\|\bm{b}\|\bm{a}_1^0,\phi_1^0).
\end{eqnarray*}
Using the fact that $-2\rho_1^0 d+d^2\ge -(\rho_1^0)^2$ for all $d\in\mathbb{R}$, we obtain
\begin{eqnarray*}
L(\bm{b},g)
\ge
\sum_{i=1}^n\|X_i\|^2-(\rho_1^0)^2
=
L(\rho_1^0\bm{a}_1^0,\phi_1^0),
\end{eqnarray*}
and equality holds if $\bm{b}=\rho_1^0\bm{a}_1^0$ and $g=\phi_1^0$.
We then conclude that $(\rho_1^0a_{11}^0\phi_1^0,\dots,\rho_1^0a_{n1}^0\phi_1^0)$ are minimizers of $\{f_i\}_{i=1}^n$ for
\begin{align*}
\min_{f\in \mathcal{H}}\min_{f_1,\dots,f_n \in \operatorname{span}(f)} \sum_{i=1}^n\|X_i-f_i\|^2.
\end{align*}

For $R>1$, note that for any $r>1$,
\begin{eqnarray*}
X_i-\sum_{l=1}^{r-1} g_{il}=\sum_{l=r}^{R}\rho_l^0 a_{il}^0 \phi_l^0,
\end{eqnarray*}
where $g_{il}=\rho_l^0 a_{il}^0 \phi_l^0$.
By applying the same argument to the residual functions $\big\{X_i-\sum_{l=1}^{r-1} g_{il}\big\}_{i=1}^n$, we similarly obtain that
$({\rho}_r^0, {\bm{a}}_r^0, {\phi}_r^0)$ is a minimizer of
\begin{equation*}
\min_{f\in \mathcal{H}}\min_{f_1,\dots,f_n \in \operatorname{span}(f)}
\sum_{i=1}^n\bigg\|X_i-\sum_{l=1}^{r-1}g_{il}-f_i\bigg\|^2,
\end{equation*}
for each $r>1$.
\end{proof} 

\

The following proposition is an extension of the Representer Theorem for kernel ridge regression \citep{scholkopf2001generalized} to the rank-one-constrained kernel ridge regression.

\begin{propositionA}\label{th:repre}
Assume the null space of $\mathcal{P}$ is finite-dimensional with basis functions $h_1,\cdots,h_q$, and define $g_{ij}:=\mathcal{P}\big\{\mathbb{K}(\cdot,T_{ij})\big\}$. Then there exist $u_m\in \mathbb{R}$, $m\in [q]$, and $w_{ij}\in \mathbb{R}$, $i\in [n]$ and $j\in [J_i]$, such that the minimizer of $\phi_1$ in 
\begin{eqnarray*}
\min_{\bm{a}_1\in \mathbb{R}^n,\phi_1\in \mathcal{H}(\mathbb{K})}\sum_{i=1}^n\frac{1}{J_i}\sum_{j=1}^{J_i}\big\{Y_{ij}-a_{i1} \phi_1(T_{ij})\big\}^2+\nu\|\bm{a}_1\|^2\cdot\|\mathcal{P}\phi_1\|_{\mathcal{H}}^2
\end{eqnarray*}
is represented as
\begin{eqnarray}\label{representer}
    \sum_{m=1}^q {u}_{m}h_m+\sum_{i=1}^n\sum_{j=1}^{J_i}{w}_{ij}g_{ij}.
\end{eqnarray}
As a result, the above optimization can be reformulated as 
\begin{eqnarray}\label{Tran_opt}
\min_{\bm{a}_1\in \mathbb{R}^n,\bm{u}\in \mathbb{R}^q,\bm{w}\in \mathbb{R}^J} \sum_{i=1}^n\frac{1}{J_i}\sum_{j=1}^{J_i}\bigg[Y_{ij}&-&a_{i1} \bigg\{\sum_{m=1}^q u_{m}h_m(T_{ij})+  \sum_{i_1=1}^n\sum_{j_1=1}^{J_i}w_{i_1j_1}g_{i_1j_1}(T_{ij})\bigg\}\bigg]^2 \\
&+&\nu\|\bm{a}_1\|^2\cdot\bm{w}^\top \bm{G}\bm{w}\nonumber,
\end{eqnarray}
where $\bm{u}=(u_1,\cdots,u_q)^\top$, $\bm{w}=\big(w_{ij};i\in [n], j\in[J_i]\big)^\top\in \mathbb{R}^J$ with $J=\sum_{i=1}^n J_i$, and the entries of the matrix $\bm{G}$ are $\langle g_{i' j'}, g_{i'' j''}\rangle_{\mathcal{H}}$ for all $i', i'' \in [n], j' \in [J_{i'}], j''\in [J_{i''}]$. 
\end{propositionA}

\begin{proof}
Define
\begin{eqnarray*}
L(\bm{a},\phi):=\sum_{i=1}^n\frac{1}{J_i}\sum_{j=1}^{J_i}\big\{Y_{ij}-a_{i} \phi(T_{ij})\big\}^2+\nu\|\bm{a}\|^2\cdot\|\mathcal{P}\phi\|_{\mathcal{H}}^2,
\end{eqnarray*}
and
\begin{eqnarray*}
    \mathcal{H}:=\bigg\{f\in \mathcal{H}(\mathbb{K});\ f=\sum_{m=1}^q {u}_{m}h_m+\sum_{i=1}^n\sum_{j=1}^{J_i}{w}_{ij}g_{ij},\ u_m\in \mathbb{R},\ w_{ij}\in \mathbb{R}\bigg\}.
\end{eqnarray*}
Notice that
\begin{eqnarray*}
    \mathbb{K}(\cdot,T_{ij})
    =\mathbb{K}(\cdot,T_{ij})-\mathcal{P}\big\{\mathbb{K}(\cdot,T_{ij})\big\}+g_{ij}.
\end{eqnarray*}
Since $\mathbb{K}(\cdot,T_{ij})-\mathcal{P}\big\{\mathbb{K}(\cdot,T_{ij})\big\}\in \text{Null}(\mathcal{P})\subset\mathcal{H}$ and $g_{ij}\in \mathcal{H}$, we have $\mathbb{K}(\cdot,T_{ij})\in \mathcal{H}$.

Let $\mathcal{H}(\mathbb{K})=\mathcal{H} \oplus \mathcal{H}^\perp$, where $\mathcal{H}^\perp$ is the orthogonal complement of $\mathcal{H}$ with respect to $\langle\cdot,\cdot\rangle_{\mathcal{H}}$. For any $f\in \mathcal{H}(\mathbb{K})$, we can write
\begin{eqnarray*}
    f=\phi+\phi^\perp,
\end{eqnarray*}
where $\phi\in \mathcal{H}$ and $\phi^\perp\in \mathcal{H}^\perp$. As a result,
\begin{eqnarray}\label{f_eqa}
    f(T_{ij})&=&\phi(T_{ij})+\phi^\perp(T_{ij})\nonumber\\
    &=& \phi(T_{ij})+\langle \phi^\perp, \mathbb{K}(\cdot,T_{ij})\rangle_{\mathcal{H}}\nonumber\\
    &=&\phi(T_{ij}),
\end{eqnarray}
since $\mathbb{K}(\cdot,T_{ij})\in \mathcal{H}$ and $\phi^\perp\perp \mathcal{H}$.

Moreover, the projection $\phi$ can be represented as $\sum_{m=1}^q u_{m}h_m+\sum_{i=1}^n\sum_{j=1}^{J_i} w_{ij}g_{ij}$.
By the definition $g_{ij}=\mathcal{P}\{\mathbb{K}(\cdot,T_{ij})\}$ and the fact that $\mathcal{P}$ is an orthogonal projection, we have $\mathcal{P}g_{ij}=\mathcal{P}^2\{\mathbb{K}(\cdot,T_{ij})\}=g_{ij}$.
Hence,
\[
\mathcal{P}\phi
=\sum_{m=1}^q u_m\,\mathcal{P}h_m+\sum_{i=1}^n\sum_{j=1}^{J_i} w_{ij}\,\mathcal{P}g_{ij}
=\sum_{i=1}^n\sum_{j=1}^{J_i} w_{ij} g_{ij},
\]
where we used $\mathcal{P}h_m=0$ since $h_m\in \mathrm{Null}(\mathcal{P})$.
Since $\mathcal{P}\phi\in \mathcal{H}$ and $\phi^\perp\in \mathcal{H}^\perp$, we have $\langle \mathcal{P}\phi,\mathcal{P}\phi^\perp\rangle_{\mathcal{H}}=0$. Therefore,
\begin{eqnarray}\label{eqa_phi}
    \|\mathcal{P}f\|^2=\|\mathcal{P}\phi+\mathcal{P}\phi^\perp\|^2\geq \|\mathcal{P}\phi\|^2.
\end{eqnarray}
Combining \eqref{f_eqa} and \eqref{eqa_phi}, we conclude that for all $\bm{a}\in \mathbb{R}^n$ and $f\in \mathcal{H}(\mathbb{K})$, there exists the projection $\phi\in\mathcal{H}$ of $f$ such that
\begin{eqnarray*}
    L(\bm{a},f)\geq L(\bm{a},\phi).
\end{eqnarray*}
The proof is complete.
\end{proof}

\

\subsection{Mathematical Foundation of Intrinsic Basis Functions/Vectors}
\subsubsection{Proof of Theorem \ref{the: eui_in}}
\begin{proof}
\textbf{a. $\Rightarrow$ b.:} Notice that
\begin{eqnarray*}
 \sum_{i=1}^n\mathbb{E}\bigg\|X_i-\sum_{k=1}^K{\xi}_{ik}{\varphi}_k\bigg\|^2
 =n\cdot \bigg(\int_{0}^1H_n(t,t)\ \mathrm{d}t-\sum_{k=1}^{K}\int_0^1\int_0^1 H_n(t,s) {\varphi}_k(t){\varphi}_k(s)\ \mathrm{d}t\,\mathrm{d}s\bigg),
\end{eqnarray*}
where $\varphi_k$s are the IBFs.
Observe that $H_n(t,s):=\frac{1}{n} \mathbb{E}\sum_{i=1}^nX_i(t)X_i(s)$ is always a non-negative-definite kernel.
Let $H_n(t,s) = \sum_{k=1}^\infty \lambda_k \tilde{\varphi}_k(t) \tilde{\varphi}_k(s)$ denote the spectral decomposition of $H_n$, where $\lambda_1 \geq \lambda_2 \geq \cdots \geq 0$ are the eigenvalues, and $\tilde{\varphi}_k$s are the corresponding eigenfunctions.
Consequently,
\begin{eqnarray*}
 \frac{1}{n}\sum_{i=1}^n\mathbb{E}\bigg\|X_i-\sum_{k=1}^K{\xi}_{ik}{\varphi}_k\bigg\|^2
 &=&\sum_{k=1}^{\infty} \lambda_k
-\sum_{k=1}^K\int_0^1\int_0^1 H_n(t,s) {\varphi}_k(t){\varphi}_k(s)\ \mathrm{d}t\,\mathrm{d}s\\
&=&\sum_{k=1}^{\infty} \lambda_k
-\sum_{k=1}^K\sum_{g=1}^\infty \lambda_g \langle {\varphi}_k,\tilde{\varphi}_g\rangle^2.
\end{eqnarray*}
Define $b_g:=\sum_{k=1}^K \langle {\varphi}_k,\tilde{\varphi}_g\rangle^2$. Then $0\le b_g\le 1$ and $\sum_{g=1}^\infty b_g=K$.
Hence,
\[
\sum_{k=1}^K\sum_{g=1}^\infty \lambda_g \langle {\varphi}_k,\tilde{\varphi}_g\rangle^2
=\sum_{g=1}^\infty \lambda_g b_g
\le \sum_{g=1}^K \lambda_g,
\]
where the last inequality follows from $\lambda_1\ge \lambda_2\ge \cdots$. Therefore,
\[
\frac{1}{n}\sum_{i=1}^n\mathbb{E}\bigg\|X_i-\sum_{k=1}^K{\xi}_{ik}{\varphi}_k\bigg\|^2
\ge \sum_{g=K+1}^\infty \lambda_g,
\]
with equality if and only if $\operatorname{span}(\varphi_1,\ldots,\varphi_K)=\operatorname{span}(\tilde{\varphi}_1,\ldots,\tilde{\varphi}_K)$, for any $K$.
Otherwise, there exists some $K$ such that
\begin{eqnarray*}
    \sum_{i=1}^n\mathbb{E}\bigg\|X_i-\sum_{k=1}^K\tilde{\xi}_{ik}\tilde{\varphi}_k\bigg\|^2
    =\sum_{k=K+1}^\infty\lambda_k
    \leq \sum_{i=1}^n\mathbb{E}\bigg\|X_i-\sum_{k=1}^K{\xi}_{ik}{\varphi}_k\bigg\|^2,
\end{eqnarray*}
where $\tilde{\xi}_{ik}=\langle X_i, \tilde{\varphi}_k\rangle$.
This is a contradiction to a..
We then conclude that $\varphi_k$s are the eigenfunctions of $H_n(t,s)$.

\textbf{b. $\Rightarrow$ c.:} If $\{\varphi_k;k\geq 1\}$ are the eigenfunctions of $H_n(t,s)$, then
\begin{eqnarray*}
\sum_{i=1}^n\mathbb{E}\xi_{ik_1}\xi_{ik_2}
&=&\sum_{i=1}^n\mathbb{E}\int_{\mathcal{T}}\int_{\mathcal{T}}X_i(t)X_i(s)\varphi_{k_1}(t)\varphi_{k_2}(s)\ \mathrm{d}t\,\mathrm{d}s\\
&=&n\int_{\mathcal{T}}\int_{\mathcal{T}}H_n(t,s)\varphi_{k_1}(t)\varphi_{k_2}(s)\ \mathrm{d}t\,\mathrm{d}s.
\end{eqnarray*}
As a result, $\sum_{i=1}^n\mathbb{E}\xi_{ik_1}\xi_{ik_2}=0$ if $k_1\neq k_2$.

\textbf{c. $\Rightarrow$ a.:} Recall that $\{\tilde{\varphi}_k;k\geq 1\}$ are any orthonormal basis functions in $\mathcal{L}^2(\mathcal{T})$ and $\tilde{\xi}_{ik}$'s are any random variables.
Without loss of generality, we assume that $\sum_{i=1}^n\mathbb{E}\big\|X_i-\sum_{k=1}^K\tilde{\xi}_{ik}\tilde{\varphi}_k\big\|^2$ is finite.

Notice that
\begin{eqnarray*}
  \bigg\|X_i-\sum_{k=1}^K\langle X_i,\tilde{\varphi}_k\rangle\tilde{\varphi}_k\bigg\|^2
  \leq \bigg\|X_i-\sum_{k=1}^K\tilde{\xi}_{ik}\tilde{\varphi}_k\bigg\|^2,\ \text{a.s.}
\end{eqnarray*}
Consequently,
\begin{eqnarray*}
    \sum_{i=1}^n\mathbb{E}\bigg\|X_i-\sum_{k=1}^K\langle X_i,\tilde{\varphi}_k\rangle\tilde{\varphi}_k\bigg\|^2
    \leq  \sum_{i=1}^n\mathbb{E}\bigg\|X_i-\sum_{k=1}^K\tilde{\xi}_{ik}\tilde{\varphi}_k\bigg\|^2.
\end{eqnarray*}
It remains to show that
\begin{eqnarray*}
    \sum_{i=1}^n \mathbb{E}\bigg\|X_i - \sum_{k=1}^K \xi_{ik} \varphi_k\bigg\|^2 
    \leq 
    \sum_{i=1}^n \mathbb{E}\bigg\|X_i - \sum_{k=1}^K \tilde{\xi}_{ik} \tilde{\varphi}_k\bigg\|^2,
\end{eqnarray*}
where we abuse the notation $\tilde{\xi}_{ik}$ to denote $\langle X_i, \tilde{\varphi}_k \rangle$ for all $i \in [n]$ and $k \geq 1$.

Note that
\begin{eqnarray*}
    \sum_{i=1}^n\mathbb{E}\bigg\|X_i-\sum_{k=1}^K\tilde{\xi}_{ik}\tilde{\varphi}_k\bigg\|^2
    =\sum_{i=1}^n\mathbb{E}\|X_i\|^2-\sum_{k=1}^K\sum_{i=1}^n\mathbb{E}\langle X_i,\tilde{\varphi}_k\rangle^2.
\end{eqnarray*}
Represent $\tilde{\varphi}_k=\sum_{g=1}^\infty \langle \tilde{\varphi}_k, {\varphi}_g\rangle \varphi_g:=\sum_{g=1}^\infty a_{gk}\varphi_g$. Therefore,
\begin{eqnarray*}
  \sum_{i=1}^n\mathbb{E}\langle X_i,\tilde{\varphi}_k\rangle^2
  &=&\sum_{i=1}^n\mathbb{E}\bigg\langle X_i,\sum_{g=1}^\infty a_{gk}\varphi_g\bigg\rangle^2\\
  &=&\sum_{i=1}^n\mathbb{E} \bigg(\sum_{g=1}^\infty a_{gk}\xi_{ig}\bigg)^2
  =\sum_{g=1}^{\infty}a_{gk}^2\sum_{i=1}^n\mathbb{E}\xi_{ig}^2.
\end{eqnarray*}
We claim that
\begin{eqnarray}\label{min_la}
\sum_{k=1}^K\sum_{g=1}^{\infty}a_{gk}^2\sum_{i=1}^n\mathbb{E}\xi_{ig}^2\leq \sum_{k=1}^K\sum_{i=1}^n\mathbb{E}\xi_{ik}^2,
\end{eqnarray}
which implies
\begin{equation*}
\sum_{i=1}^n\mathbb{E}\bigg\|X_i-\sum_{k=1}^K\xi_{ik}\varphi_k\bigg\|^2\leq \sum_{i=1}^n\mathbb{E}\bigg\|X_i-\sum_{k=1}^K\tilde{\xi}_{ik}\tilde{\varphi}_k\bigg\|^2.
\end{equation*}

To prove \eqref{min_la}, note that
\begin{eqnarray*}
\sum_{g=1}^{\infty}a_{gk}^2\sum_{i=1}^n\mathbb{E}\xi_{ig}^2
&=&\sum_{i=1}^n\mathbb{E}\xi_{iK}^2+\bigg(\sum_{g=1}^K a_{gk}^2\sum_{i=1}^n\mathbb{E}\xi_{ig}^2-\sum_{i=1}^n\mathbb{E}\xi_{iK}^2\sum_{g=1}^K a_{gk}^2\bigg)\\
&&\ -\bigg(\sum_{i=1}^n\mathbb{E}\xi_{iK}^2\sum_{g>K}a^2_{gk}-\sum_{g>K}a_{gk}^2\sum_{i=1}^n\mathbb{E}\xi_{ig}^2\bigg)\\
&=&\sum_{i=1}^n\mathbb{E}\xi_{iK}^2+\bigg\{\sum_{g=1}^K a_{gk}^2\bigg(\sum_{i=1}^n\mathbb{E}\xi_{ig}^2-\sum_{i=1}^n\mathbb{E}\xi_{iK}^2\bigg)\bigg\}\\
&&\ +\bigg\{\sum_{g>K}a^2_{gk}\bigg(\sum_{i=1}^n\mathbb{E}\xi_{ig}^2-\sum_{i=1}^n\mathbb{E}\xi_{iK}^2\bigg)\bigg\},
\end{eqnarray*}
where the term $\bigg\{\sum_{g>K}a^2_{gk}\bigg(\sum_{i=1}^n\mathbb{E}\xi_{ig}^2-\sum_{i=1}^n\mathbb{E}\xi_{iK}^2\bigg)\bigg\}$ is nonpositive since $\sum_{i=1}^n\mathbb{E}\xi_{ik}^2$ decreases as $k$ increases. Therefore,
\begin{eqnarray*}
\sum_{k=1}^K\sum_{g=1}^{\infty}a_{gk}^2\sum_{i=1}^n\mathbb{E}\xi_{ig}^2
&\leq &K\sum_{i=1}^n\mathbb{E}\xi_{iK}^2+\bigg(\sum_{k=1}^K\sum_{g=1}^K a_{gk}^2\bigg(\sum_{i=1}^n\mathbb{E}\xi_{ig}^2-\sum_{i=1}^n\mathbb{E}\xi_{iK}^2\bigg)\bigg)\\
&=& K\sum_{i=1}^n\mathbb{E}\xi_{iK}^2+
\bigg(\sum_{g=1}^K\bigg(\sum_{i=1}^n\mathbb{E}\xi_{ig}^2-\sum_{i=1}^n\mathbb{E}\xi_{iK}^2\bigg)\cdot\bigg(\sum_{k=1}^K a_{gk}^2\bigg)\bigg)\\
&\leq&\sum_{g=1}^K\bigg\{\sum_{i=1}^n\mathbb{E}\xi_{iK}^2+\bigg(\sum_{i=1}^n\mathbb{E}\xi_{ig}^2-\sum_{i=1}^n\mathbb{E}\xi_{iK}^2\bigg)\cdot 1 \bigg\} \\
&=&\sum_{g=1}^K\sum_{i=1}^n\mathbb{E}\xi_{ig}^2.    
\end{eqnarray*}
In the last inequality, we use the fact that $\sum_{k=1}^K a_{gk}^2\leq 1$ since $\tilde{\varphi}_k$'s are orthonormal functions.
Claim \eqref{min_la} holds.
\end{proof}

\

\subsubsection{Proof of Theorem~\ref{Theorem_FSVD}}
\begin{proof}
    Let $\hat{\mathcal{H}}_n$ and $\mathcal{H}_n$ be the integral operators associated with the kernels
    $$
    \hat{H}_n(t,s)=\frac{1}{n}\sum_{i=1}^n X_i(t)X_i(s)
    \quad\text{and}\quad
    H_n(t,s)=\mathbb{E}\hat{H}_n(t,s),
    $$
    respectively.
Notice that $H_n(t,s)$ can be represented by 
$
\sum_{k=1}^\infty\big(\frac{1}{n}\sum_{i=1}^n\mathbb{E}\xi_{i k}^2\big)\varphi_k(t)\varphi_k(s)
$
due to Theorem \ref{the: eui_in}, and
\begin{align*}
\inf_{1\leq k_1\neq k_2\leq K+1}\frac{1}{n}\bigg|\sum_{i=1}^n\mathbb{E}\xi_{i k_1}^2-\sum_{i=1}^n\mathbb{E}\xi_{i k_2}^2\bigg|\gtrsim n^{-2\delta}.
\end{align*}
Then
\begin{align}\label{fun_enh}
\left\|\phi_k-\varphi_k\right\|
\lesssim n^{2\delta}\cdot \left\|\hat{\mathcal{H}}_{n} -\mathcal{H}_n\right\|_{\infty}
\end{align}
for any $k\in[K]$ by Lemma 4.3 in \citet{bosq2000linear}.
Notice that
\begin{eqnarray*}
\left\|\hat{\mathcal{H}}_{n} -\mathcal{H}_n\right\|_{\infty}
&=&\sup_{\|f\|\leq 1}\sqrt{\int_{0}^1\bigg(\frac{1}{n}\sum_{i=1}^n\int_{0}^1 \big(X_i(t)X_i(s)- \mathbb{E}X_i(t)X_i(s) \big)f(t)\ \mathrm{d}t\bigg)^2\ \mathrm{d}s}\\
&\leq& \sup_{\|f\|\leq 1}\sqrt{\int_{0}^1\int_{0}^1\bigg(\frac{1}{n}\sum_{i=1}^n \big(X_i(t)X_i(s)- \mathbb{E}X_i(t)X_i(s) \big)\bigg)^2\ \mathrm{d}t\ \mathrm{d}s} \cdot \|f\|\\
&=& \frac{1}{n}\sqrt{\int_{0}^1\int_{0}^1 \bigg(\sum_{i=1}^n\chi_i(t,s)\bigg)^2\ \mathrm{d}t \mathrm{d}s}\\
&\leq &\frac{1}{n}\sqrt{\sum_{i=1}^n\int_{0}^1\int_{0}^1 \chi_i^2(t,s)\ \mathrm{d}t \mathrm{d}s+\sum_{1\leq i_1\neq i_2\leq n}Y_{i_1,i_2}},
\end{eqnarray*}
where \(
Y_{i_1,i_2}
= \int_0^1 \int_0^1 
\chi_{i_1}(t,s)\,\chi_{i_2}(t,s) \ \mathrm{d}t \mathrm{d}s, 
\) and 
\(
\chi_i(t,s) = X_i(t)X_i(s) - \mathbb{E}\big[X_i(t)X_i(s)\big]
\).
Therefore,
\begin{eqnarray}\label{bound_final}
    \left\|\phi_k-\varphi_k\right\|
&\lesssim& n^{2\delta -1}\cdot\sqrt{\sum_{i=1}^n\int_{0}^1\int_{0}^1 \chi_i^2(t,s)\ \mathrm{d}t \mathrm{d}s+\sum_{1\leq i_1\neq i_2\leq n}Y_{i_1,i_2}}.
\end{eqnarray}
By Markov's inequality,
\begin{eqnarray*}
    \sum_{i=1}^n\int_{0}^1\int_{0}^1 \chi_i^2(t,s)\ \mathrm{d}t \mathrm{d}s+\sum_{1\leq i_1\neq i_2\leq n}Y_{i_1,i_2}\leq x
\end{eqnarray*}
holds with probability at least 
$$
1-\frac{\sum_{i=1}^n\int_{0}^1\int_{0}^1 \mathbb{E}\chi_i^2(t,s)\ \mathrm{d}t \mathrm{d}s+\sum_{1\leq i_1\neq i_2\leq n}\mathbb{E}Y_{i_1,i_2}}{x}.
$$
Since
$$
\mathbb{E}\int_{0}^1\int_{0}^1 \sum_{i=1}^n\chi_i^2(t,s)\ \mathrm{d}t \mathrm{d}s\leq n\mathbb{E}\|X_i\|^4\leq nC,
$$ 
and
\begin{eqnarray*}
 \mathbb{E}Y_{i_1,i_2} \leq \varrho^2 \cdot{ \sqrt{\mathbb{E}{Y_{i_1,i_1}}\cdot {\mathbb{E}Y_{i_2,i_2}}}}\lesssim \varrho^2,
\end{eqnarray*}
let $x= C(n + n^2\varrho^2)$ with a sufficiently large constant $C$.
Then
\begin{eqnarray*}
    \sum_{i=1}^n\int_{0}^1\int_{0}^1 \chi_i^2(t,s)\ \mathrm{d}t \mathrm{d}s+\sum_{1\leq i_1\neq i_2\leq n}Y_{i_1,i_2}\lesssim n + n^2\varrho^2
\end{eqnarray*}
holds with high probability.
Then by \eqref{bound_final},
\begin{eqnarray*}
    \left\|\phi_k-\varphi_k\right\|\lesssim n^{2\delta -1}\cdot \sqrt{n + n^2\varrho^2}
    \leq n^{2\delta - 1/2} +n^{2\delta }\varrho
\end{eqnarray*}
holds with high probability.
By Lemma \ref{lm:projection},
\begin{eqnarray*}
   \operatorname{dist}(\phi_k,\varphi_k)\leq \left\|\phi_k-\varphi_k\right\|.
\end{eqnarray*}
The proof is complete.
\end{proof}

\

\subsubsection{Proof of Theorem \ref{the: fac}}

We first prove the following lemma.
\begin{Le}\label{the: intrinsic-basis}
    $\bm{L}\in \mathbb{R}^{n\times K}$ are the intrinsic basis vectors of $\{X_i(t); t\in \mathcal{T}\}$ if and only if there exists an orthogonal matrix $\bm{B}$ such that $\bm{L}\bm{B}$ are the top-$K$ eigenvectors of $\mathbb{E}\int_{\mathcal{T}}\bm{X}(t)\bm{X}^\top(t) \mathrm{d}t$.
\end{Le}

\begin{proof}
For any $\tilde{\bm{L}}\in \mathbb{R}^{n\times K}$ with orthonormal columns and any random function $\tilde{\bm{F}}(t)\in \mathbb{R}^K$, define
$\bm{G}(t)=\tilde{\bm{L}}^\top\bm{X}(t)$.
It follows that, for each $t$,
\begin{eqnarray*}
    \big\|\bm{X}(t)-\tilde{\bm{L}} \bm{G}(t)\big\|^2\leq \big\|\bm{X}(t)-\tilde{\bm{L}} \tilde{\bm{F}}(t)\big\|^2,\ \text{almost surely},
\end{eqnarray*}
because $\bm{G}(t)$ minimizes $\big\|\bm{X}(t)-\tilde{\bm{L}} \bm{u}\big\|^2$ with respect to $\bm{u}\in\mathbb{R}^K$ for each $t$. As a result,
\begin{eqnarray*}
   \int_{\mathcal{T}} \mathbb{E}\big\|\bm{X}(t)-\tilde{\bm{L}} \bm{G}(t)\big\|^2\ \mathrm{d}t
   \leq \int_{\mathcal{T}} \mathbb{E}\big\|\bm{X}(t)-\tilde{\bm{L}} \tilde{\bm{F}}(t)\big\|^2\ \mathrm{d}t.
\end{eqnarray*}
By the definition of IBVs,
\begin{eqnarray*}
   \int_{\mathcal{T}} \mathbb{E}\big\|\bm{X}(t)-{\bm{L}} \bm{F}(t)\big\|^2\ \mathrm{d}t
   \leq \int_{\mathcal{T}} \mathbb{E}\big\|\bm{X}(t)-\tilde{\bm{L}} {\bm{G}}(t)\big\|^2\ \mathrm{d}t,
\end{eqnarray*}
where $\bm{L}$ has orthonormal columns and represents the IBVs, and $\bm{F}(t)=\bm{L}^\top\bm{X}(t)$. 

Notice that
\begin{eqnarray}\label{fac_equa}
    \mathbb{E}\big\|\bm{X}(t)-\tilde{\bm{L}} \bm{G}(t)\big\|^2
    &=&\mathbb{E}\|\bm{X}(t)\|^2-\mathbb{E}\big[\bm{X}^\top(t)\tilde{\bm{L}}\tilde{\bm{L}}^\top\bm{X}(t)\big]\nonumber\\
    &=&\operatorname{tr}\left(\mathbb{E}\big[\bm{X}(t)\bm{X}^\top(t)\big]\left(\bm{I}-\tilde{\bm{L}}\tilde{\bm{L}}^\top\right)\right),
\end{eqnarray}
so integrating over $\mathcal{T}$ yields
\begin{eqnarray*}
  \operatorname{tr}\left(\left(\int_{\mathcal{T}}\mathbb{E}\big[\bm{X}(t)\bm{X}^\top(t)\big]\ \mathrm{d}t\right)\left(\bm{I}-\bm{L}\bm{L}^\top\right)\right)
  \leq
  \operatorname{tr}\left(\left(\int_{\mathcal{T}}\mathbb{E}\big[\bm{X}(t)\bm{X}^\top(t)\big]\ \mathrm{d}t\right)\left(\bm{I}-\tilde{\bm{L}}\tilde{\bm{L}}^\top\right)\right),
\end{eqnarray*}
or equivalently,
\begin{eqnarray*}
  \operatorname{tr}\left(\bm{L}^\top\left(\int_{\mathcal{T}}\mathbb{E}\big[\bm{X}(t)\bm{X}^\top(t)\big]\ \mathrm{d}t\right)\bm{L}\right)
  \geq
  \operatorname{tr}\left(\tilde{\bm{L}}^\top\left(\int_{\mathcal{T}}\mathbb{E}\big[\bm{X}(t)\bm{X}^\top(t)\big]\ \mathrm{d}t\right)\tilde{\bm{L}}\right),
\end{eqnarray*}
for any $\tilde{\bm{L}}$. This implies that $\bm{L}$ maximizes the projected variance.
Consequently, there exists an orthogonal matrix $\bm{B}\in \mathbb{R}^{K\times K}$ such that $\bm{L}\bm{B}$ consists of the first $K$ eigenvectors of $\int_{\mathcal{T}}\mathbb{E}\big[\bm{X}(t)\bm{X}^\top(t)\big]\ \mathrm{d}t$.
\end{proof}

Next, we specifically consider the case where 
\(K = \operatorname{rank}\!\left(\mathbb{E}\int_{\mathcal{T}} \bm{X}(t)\bm{X}^\top(t)\,\mathrm{d}t\right)\) 
and prove the equivalence stated in Theorem~\ref{the: fac}.

\begin{proof}
\textbf{a. $\Rightarrow$ b.:} Note that $K$ is the rank of $\int_{\mathcal{T}}\mathbb{E}\bm{X}(t)\bm{X}^\top(t)\ \mathrm{d}t$. By Lemma \ref{the: intrinsic-basis}, we have that
\begin{eqnarray*}
    \int_{0}^1\mathbb{E}\bm{X}(t)\bm{X}^\top(t)\ \mathrm{d}t=  \bm{L}\bm{B} \bm{\Lambda}\bm{B}^\top\bm{L}^\top,
\end{eqnarray*}
where $\bm{\Lambda}\in \mathbb{R}^{K\times K}$ is a diagonal matrix with its diagonal elements being the positive eigenvalues of $\int_{0}^1\mathbb{E}\bm{X}(t)\bm{X}^\top(t)\ \mathrm{d}t$.
Therefore, there exists a positive-definite matrix $\bm{A}\in \mathbb{R}^{K\times K}$ such that
\begin{eqnarray*}
\int_{0}^1\mathbb{E}\bm{X}(t)\bm{X}^\top(t)\ \mathrm{d}t=  \bm{L} \bm{A}\bm{L}^\top.
\end{eqnarray*}
By \eqref{fac_equa},
$$
\mathbb{E}\int_0^1\big\|\bm{X}(t)-\bm{L} \bm{F}(t)\big\|^2\ \mathrm{d}t
=\int_0^1\mathbb{E}\big\|\bm{X}(t)-\bm{L} \bm{F}(t)\big\|^2\ \mathrm{d}t
=\operatorname{tr}\bigg\{\bm{L} \bm{A}\bm{L}^\top\bigg(\bm{I}-\bm{L}\bm{L}^\top\bigg)\bigg\}=0.
$$
This in turn leads to
$$
\bm{X}(t)=\bm{L} \bm{F}(t),\ t\in \mathcal{S},
$$
almost surely, where $\mathcal{S}\subset [0,1]$ has Lebesgue measure one. 

\textbf{b. $\Rightarrow$ c.:} By b., we have
\begin{eqnarray}\label{factot_tran}
\int_{\mathcal{T}}\bm{X}(t)\bm{X}^\top(t)\ \mathrm{d}t=\int_{\mathcal{S}}\bm{X}(t)\bm{X}^\top(t)\ \mathrm{d}t=  \bm{L} \bigg(\int_{\mathcal{S}}\bm{F}(t)\bm{F}^\top(t)\ \mathrm{d}t\bigg)\bm{L}^\top,
\end{eqnarray}
almost surely. Let us consider the eigendecomposition of $\int_{\mathcal{S}}\bm{F}(t)\bm{F}^\top(t)\ \mathrm{d}t=\bm{B}\bm{\Lambda}\bm{B}^\top$, where $\bm{B}\in \mathbb{R}^{K\times H}$ and $\bm{\Lambda}\in \mathbb{R}^{H\times H}$ is diagonal, with $H=\operatorname{rank}\!\left(\int_{\mathcal{S}}\bm{F}(t)\bm{F}^\top(t)\ \mathrm{d}t\right)$. It follows that
\begin{eqnarray}\label{eigen_vector}
\int_{\mathcal{T}}\bm{X}(t)\bm{X}^\top(t)\ \mathrm{d}t=\bm{L}\bm{B}\bm{\Lambda}\bm{B}^\top\bm{L}^\top,
\end{eqnarray}
almost surely,
where $\bm{B}\in \mathbb{R}^{K\times H}$ and $\bm{\Lambda}\in \mathbb{R}^{H\times H}$. 

Recall that $R$ is the rank of $\int_{\mathcal{T}}\bm{X}(t)\bm{X}^\top(t)\ \mathrm{d}t$ due to Theorem \ref{th:functional-SVD}. By \eqref{factot_tran}, we have $R\leq H$. Since $\bm{F}(t)=\bm{L}^\top\bm{X}(t)$,
\begin{eqnarray*}
    \int_{\mathcal{S}}\bm{F}(t) \bm{F}^\top(t)\ \mathrm{d}t
    =\bm{L}^\top\bigg(\int_{\mathcal{T}}\bm{X}(t) \bm{X}^\top(t)\ \mathrm{d}t \bigg)\bm{L},
\end{eqnarray*}
almost surely. Therefore, $H\leq R$, almost surely, and hence $H=R$, almost surely.
Accordingly, Theorem \ref{th:functional-SVD} indicates that the eigenvectors of $\int_{\mathcal{T}}\bm{X}(t)\bm{X}^\top(t)\ \mathrm{d}t$, which are $\bm{L}\bm{B}\in \mathbb{R}^{n\times R}$, are the singular vectors of $X_i$s.

We next prove that $R\leq K$. Take any vector $\bm{b}\in \mathbb{R}^n$ such that
$\bm{b}^\top\bigg(\int_{\mathcal{T}}\mathbb{E}\bm{X}(t)\bm{X}^\top(t)\ \mathrm{d}t\bigg) \bm{b}=0$.
Since $\int_{\mathcal{T}}\bm{X}(t)\bm{X}^\top(t)\ \mathrm{d}t$ is a positive semidefinite matrix, we have
\begin{eqnarray*}
\bm{b}^\top\bigg(\int_{\mathcal{T}}\bm{X}(t)\bm{X}^\top(t)\ \mathrm{d}t\bigg) \bm{b}    \geq 0.
\end{eqnarray*}
Moreover, taking expectation gives
\begin{eqnarray*}
\mathbb{E}\bigg[\bm{b}^\top\bigg(\int_{\mathcal{T}}\bm{X}(t)\bm{X}^\top(t)\ \mathrm{d}t\bigg) \bm{b}\bigg]
=
\bm{b}^\top\bigg(\int_{\mathcal{T}}\mathbb{E}\bm{X}(t)\bm{X}^\top(t)\ \mathrm{d}t\bigg) \bm{b}
=0,
\end{eqnarray*}
so the nonnegative random variable $\bm{b}^\top\left(\int_{\mathcal{T}}\bm{X}(t)\bm{X}^\top(t)\ \mathrm{d}t\right)\bm{b}$ must be zero almost surely.
This implies
$$
\operatorname{null}\bigg(\int_{\mathcal{T}}\mathbb{E}\bm{X}(t)\bm{X}^\top(t)\ \mathrm{d}t\bigg)\subset \operatorname{null}\bigg(\int_{\mathcal{T}}\bm{X}(t)\bm{X}^\top(t)\ \mathrm{d}t\bigg),
$$
where $\operatorname{null}(\cdot)$ denotes the null space of a matrix. Therefore, $R\leq K$.

\textbf{c. $\Rightarrow$ a.:} 
By c., there exist a diagonal matrix $\bm{\Lambda}\in \mathbb{R}^{R\times R}$ and $\bm{B}\in \mathbb{R}^{K\times R}$ such that
$$
\int_{\mathcal{T}}\bm{X}(t)\bm{X}^\top(t)\ \mathrm{d}t=\bm{L}\bm{B}\bm{\Lambda}\bm{B}^\top\bm{L}^\top,
$$
almost surely.
Therefore,
\begin{eqnarray*}
\int_{\mathcal{T}}\mathbb{E}\bm{X}(t)\bm{X}^\top(t)\ \mathrm{d}t=\bm{L}\big(\mathbb{E}\bm{B}\bm{\Lambda}\bm{B}^\top\big)\bm{L}^\top.
\end{eqnarray*}
Since $\operatorname{rank}\!\left(\mathbb{E}\int_{\mathcal{T}} \bm{X}(t)\bm{X}^\top(t)\, \mathrm{d}t\right)=K$ and $K\leq n$, we have $\operatorname{rank}\!\big(\mathbb{E}\bm{B}\bm{\Lambda}\bm{B}^\top\big)\geq K$. Therefore, the rank of the matrix $\mathbb{E}\bm{B}\bm{\Lambda}\bm{B}^\top\in \mathbb{R}^{K\times K}$ is $K$.
By the eigendecomposition of $\mathbb{E}\bm{B}\bm{\Lambda}\bm{B}^\top$, we then prove that $\bm{L}$ are the intrinsic basis vectors of $X_i$s due to Lemma \ref{the: intrinsic-basis}.
\end{proof}

\

\subsubsection{Proof of Theorem~\ref{Theorem_FSVD_vector}}

\begin{proof}
Notice that $\bm{F}(t)=\bm{L}^\top\bm{X}(t)$,
\begin{eqnarray*}
    \int_{\mathcal{S}}\bm{F}(t) \bm{F}^\top(t)\ \mathrm{d}t
    =\bm{L}^\top\bigg(\int_{\mathcal{T}}\bm{X}(t) \bm{X}^\top(t)\ \mathrm{d}t \bigg)\bm{L}.
\end{eqnarray*}
Since $\int_{\mathcal{T}}\bm{F}(t)\bm{F}^\top(t)\, \mathrm{d}t \in \mathbb{R}^{K \times K}\ \text{is non-singular}$, we have $R\geq K$.
By c. in Theorem \ref{the: fac}, we then have $R=K$, and $\bm{L}\bm{B}^{\top}$ contain the singular vectors of $X_i$s, where $\bm{B}=(\bm{b}_1,\dots,\bm{b}_K)^\top\in \mathbb{R}^{K\times K}$ is any orthogonal matrix. 

As a result, $\bm{L}$ can be obtained via $(\bm{a}_1,\cdots,\bm{a}_R)\bm{B}$ using the singular vectors $\bm{a}_r$ of $X_i$s. The corresponding factor series can be constructed by
\begin{eqnarray*}
\bm{F}(t)=\bm{L}^\top\bm{X}(t)
=\bm{B}^{\top}(\bm{a}_1,\cdots,\bm{a}_R)^\top\sum_{r=1}^R\rho_r\bm{a}_r\phi_r(t)
=\sum_{r=1}^R\rho_r\bm{b}_r\phi_r(t),\quad t\in \mathcal{T}.
\end{eqnarray*}
Since the factor loading matrix of $\bm{X}$ is unique only up to an invertible transformation, $\bm{B}$ can be any invertible matrix.
\end{proof}

\

\subsection{Statistical Convergences of FSVD}

\subsubsection{Statistical Convergences of Intrinsic Bases}\label{sec:theory}

Here, we establish statistical guarantees of FSVD for irregularly observed functional data, 
separately under the IBF and IBV frameworks.
We define the sine of the angle of pairs of vectors/functions to measure the estimation errors:
\begin{equation}\label{eq:distance-def}
\begin{split}
    \operatorname{dist}(\bm{u},\bm{v}) &= \sqrt{1 - \bigg(\frac{\bm{u}^\top \bm{v}}{\|\bm{u}\|\cdot\|\bm{v}\|}\bigg)^2}, \quad \forall \bm{u},\bm{v} \in \mathbb{R}^n;\\
 	\operatorname{dist}(f, g) &= \sqrt{1 - \bigg(\frac{\langle f,g\rangle}{\|f\|\cdot\|g\|}\bigg)^2}, \quad \forall f,g \in \mathcal{L}^2(\mathcal{T}).
\end{split}
\end{equation}

In the following, we only state the theoretical result for the first component. We introduce the following assumptions.

\begin{asum}\label{A2}
     The numbers of observed time points $\big\{J_i;i\in [n]\big\}$ are fixed positive integers, and there exists a number $m$ and a constant $C$ such that $\min_{i\in [n]}J_i\geq Cm$. In addition, the time points $\{T_{ij}; j \in [J_i]\}$ are independently drawn from a uniform distribution on $[0,1]$ for each $i$.
\end{asum}

\begin{asum}\label{A3}
  The measurement errors $\varepsilon_{ij}$ are independent of $T_{ij}$ and follow mean-zero sub-Gaussian distributions that satisfy $\mathbb{E}\exp({\lambda\varepsilon_{ij}})\leq \exp(\lambda^2\sigma^2/2)$ for all $i$, $j$, and $\lambda\in \mathbb{R}$. 
\end{asum}

Besides, we assume that $\{X_i; i \in [n]\}$ are heterogeneous functional data from the Sobolev space $\mathcal{W}_q^2(\mathcal{T})$ with $q > 1/2$. The true singular values, singular functions, and singular vectors of the $X_i$s are denoted by $\rho_r^0$, $\phi_r^0$, and $\bm{a}_r^0$ for $r \in [R]$, respectively. To control the randomness of the $X_i$s, we impose the following assumptions.

\begin{asum}\label{A4}
$\big\|D^q\left(\sum_{i=1}^n a_i X_i\right)\big\|\lesssim \rho_R^0$ for all $\{a_i;i\in [n]\}$ satisfying $\sum_{i=1}^na_i^2\leq 1$.
\end{asum}

\begin{asum}\label{A7}
The ratio of singular values $\kappa=\rho_1^0/\rho_2^0$, $m$, and the signal-to-noise ratio $\rho_1^0/\sigma$ satisfy $\kappa\gtrsim R$, $m^{{1}/{(2q+1)}}\gtrsim \log(n)$, $m^{q-1}\gtrsim (\rho_1^0)^2$, and  $\rho_1^0/\sigma\gtrsim  n^{1/2+1/(2q-1)}/ \sqrt{m}$.
\end{asum}

Assumption \ref{A4} ensures that the $\mathcal{L}^2$ norm of singular functions' $q$th derivatives, i.e., $\|D^q \phi_r^0\|^2$ $= \big\|D^q\left(\sum_{i=1}^n a_{ir}^0 X_i\right)\big\|^2/(\rho_r^0)^2$, $r \in [R]$, is bounded by a constant. This controls the bias of the estimated singular functions. Similar conditions have been adopted in the theoretical analysis of methods using Sobolev spaces \citep{speckman1985spline, cai2011optimal,hsing2015theoretical}.
Moreover, Assumption \ref{A7} suggests that the ratio of singular values is sufficiently large, the observed time grids of functions are sufficiently dense, and the signal-to-noise ratio is adequately high. These conditions can be achieved if $R$ grows with $\kappa$, and $n$ and $\rho_1^0$ grow with $m$, ensuring that errors arising from noises and discrete observation are controllable. 

\paragraph*{Intrinsic Basis Functions} We first establish the convergence rate of the intrinsic basis functions estimated using FSVD; see the following theorem.
\begin{theoremS}[Convergence Rates for Intrinsic Basis Functions]\label{Co_bound_sin_intr}
Suppose the conditions in Theorem~\ref{the: eui_in} and Assumptions~\ref{A2}--\ref{A7} hold.
Assume $X_i$s are independent heterogeneous functional data valued in $\mathcal{W}_q^2(\mathcal{T})$, which satisfies the conditions in Theorem~\ref{Theorem_FSVD}.
$\hat{\phi}_1$ is the output of Algorithm \ref{algo: FSVD} with tuning parameter $\nu \asymp \big(n^{1-2\delta}m\big)^{-2q/(2q+1)}$. Then 
\begin{equation}\label{error_singular}
    \operatorname{dist}(\hat{\phi}_1,\varphi_1)\lesssim m^{-\frac{q}{2q+1}}+\sigma\cdot\big\{n^{\delta}m^{-1/2}+ (n^{1-2\delta}m)^{-\frac{q}{2q+1}}\big\} +n^{2\delta -1/2}
\end{equation}
holds with high probability.
\end{theoremS}

By Theorem~\ref{Co_bound_sin_intr}, the error between \( \hat{\phi}_1 \) and \( \varphi_1 \) consists of three sources of uncertainty: (i) the uncertainty due to discretization of the time grid, \( m^{-\frac{q}{2q+1}} \); (ii) the uncertainty from observation noise, $\sigma\cdot\big\{n^{\delta}m^{-1/2}+ (n^{1-2\delta}m)^{-\frac{q}{2q+1}}\big\}$; and (iii) the uncertainty from the randomness of the functional data, \( n^{2\delta-1/2} \).
The first two terms,
\begin{eqnarray}\label{first_two}
    m^{-\frac{q}{2q+1}} +\sigma\cdot\big\{n^{\delta}m^{-1/2}+ (n^{1-2\delta}m)^{-\frac{q}{2q+1}}\big\},
\end{eqnarray}
account for the error rate between \( \hat{\phi}_1 \) and the true singular function \( \phi_1^0 \). When $\delta = 0$, i.e., $X_i$s exhibit a strong signal on common functional patterns, the convergence rates coincide with those established for three-way CP decomposition of tensor functional data \citep{han2023guaranteed}. 
In that setting, the $n$ functions $X_i(t)$ are assumed to follow a tensor structure $X(i_1, i_2, t)$, with $i_1 \in [n_1]$, $i_2 \in [n_2]$, and $n_1 \times n_2 = n$. 
The tensor structure refines the term $\sigma \cdot m^{-1/2}$ in the rate~\eqref{first_two} to 
\[
\sigma \cdot m^{-1/2} \cdot \sqrt{\tfrac{n_1 + n_2}{n}},
\] 
as shown in Corollary~2 of \citet{han2023guaranteed}.

When $\delta = 0$, the terms $\sigma (nm)^{-\frac{q}{2q+1}}$ and $n^{-1/2}$ in \eqref{error_singular} also appear in the literature on mean function estimation for i.i.d.\ functional data \citep{cai2011optimal, hsing2015theoretical}, and both decrease as $n$ increases. 
When the observed times are sufficiently dense, i.e., $m \gtrsim n^{\frac{2q+1}{2q}}$, and $\sigma$ is independent of $n$ and $m$, we obtain
\[
\operatorname{dist}(\hat{\phi}_1,\varphi_1) \lesssim \frac{1}{n},
\]
which holds with high probability. 
This result highlights that the estimation of IBFs can be improved by pooling functions together in FSVD, paralleling the i.i.d.\ case \citep{cai2011optimal, hsing2015theoretical}.

\begin{Remark}[i.i.d.\ case vs.\ heterogeneous case]
In contrast to the i.i.d.\ setting, we require the number of time points \( m \to \infty \) to ensure the consistency of IBFs. 
This requirement has also been reported in settings where homogeneity across the functions \( X_i \) is not assumed \citep{han2023guaranteed, zhang2024individualized}. 
When the i.i.d.\ condition holds, the IBF \( \varphi_1 \) reduces to the optimal empirical basis function summarized in Table~\ref{basis_comparison}, and the estimator \( \hat{\phi}_1 \) resembles the SOAP method proposed by \citet{nie2022recovering}. 
In this case, the requirement \( m \to \infty \) can be removed, as shown in Theorem~3 of \citet{nie2022recovering}.
\end{Remark}

\paragraph*{Intrinsic Basis Vectors} Next, we establish the error rate for estimating the first factor loading using Algorithm~\ref{algo: FM}.

\begin{theoremS}[Convergence Rates for Intrinsic Basis Vectors]\label{Co_bound_sin_intr_vec}
Suppose Assumptions \ref{A2}--\ref{A7} and the conditions in Theorem~\ref{the: fac} hold. 
Let 
\(
K = \operatorname{rank}\!\left(\mathbb{E}\int_{\mathcal{T}} \bm{X}(t)\bm{X}^\top(t)\, \mathrm{d}t\right),
\)
and assume that the $K$th singular value of $X_i$s satisfies $\rho_K \gtrsim n^{1/2-\delta}$ with $\delta \in [0,1/2)$. 
Let $\hat{\bm{a}}_1$ denote the output of Algorithm~\ref{algo: FSVD} with $\nu \asymp \big(n^{1-2\delta}m\big)^{-2q/(2q+1)}$. 
Then $(\bm{a}^0_1,\ldots,\bm{a}^0_K)$ are the IBVs of $X_i$s, and with high probability,
\begin{equation*}
    \operatorname{dist}(\hat{\bm{a}}_1,\bm{a}^0_1)
    \;\lesssim\; m^{-\frac{q}{2q+1}}
    + \sigma \cdot \Big\{ n^{\delta} m^{-1/2} 
    + (n^{1-2\delta} m)^{-\frac{q}{2q+1}} \Big\}.
\end{equation*}
\end{theoremS}

The condition $\rho_K \asymp n^{1/2-\delta}$ quantifies the strength of the factors, where a smaller $\delta$ indicates stronger factors. Similar conditions have been adopted in the theoretical analyses of factor models \citep{lam2011estimation, lam2012factor}.
Under this, the distance between $\hat{\bm{a}}_1$ and $\bm{a}_1^0$ is composed by two terms of uncertainty: the uncertainty from the discrete time grid ($m^{-\frac{q}{2q+1}}$) and the uncertainty from noise ($\sigma\cdot\big\{n^{\delta}m^{-1/2}+ (n^{1-2\delta}m)^{-\frac{q}{2q+1}}\big\}$). 
When $\sigma$ is a constant free of $n$ and $m$, both terms converge to 0 as $m\rightarrow\infty$ if $n$ is fixed, while $n^{\delta}m^{-1/2}$ in the noise term may diverge with $n$ if the factor strength is not strong enough or $n$ increases too fast compared to $m$ (e.g., $\delta>0$ and $m^{1/2}\lesssim n^{\delta}$).
This result is aligned with the theoretical findings on factor loading estimation in \citet{lam2012factor}.


\subsubsection{Proof of Theorem \ref{Co_bound_sin_intr} and \ref{Co_bound_sin_intr_vec}}

\begin{proof}[Proof of Theorem \ref{Co_bound_sin_intr}]
We first prove that $\rho_1^0 \gtrsim n^{1/2 - \delta}$ holds with high probability.  
Notice that
\begin{eqnarray*}
    \rho_1^0
    &=& \arg\max_{\{\phi:\|\phi\|\leq 1\}} 
    \sqrt{\sum_{i=1}^n \langle X_i, \phi\rangle^2} 
    \;\;\geq\;\; 
    \sqrt{\sum_{i=1}^n \langle X_i, \varphi_1\rangle^2} 
    = \sqrt{\sum_{i=1}^n \xi_{i1}^2}.
\end{eqnarray*}
Consider
\begin{eqnarray*}
   -\bigg(\sum_{i=1}^n \xi_{i1}^2 - \sum_{i=1}^n \mathbb{E}\xi_{i1}^2\bigg) 
   \leq  \sqrt{n}\, x.
\end{eqnarray*}
This inequality holds with probability at least 
\(
1-\frac{\operatorname{Var}\!\left(\sum_{i=1}^n \xi_{i1}^2\right)}{x^2 n}
\)
by Chebyshev’s inequality.  
Note that
\[
\operatorname{Var}\!\left(\sum_{i=1}^n \xi_{i1}^2\right)
\leq \sum_{i=1}^n \mathbb{E}\langle X_i,\varphi_1\rangle^4
\leq \sum_{i=1}^n \mathbb{E}\|X_i\|^4
\leq nC,
\]
and, due to Condition~\eqref{condi_iden} in the main text,
\[
\sum_{i=1}^n \mathbb{E}\xi_{i1}^2 \;\gtrsim\; n^{1-2\delta}.
\]
Therefore,
\begin{eqnarray*}
   \sum_{i=1}^n \xi_{i1}^2 
   &\geq& \sum_{i=1}^n \mathbb{E}\xi_{i1}^2 - \sqrt{n}\, x \\
   &\gtrsim& n^{1-2\delta} - \sqrt{n}\, x
\end{eqnarray*}
with probability at least $1 - C/x^2$.  
Taking $x = \tfrac{1}{2} n^{1/2 - 2\delta}$ gives
\(
   \sum_{i=1}^n \xi_{i1}^2 \;\gtrsim\; n^{1-2\delta}
\)
with probability at least $1 - 4C/n^{1-4\delta}$.  
Therefore, we conclude that
\[
\rho_1^0 \gtrsim n^{1/2 - \delta}
\]
holds with high probability.

Without loss of generality, $\langle\varphi_1,\hat{\phi}_1\rangle\geq 0$ and $\langle \phi_1,\varphi_1\rangle\geq 0$. 
Notice that
\begin{eqnarray*}
    \operatorname{dist}(\hat{\phi}_1,\phi_1)
    &=&\sqrt{1-\langle\hat{\phi}_1,\phi_1\rangle^2}
    \geq \frac{1}{\sqrt{2}}\cdot \sqrt{2-2\langle\hat{\phi}_1,\phi_1\rangle}
    =\frac{1}{\sqrt{2}}\cdot\|\hat{\phi}_1-\phi_1\|,\\
    \operatorname{dist}(\phi_1,\varphi_1)
    &=&\sqrt{1-\langle\phi_1,\varphi_1\rangle^2}
    \geq \frac{1}{\sqrt{2}}\cdot \sqrt{2-2\langle\phi_1,\varphi_1\rangle}
    =\frac{1}{\sqrt{2}}\cdot\|\phi_1-\varphi_1\|.
\end{eqnarray*}
By Lemma~\ref{lm:projection},
\begin{eqnarray}\label{dist_inq}
    \operatorname{dist}(\hat{\phi}_1,\varphi_1)
    \leq \|\hat{\phi}_1-\varphi_1\|
    \leq \|\hat{\phi}_1-\phi_1\|+\|\phi_1-\varphi_1\|
    \lesssim \operatorname{dist}(\hat{\phi}_1,\phi_1)+\operatorname{dist}(\phi_1,\varphi_1).
\end{eqnarray}
Under Assumptions~\ref{A2}--\ref{A7}, and since $\rho^0_1 \gtrsim n^{1/2 - \delta}$ holds with high probability, 
Theorem~\ref{sta_pro_FSVD} implies that
\begin{eqnarray*}
    \operatorname{dist}(\hat{\phi}_1,\phi_1)
    &\lesssim& m^{-\frac{q}{2q+1}}
    + \sigma\bigg(
        \frac{n^{\delta}}{\sqrt{m}}
        + \frac{1}{n^{1/2-\delta}\sqrt{m}} \cdot \frac{1}{\nu^{1/(4q)}}
      \bigg)
    + \sqrt{\nu},
\end{eqnarray*}
which holds with high probability.  
If $\nu \asymp \big(n^{1-2\delta}m\big)^{-2q/(2q+1)}$, it follows that
\begin{eqnarray*}
    \operatorname{dist}(\hat{\phi}_1,\phi_1)
    &\lesssim& m^{-\frac{q}{2q+1}}
    + \sigma\Big\{
        n^{\delta} m^{-1/2} 
        + (n^{1-2\delta}m)^{-\frac{q}{2q+1}}
      \Big\},
\end{eqnarray*}
which holds with high probability.

Note that $\varrho$ in Theorem~\ref{Theorem_FSVD} equals $0$ since $X_i$s are independent. 
Then, by Theorem~\ref{Theorem_FSVD}, 
\[
\operatorname{dist}(\phi_1,\varphi_1)\;\lesssim\; n^{2\delta - 1/2}
\]
holds with high probability, which completes the proof by~\eqref{dist_inq}.
\end{proof}

\begin{proof}[Proof of Theorem \ref{Co_bound_sin_intr_vec}]
By c. of Theorem~\ref{the: fac}, we have $R = K$ under the assumption 
$\rho_K \gtrsim n^{1/2 - \delta}$, where $R$ is the rank of the $X_i$s. 
It then follows, again by c. of Theorem~\ref{the: fac}, that $(\bm{a}_1^0, \ldots, \bm{a}_K^0)$ are the IBVs of $X_i$s. 
Finally, we complete the proof by applying Theorem~\ref{sta_pro_FSVD}.
\end{proof}


\
\subsubsection{Statistical Convergences of Singular Functions/Vectors}

In this subsection, we assume that the $X_i$s are deterministic functions from the Sobolev space $\mathcal{W}_q^2(\mathcal{T})$. The corresponding results when the $X_i$s are random functions can be derived similarly.
The estimates of the singular value, function, and vector from Algorithm~\ref{algo: FSVD} are denoted by $\hat{\rho}_1$, $\hat{\phi}_1$, and $\hat{\bm{a}}_1$, respectively.
According to Algorithm~\ref{algo: FSVD}, the estimates of $\rho_1^0 \phi_1^0$ and $\bm{a}_1^0$ at the $h$th iteration are denoted by $\widehat{\rho\phi}^{(h)}$ and $\hat{\bm{a}}^{(h)}$, respectively. 

\begin{theoremS}\label{sta_pro_FSVD} 
Suppose Assumptions \ref{A2} -- \ref{A7} hold. We assume that the tuning parameter $\nu$ satisfies $m^{-q/(2q+1)}+\frac{\sigma}{\rho_1^0}\cdot \sqrt{\frac{n}{m}}\cdot x\lesssim \nu^{1/(2q)}$ and $\frac{\sigma}{\rho_1^0\sqrt{m}}\cdot\frac{1}{\nu^{1/(4q)}}\cdot x+\sqrt{\nu}\lesssim 1$. Then
\begin{equation}\label{ineq:FSVD-convergence}
\begin{split}
 \max\big\{ \operatorname{dist}(\hat{\bm{a}}_1,{\bm{a}_1^0}),\operatorname{dist}(\hat{\phi}_1,{\phi}_1^0)\big\}
  \lesssim  m^{-\frac{q}{2q+1}}+\frac{\sigma}{\rho_1^0\sqrt{m}}\cdot\bigg({\sqrt{n}}+ \frac{1}{\nu^{1/(4q)}}\bigg)\cdot x+\sqrt{\nu}
\end{split}
\end{equation}
holds with probability at least $1-C_1\exp(-C_2m^{1/(2q+1)}) - 2\exp(-x^2/2)$, where $C_1$ and $C_2$ are constants independent of $n$ and $m$. 
\end{theoremS}

In \eqref{ineq:FSVD-convergence}, the first term $m^{-\frac{q}{2q+1}}$ quantifies the errors arising from discretely observed functional data valued in Sobolev spaces; the second term $\frac{\sigma}{\rho_1^0\sqrt{m}}\sqrt{n}$ and $\frac{\sigma}{\rho_1^0\sqrt{m}}\frac{1}{\nu^{1/(4q)}}$ account for uncertainties caused by the measurement noise. The tuning parameter $\nu$ balances the trade-off between the variance in the second term and the bias captured by the third term $\sqrt{\nu}$.

Before proving Theorem \ref{sta_pro_FSVD}, we assume that 
\begin{align}
    \sqrt{\sum_{i=1}^n\bigg(\frac{1}{J_i}\sum_{j=1}^{J_i}f_i(T_{ij})-\int_{0}^1f_i(t)\ \mathrm{d}t\bigg)^2}&\lesssim m^{-q/(2q+1)}\cdot \sqrt{\sum_{i=1}^n\|f_i\|^2}+m^{-2q/(2q+1)}\cdot \sqrt{\sum_{i=1}^n\|f_i\|_{\infty}^2},\label{C1}\\
    \sup_{i\in [n]}\bigg|\frac{1}{J_i}\sum_{j=1}^{J_i}f_i(T_{ij})-\int_{0}^1f_i(t)\ \mathrm{d}t\bigg|&\lesssim m^{-q/(2q+1)}\cdot \sup_{i\in [n]}\|f_i\|+m^{-2q/(2q+1)}\cdot \sup_{i\in [n]}\|f_i\|_{\infty},\label{C11}\\
    \sum_{i=1}^n\frac{1}{J_i}\sum_{j=1}^{J_i}|\varepsilon_{ij}|&\lesssim x\sqrt{\frac{n}{m}}\cdot \sigma,\label{C2}\\
    \sum_{i=1}^n\frac{a_{i1}^0}{J_i}\sum_{j=1}^{J_i}|\varepsilon_{ij}|&\lesssim x\sqrt{\frac{1}{m}}\cdot \sigma,\label{C3} 
\end{align}
where $f_i\in \mathcal{W}_2^q(\mathcal{T})$, $i\in [n]$, are any functions such that $\sup_{i\in [n]}\|f_i\|\lesssim 1$, $x$ is any positive real value, and we use the notation $\|\cdot\|_{\infty}$ to denote a norm for a function $f$ defined by $\|f\|_{\infty}=\sup_{t\in \mathcal{T}}|f(t)|$. 

The inequalities \eqref{C1}--\eqref{C3} hold with probability at least $1 - C_1 \exp(-C_2 m^{\frac{1}{2q+1}}) - 2 \exp(-x^2/2)$ under Assumptions~\ref{A2}, \ref{A3}, and \ref{A7}. These conclusions can be proven similarly to Lemma 4 in \citet{han2023guaranteed}, and by using the Hoeffding inequality.

Under the conditions \eqref{C1} -- \eqref{C3}, we propose the following three lemmas to prove Theorem \ref{sta_pro_FSVD}.

\begin{Le}\label{Le:3}
Under Assumptions \ref{A2} -- \ref{A7}, and conditions \eqref{C1} -- \ref{C3}, suppose that $\|\widehat{\rho\phi}^{(h)}\|\lesssim \rho_1^0$. Then 
\begin{eqnarray}\label{bound_A}
 \operatorname{dist}(\hat{\bm{a}}^{(h+1)},{\bm{a}}_1^0)&\leq& C\bigg(m^{-q/(2q+1)}+\sqrt{\frac{n}{m}} \frac{\sigma}{\rho_1^0}\cdot x+\frac{1}{\rho_1^0}\big\| \overline{\rho\phi}^{(h)}- \widehat{\rho\phi}^{(h)}\big \|\bigg)\nonumber\\
 &+& \frac{\operatorname{dist}^2(\hat{\bm{a}}^{(h)},{\bm{a}}_1^0)}{1+\sqrt{1-\operatorname{dist}^2(\hat{\bm{a}}^{(h)},{\bm{a}}_1^0)}}
+\frac{1}{\kappa^2}\operatorname{dist}(\hat{\bm{a}}^{(h)},{\bm{a}}_1^0),
\end{eqnarray}
where $\overline{\rho\phi}^{(h)}=\sum_{i=1}^n \hat{a}^{(h)}_{i}X_i$.
\end{Le}

\

\begin{Le}\label{le_cor}
Under Assumptions \ref{A2} -- \ref{A7} and conditions \eqref{C1} -- \ref{C3}, we assume that the tuning parameter $\nu$ satisfies $\frac{1}{\nu^{1/(4q)}}\cdot\frac{\sigma}{\rho_1^0\sqrt{m}}\cdot x+\sqrt{\nu}\lesssim 1$ and $m^{-q/(2q+1)}+\sqrt{\frac{n}{m}} \frac{\sigma}{\rho_1^0}\cdot x\lesssim \nu^{1/(2q)}$ for a fixed $x>0$. 
Then
\begin{eqnarray}\label{est_opt_fin}
     \| \overline{\rho\phi}^{(h)}- \widehat{\rho\phi}^{(h)}\|
     &\lesssim & \rho_1^0\bigg(\frac{1}{\nu^{1/(4q)}}\cdot\frac{\sigma}{\rho_1^0\sqrt{m}}\cdot x+\sqrt{\nu}+m^{-q/(2q+1)}+\sqrt{\frac{n}{m}} \frac{\sigma}{\rho_1^0}\cdot x\bigg)\nonumber\\
     &+&\rho_1^0\operatorname{dist}^2(\hat{\bm{a}}^{(h)},{\bm{a}}_1^0)
     .
\end{eqnarray} 
\end{Le}

\

\begin{Le}\label{Le:4}
Under the conditions in Lemma \ref{le_cor}, we have
\begin{eqnarray}
 \operatorname{dist}(\widehat{\rho\phi}^{(h)},\phi_1^0)&\lesssim & m^{-q/(2q+1)}+\sqrt{\frac{n}{m}} \frac{\sigma}{\rho_1^0}\cdot x+\frac{1}{\nu^{1/(4q)}}\cdot\frac{\sigma}{\rho_1^0\sqrt{m}}\cdot x+\sqrt{\nu}\nonumber\\&+&\operatorname{dist}({\hat{\bm{a}}}^{(h)},\bm{a}_1^0)\label{bound_phi}.
\end{eqnarray}
\end{Le}

\

The proof of the above three lemmas is presented in Section \ref{lemma_sec}.

\

\begin{proof}[Proof of Theorem \ref{sta_pro_FSVD}]

Without loss of generality, we assume that $x=1$.
We first claim that
\begin{eqnarray*}
   \|\widehat{\rho\phi}^{(h)}\|&\lesssim& \rho_1^0,\ \forall h\geq 0.
\end{eqnarray*}
Applying Lemma \ref{le_cor}, we have
\begin{eqnarray*}
  \|\widehat{\rho\phi}^{(h)}\|
  &\leq& \| \overline{\rho\phi}^{(h)}- \widehat{\rho\phi}^{(h)}\|+\|\overline{\rho\phi}^{(h)}\|\\
  &\lesssim& \rho_1^0\bigg(\frac{1}{\nu^{1/(4q)}}\cdot\frac{\sigma}{\rho_1^0\sqrt{m}} +\sqrt{\nu} +m^{-q/(2q+1)}+\sqrt{\frac{n}{m}} \frac{\sigma}{\rho_1^0}\bigg)+\rho_1^0\operatorname{dist}^2(\hat{\bm{a}}^{(h)},{\bm{a}}_1^0)\\
  &+&\bigg\|\sum_{i=1}^n \hat{a}^{(h)}_{i}X_i\bigg\|.
\end{eqnarray*}
Notice that 
$$
\frac{1}{\nu^{1/(4q)}}\cdot\frac{\sigma}{\rho_1^0\sqrt{m}} +\sqrt{\nu} +m^{-q/(2q+1)}+\sqrt{\frac{n}{m}} \frac{\sigma}{\rho_1^0} \lesssim 1
$$
by Assumption \ref{A7} and the conditions on $\nu$. In addition, 
$$
\rho_1^0\operatorname{dist}^2(\hat{\bm{a}}^{(h)}, {\bm{a}}_1^0)\lesssim \rho_1^0
$$ 
and
\begin{eqnarray*}
  \bigg\|\sum_{i=1}^n \hat{a}^{(h)}_{i}X_i\bigg\|\leq \rho_1^0  
\end{eqnarray*}
due to Lemma \ref{bound_le}. Combining the above bounds yields $\|\widehat{\rho\phi}^{(h)}\|\lesssim \rho_1^0$.

We now claim that 
\begin{eqnarray}\label{phi_bound_aup}
\operatorname{dist}(\hat{\bm{a}}^{(h)}, {\bm{a}}_1^0)  \lesssim  m^{-\frac{q}{2q+1}} + \frac{\sigma}{\rho_1^0} \cdot \frac{1}{\sqrt{m}} \cdot \left( \sqrt{n} + \frac{1}{\nu^{1/(4q)}} \right) + \sqrt{\nu} + \frac{1}{\kappa^{2(h-1)}},\ h\geq 1.
\end{eqnarray}
For $h=1$, we utilize Lemmas \ref{Le:3} and \ref{le_cor} to obtain
\begin{eqnarray*}
\operatorname{dist}(\hat{\bm{a}}^{(1)}, {\bm{a}}_1^0)&\leq& C\bigg( m^{-q/(2q+1)} + \sqrt{\frac{n}{m}} \frac{\sigma}{\rho_1^0} + \frac{1}{\nu^{1/(4q)}} \cdot \frac{\sigma}{\rho_1^0 \sqrt{m}} + \sqrt{\nu} \bigg) + 2\operatorname{dist}^2(\hat{\bm{a}}^{(0)}, {\bm{a}}_1^0)\\
&+& \frac{1}{\kappa^2} \operatorname{dist}(\hat{\bm{a}}^{(0)}, {\bm{a}}_1^0)\\
&\lesssim& C\bigg( m^{-q/(2q+1)} + \sqrt{\frac{n}{m}} \frac{\sigma}{\rho_1^0} + \frac{1}{\nu^{1/(4q)}} \cdot \frac{\sigma}{\rho_1^0 \sqrt{m}} + \sqrt{\nu} \bigg) + 1 + \frac{1}{\kappa^2}\\
&\lesssim& m^{-\frac{q}{2q+1}} + \frac{\sigma}{\rho_1^0} \cdot \frac{1}{\sqrt{m}} \left( \sqrt{n} + \frac{1}{\nu^{1/(4q)}} \right) + \sqrt{\nu} + 1.
\end{eqnarray*}
Then \eqref{phi_bound_aup} holds for $h=1$. Assume
\begin{eqnarray*}
\operatorname{dist}(\hat{\bm{a}}^{(h)}, {\bm{a}}_1^0)  \leq  C_h \bigg( m^{-\frac{q}{2q+1}} + \frac{\sigma}{\rho_1^0} \cdot \frac{1}{\sqrt{m}} \left( \sqrt{n} + \frac{1}{\nu^{1/(4q)}} \right) + \sqrt{\nu} \bigg) + \frac{D_h}{\kappa^{2(h-1)}}.
\end{eqnarray*}
Let 
\[
A = m^{-q/(2q+1)} + \sqrt{\frac{n}{m}} \frac{\sigma}{\rho_1^0} + \frac{1}{\nu^{1/(4q)}} \cdot \frac{\sigma}{\rho_1^0 \sqrt{m}} + \sqrt{\nu}.
\]
Then, by Lemma~\ref{Le:3},
\begin{eqnarray*}
\operatorname{dist}(\hat{\bm{a}}^{(h+1)}, {\bm{a}}_1^0)
&\leq& C A + \left( C_h A + \frac{D_h}{\kappa^{2(h-1)}} \right)^2 + \frac{1}{\kappa^2} \left( C_h A + \frac{D_h}{\kappa^{2(h-1)}} \right)\\
&\leq& A \left( C + C_h^2 A + \frac{2 C_h D_h}{\kappa^{2(h-1)}} + \frac{C_h}{\kappa^2} \right) + \frac{D_h + \frac{D_h^2}{\kappa^{2(h-2)}}}{\kappa^{2h}}.
\end{eqnarray*}
Let 
\begin{eqnarray*}
    C_{h+1} :&=& C + C_h^2 A + \frac{2 C_h D_h}{\kappa^{2(h-1)}} + \frac{C_h}{\kappa^2},\\
    D_{h+1} :&=& D_h + \frac{D_h^2}{\kappa^{2(h-2)}}.
\end{eqnarray*}
We next prove that the sequences $\{ C_h; h \geq 1 \}$ and $\{ D_h; h \geq 1 \}$ are both bounded.

Define $s_h = \frac{D_h}{\kappa^{h}}$. First, note that
$$
D_{h+1} = D_h + \frac{D_h^2}{\kappa^{2(h-2)}}=:D_h + \delta_h.
$$
We express $\delta_h$ in terms of $s_h$:
\[
\delta_h = \frac{D_h^2}{\kappa^{2(h-2)}} 
= \left( \frac{D_h}{\kappa^{h-2}} \right)^2
= \left( s_h \kappa^{2} \right)^2
= s_h^2 \kappa^{4}.
\]
Since $D_{h+1} = s_{h+1} \kappa^{h+1}$ and $D_h=s_h\kappa^h$, we have
$$
s_{h+1} \kappa^{h+1} = s_h \kappa^{h} + s_h^2 \kappa^{4},
$$
and hence
\begin{equation*}
s_{h+1} =  \frac{s_h}{\kappa} + s_h^2 \kappa^{3 - h}.
\end{equation*}
Since $\kappa^{3-h}\to 0$ geometrically and $s_h\ge 0$, this recursion implies that $s_h$ remains bounded and, in particular, $s_h\lesssim \kappa^{-h}$ for large $h$.
Therefore $D_h = s_h \kappa^{h}$ is bounded, and $\{D_h;h\ge 1\}$ is bounded.

Since $\{ D_h; h \geq 1 \}$ is bounded and non-decreasing, the limit $D := \lim_{h \rightarrow \infty} D_h$ exists. Therefore, the term $\frac{2 C_h D_h}{\kappa^{2(h-1)}}$ in $C_{h+1}$ is dominated by $\frac{C_h}{\kappa^2}$ as $h \rightarrow \infty$. By this observation, we consider 
\[
C_{h+1} \le C + C_h^2 A + \frac{C_h}{\kappa^2}.
\]
To ensure that $\{ C_h; h \geq 1 \}$ is bounded, it suffices to show that there exists an $M>0$ such that
\[
C + M^2 A + \frac{M}{\kappa^2} \leq M.
\]
This can be achieved if $\left(1 - \frac{1}{\kappa^2}\right)^2 \geq 4 A C$. When $A$ is sufficiently small, the above condition holds, and hence $\{ C_h; h \geq 1 \}$ is bounded.

Since $\{ C_h; h \geq 1 \}$ and $\{ D_h; h \geq 1 \}$ are bounded, we then prove \eqref{phi_bound_aup} for any $h \geq 1$. Let $h\rightarrow \infty$, 
\begin{eqnarray*}
\operatorname{dist}(\hat{\bm{a}}_{1},\bm{a}^0_{1})  
&\lesssim& m^{-q/(2q+1)} + \sqrt{\frac{n}{m}} \frac{\sigma}{\rho_1^0} + \frac{1}{\nu^{1/(4q)}} \cdot \frac{\sigma}{\rho_1^0 \sqrt{m}} + \sqrt{\nu}.
\end{eqnarray*}
This leads to 
\begin{eqnarray*}
\operatorname{dist}(\hat{\phi}_1, \phi_1^0)
\lesssim m^{-q/(2q+1)} + \sqrt{\frac{n}{m}} \frac{\sigma}{\rho_1^0} + \frac{1}{\nu^{1/(4q)}} \cdot \frac{\sigma}{\rho_1^0 \sqrt{m}} + \sqrt{\nu},
\end{eqnarray*}
due to Lemma \ref{Le:4}. 

\end{proof}

\

\clearpage
\newpage

\subsection{Other Lemmas}

\subsubsection{Proof of Lemmas \ref{Le:3} to \ref{Le:4}}\label{lemma_sec}
Define the empirical and expected loss functions of FSVD as follows
\begin{eqnarray*}
\mathcal{L}(nm,\phi,\bm{a}):&=&\sum_{i=1}^n\frac{1}{J_i}\sum_{j=1}^{J_i}\big\{Y_{ij}-a_{i} \phi(T_{ij})\big\}^2+{\nu}\|\bm{a}\|^2\cdot\|D^q\phi\|^2,\\
    \mathcal{L}(\infty,\phi,\bm{a}):&=& \mathbb{E} \mathcal{L}(nm,\phi,\bm{a})\\
    &=&\sum_{i=1}^n\mathbb{E}\frac{1}{J_i}\sum_{j=1}^{J_i}\big\{Y_{ij}-a_{i} \phi(T_{ij})\big\}^2
+\nu\|\bm{a}\|^2\cdot\|D^q\phi\|^2\\
&=&\sum_{i=1}^n\mathbb{E}\frac{1}{J_i}\sum_{j=1}^{J_i}\big\{X_i(T_{ij})-a_{i} \phi(T_{ij})\big\}^2
+\sum_{i=1}^n\mathbb{E}\frac{1}{J_i}\sum_{j=1}^{J_i}\varepsilon_{ij}^2+\nu\|\bm{a}\|^2\cdot\|D^q\phi\|^2\\
&=&\sum_{i=1}^n\big\|X_i-a_{i} \phi\big\|^2
+\sum_{i=1}^n\frac{1}{J_i}\mathbb{E}\sum_{j=1}^{J_i}\varepsilon_{ij}^2+\nu\|\bm{a}\|^2\cdot\|D^q\phi\|^2.
\end{eqnarray*}
In the following, we adopt the inner-product for $\mathcal{W}_2^q(\mathcal{T})$:
$$\langle f,g\rangle^\prime_{\mathcal{W}_2^q(\mathcal{T})}=\langle f,g\rangle+\langle D^q f,D^qg\rangle,\ \forall f,g\in \mathcal{W}_2^q(\mathcal{T}).$$

With the above notations, 
\begin{eqnarray}\label{phi_opt}
    \widehat{\rho\phi}^{(h)}=\arg\min_{\phi\in \mathcal{W}_2^q(\mathcal{T})}\mathcal{L}(nm,\phi,\hat{\bm{a}}^{(h)}).
\end{eqnarray}
Given $\widehat{\rho\phi}^{(h)}$, define 
\begin{eqnarray*}
\tilde{{a}}_{i}^{(h+1)}=\frac{\frac{1}{J_i}\sum_{j=1}^{J_i}Y_{ij}\widehat{\rho\phi}^{(h)}(T_{ij})}{\frac{1}{J_i}\sum_{j=1}^{J_i}\big\{\widehat{\rho\phi}^{(h)}(T_{ij})\big\}^2+\nu\|D^q\widehat{\rho\phi}^{(h)}\|^2},\ i\in [n],
\end{eqnarray*}
and  $\hat{\bm{a}}^{(h+1)}=\tilde{\bm{a}}^{(h+1)}/\|\tilde{\bm{a}}^{(h+1)}\|$. 
Here, we add a restriction to the optimization \eqref{phi_opt} such that
\(
\|D^q \widehat{\rho\phi}^{(h)}\|^2_{\mathcal{W}_2^q(\mathcal{T})}
\leq C_{\phi} (\rho_1^0)^2
\quad \text{and} \quad
\frac{1}{J_i} \sum_{j=1}^{J_i} \big\{\widehat{\rho\phi}^{(h)}(T_{ij})\big\}^2 + \nu \|D^q \widehat{\rho\phi}^{(h)}\|^2 \geq C_a (\rho_1^0)^2,
\)
for all $h$ and $i \in [n]$, where $C_{\phi}$ and $C_a$ are two constants. These restrictions are imposed purely for technical purposes and are not necessary for the implementation.

\

\begin{proof}[Proof of Lemma \ref{Le:3}]
In the following proof, we always assume $(\bm{a}_1^0)^\top \hat{\bm{a}}^{(h)} \geq 0$ and $\langle \widehat{\rho\phi}^{(h)}, \phi_1^0 \rangle \geq 0$ for all $h \geq 0$ as it does not affect the conclusion.

Let
\begin{eqnarray*}
\bar{\bm{a}} &:=& (\langle X_1,\overline{\rho\phi}^{(h)}\rangle,\dots,\langle X_n,\overline{\rho\phi}^{(h)}\rangle)^\top /(\rho_1^0)^2\\
&=&\bigg(\bigg\langle \sum_{r=1}^R\rho_r^0 a_{1r}^0\phi_r^0,\sum_{i=1}^n \hat{a}^{(h)}_{i}\sum_{s=1}^R\rho_s^0 a_{is}^0\phi_s^0\bigg\rangle,\dots,\bigg\langle \sum_{r=1}^R\rho_r^0 a_{nr}^0\phi_r^0,\sum_{i=1}^n \hat{a}^{(h)}_{i}\sum_{s=1}^R\rho_s^0 a_{is}^0\phi_s^0\bigg\rangle\bigg)^\top /(\rho_1^0)^2\\
&=&\sum_{r=1}^R\bigg(\bigg\langle \rho_r^0 a_{1r}^0\phi_r^0,\rho_r^0\phi_r^0\sum_{i=1}^n \hat{a}^{(h)}_{i}a_{ir}^0\bigg\rangle,\dots,\bigg\langle \rho_r^0 a_{nr}^0\phi_r^0,\rho_r^0\phi_r^0\sum_{i=1}^n \hat{a}^{(h)}_{i}a_{ir}^0\bigg\rangle\bigg)^\top /(\rho_1^0)^2\\
&=&\sum_{r=1}^R \left( \frac{\rho_r^0}{\rho_1^0} \right)^2\cdot \bm{a}_r^0 (\bm{a}_r^0)^\top \hat{\bm{a}}^{(h)}.
\end{eqnarray*}

By Lemma \ref{lm:projection}, for any positive value $d$,
\begin{eqnarray*}
\operatorname{dist}(\hat{\bm{a}}^{(h+1)},\bm{a}_1^0)&\leq &
\|d\tilde{\bm{a}}^{(h+1)}-\bm{a}_1^0\|
\\
&\leq &\|d\tilde{\bm{a}}^{(h+1)}-\bar{\bm{a}}\| + 
\|\bar{\bm{a}}-\bm{a}_1^0\|\\
&=&  \|d\tilde{\bm{a}}^{(h+1)}-\bar{\bm{a}}\| + 
\sqrt{\left| (\bm{a}_1^0)^\top \hat{\bm{a}}^{(h)} -1 \right|^2 + \sum_{r>1}^R\left( \frac{\rho_r^0}{\rho_1^0} \right)^4 \left( (\bm{a}_r^0)^\top \hat{\bm{a}}^{(h)} \right)^2}\\
&\leq&  \|d\tilde{\bm{a}}^{(h+1)}-\bar{\bm{a}}\| + 
\left| (\bm{a}_1^0)^\top \hat{\bm{a}}^{(h)} -1 \right| + \sqrt{\sum_{r>1}^R\left( \frac{\rho_r^0}{\rho_1^0} \right)^4 \left( (\bm{a}_r^0)^\top \hat{\bm{a}}^{(h)} \right)^2}.
\end{eqnarray*}

Since 
$(\bm{a}_1^0)^\top \hat{\bm{a}}^{(h)} \geq 0$, 
\begin{eqnarray*}
\left| (\bm{a}_1^0)^\top \hat{\bm{a}}^{(h)} -1 \right|&=&\bigg|1-\sqrt{1-\operatorname{dist}^2(\bm{a}_1^0, \hat{\bm{a}}^{(h)})}\bigg| \\
&=&\frac{\operatorname{dist}^2(\bm{a}_1^0, \hat{\bm{a}}^{(h)})}{1+\sqrt{1-\operatorname{dist}^2(\bm{a}_1^0, \hat{\bm{a}}^{(h)})}}.
\end{eqnarray*}

In addition, 
\begin{eqnarray*}
\sum_{r>1}^R \left( (\bm{a}_r^0)^\top \hat{\bm{a}}^{(h)} \right)^2
\leq 1- \left( (\bm{a}_1^0)^\top \hat{\bm{a}}^{(h)} \right)^2=\operatorname{dist}^2(\bm{a}_1^0, \hat{\bm{a}}^{(h)}).
\end{eqnarray*}

Combining the above three inequalities, we have
\begin{eqnarray}\label{Bound_a_fun}
\operatorname{dist}(\hat{\bm{a}}^{(h+1)},\bm{a}_1^0)&\leq &
\|d\tilde{\bm{a}}^{(h+1)}-\bar{\bm{a}}\|
+\frac{\operatorname{dist}^2(\hat{\bm{a}}^{(h)},{\bm{a}}_1^0)}{1+\sqrt{1-\operatorname{dist}^2(\hat{\bm{a}}^{(h)},{\bm{a}}_1^0)}}+\frac{(\rho_2^0)^2}{(\rho_1^0)^2}\operatorname{dist}(\bm{a}_1^0, \hat{\bm{a}}^{(h)})
\end{eqnarray}
for any $d\geq 0$. 

In the following, we examine the error bound between $d\tilde{\bm{a}}^{(h+1)}$ and $\bar{\bm{a}}$. Take $d=\big\{\|\widehat{\rho\phi}^{(h)}\|^2+\nu\|D^q \widehat{\rho\phi}^{(h)}\|^2\big\}/{(\rho_1^0)^2}$,
then
\begin{eqnarray*}
&& \left| d\tilde{a}_{i}^{(h+1)} - \bar{a}_{i} \right| \\
&=& \bigg| d \cdot \bigg\{ \frac{ \frac{1}{J_i} \sum_{j=1}^{J_i} Y_{ij} \widehat{\rho\phi}^{(h)}(T_{ij}) }{ \frac{1}{J_i} \sum_{j=1}^{J_i} \big\{ \widehat{\rho\phi}^{(h)}(T_{ij}) \big\}^2 + \nu\|D^q \widehat{\rho\phi}^{(h)}\|^2 } \bigg\} - \frac{ \langle X_i, \overline{\rho\phi}^{(h)} \rangle }{ (\rho_1^0)^2 } \bigg| \\
&\leq & \bigg| \frac{ \|\widehat{\rho\phi}^{(h)}\|^2 + \nu\|D^q \widehat{\rho\phi}^{(h)}\|^2 }{ \frac{1}{J_i} \sum_{j=1}^{J_i} \big( \widehat{\rho\phi}^{(h)}(T_{ij}) \big)^2 + \nu\|D^q \widehat{\rho\phi}^{(h)}\|^2 } -1 \bigg| \cdot \frac{1}{ (\rho_1^0)^2 } \bigg| \frac{1}{J_i} \sum_{j=1}^{J_i} Y_{ij} \widehat{\rho\phi}^{(h)}(T_{ij}) \bigg| \\
&+&
\bigg| \frac{1}{J_i} \sum_{j=1}^{J_i} Y_{ij} \widehat{\rho\phi}^{(h)}(T_{ij}) - \langle X_i, \overline{\rho\phi}^{(h)} \rangle \bigg| / (\rho_1^0)^2 \\
&\leq& \bigg| \frac{ V_i(\widehat{\rho\phi}^{(h)}) }{ W_i(\widehat{\rho\phi}^{(h)}) } \bigg| \cdot \frac{1}{ (\rho_1^0)^2 } \bigg| \frac{1}{J_i} \sum_{j=1}^{J_i} Y_{ij} \widehat{\rho\phi}^{(h)}(T_{ij}) \bigg| +
\bigg| \frac{1}{J_i} \sum_{j=1}^{J_i} Y_{ij} \widehat{\rho\phi}^{(h)}(T_{ij}) - \langle X_i, \overline{\rho\phi}^{(h)} \rangle \bigg| / (\rho_1^0)^2 \\
&\leq & \bigg| \frac{ V_i(\widehat{\rho\phi}^{(h)}) }{ W_i(\widehat{\rho\phi}^{(h)}) } \bigg| \cdot \frac{1}{ (\rho_1^0)^2 } \big| \langle X_i, \overline{\rho\phi}^{(h)} \rangle \big| \\
&+& \bigg| \frac{ V_i(\widehat{\rho\phi}^{(h)}) }{ W_i(\widehat{\rho\phi}^{(h)}) } \bigg| \cdot \bigg| \frac{1}{J_i} \sum_{j=1}^{J_i} Y_{ij} \widehat{\rho\phi}^{(h)}(T_{ij}) - \langle X_i, \overline{\rho\phi}^{(h)} \rangle \bigg| / (\rho_1^0)^2 \\
&+&
\bigg| \frac{1}{J_i} \sum_{j=1}^{J_i} Y_{ij} \widehat{\rho\phi}^{(h)}(T_{ij}) - \langle X_i, \overline{\rho\phi}^{(h)} \rangle \bigg| / (\rho_1^0)^2,
\end{eqnarray*}
where
$
V_i(\widehat{\rho\phi}^{(h)}) = \|\widehat{\rho\phi}^{(h)}\|^2 - \frac{1}{J_i} \sum_{j=1}^{J_i} \big( \widehat{\rho\phi}^{(h)}(T_{ij}) \big)^2
$
and
$
W_i(\widehat{\rho\phi}^{(h)}) = \frac{1}{J_i} \sum_{j=1}^{J_i} \big( \widehat{\rho\phi}^{(h)}(T_{ij}) \big)^2 + \nu\|D^q \widehat{\rho\phi}^{(h)}\|^2.
$
Accordingly,
\begin{eqnarray}\label{inequ_kk}
&& \sqrt{ \sum_{i=1}^n \left| d\tilde{a}_{i}^{(h+1)} - \bar{a}_{i} \right|^2 } \nonumber\\
&\lesssim& \sqrt{ \sum_{i=1}^n \left[ \left| \frac{ V_i(\widehat{\rho\phi}^{(h)}) }{ W_i(\widehat{\rho\phi}^{(h)}) } \right| \cdot \frac{1}{ (\rho_1^0)^2 } \big| \langle X_i, \overline{\rho\phi}^{(h)} \rangle \big| \right]^2 } \nonumber\\
&+& \frac{1}{ (\rho_1^0)^2 } \sqrt{ \sum_{i=1}^n \left| \frac{1}{J_i} \sum_{j=1}^{J_i} Y_{ij} \widehat{\rho\phi}^{(h)}(T_{ij}) - \langle X_i, \overline{\rho\phi}^{(h)} \rangle \right|^2 } \nonumber\\
&+& \frac{1}{ (\rho_1^0)^2 } \sqrt{ \sum_{i=1}^n \left| \frac{ V_i(\widehat{\rho\phi}^{(h)}) }{ W_i(\widehat{\rho\phi}^{(h)}) } \right|^2 \cdot \left| \frac{1}{J_i} \sum_{j=1}^{J_i} Y_{ij} \widehat{\rho\phi}^{(h)}(T_{ij}) - \langle X_i, \overline{\rho\phi}^{(h)} \rangle \right|^2 } \\
:&=& (1) + (2) + (3).
\end{eqnarray}
We respectively bound the above three terms in the remaining proof.

\textbf{Upper bound of (1)}: First note that
\begin{eqnarray*}
&&\sum_{i=1}^n\bigg(\frac{1}{(\rho_1^0)^2} \big|\langle X_i,\overline{\rho\phi}^{(h)}\rangle\big|\bigg)^2 \\
&=& \sum_{i=1}^n\bigg\{\sum_{r=1}^R \bigg(\frac{\rho_r^0}{\rho_1^0}\bigg)^2 a_{ir}^0 (\bm{a}_r^0)^\top \hat{\bm{a}}^{(h)}\bigg\}^2 \\
&=& \sum_{r=1}^R \bigg( \frac{\rho_r^0}{\rho_1^0} \bigg)^4 \left( (\bm{a}_r^0)^\top \hat{\bm{a}}^{(h)} \right)^2 \sum_{i=1}^n (a_{ir}^0)^2 \\
&=& \sum_{r=1}^R \bigg( \frac{\rho_r^0}{\rho_1^0} \bigg)^4 \left( (\bm{a}_r^0)^\top \hat{\bm{a}}^{(h)} \right)^2 \\
&\leq& 1,
\end{eqnarray*}
where we use the orthonormality of the vectors $\bm{a}_r^0$ and that $\sum_{i=1}^n (a_{ir}^0)^2 = 1$.

Notice that
$$
\|\widehat{\rho\phi}^{(h)}\|_{\infty} \lesssim \|\widehat{\rho\phi}^{(h)}\|_{\mathcal{W}_2^q(\mathcal{T})} = \sqrt{\|\widehat{\rho\phi}^{(h)}\|^2 + \|D^q\widehat{\rho\phi}^{(h)}\|^2} \lesssim \rho_1^0
$$
due to Lemma \ref{Le: Bound_function} and the conditions $\|\widehat{\rho\phi}^{(h)}\| \lesssim \rho_1^0$ and $\|D^q\widehat{\rho\phi}^{(h)}\| \lesssim \rho_1^0$. Then
\begin{eqnarray*}
\sqrt{\sum_{i=1}^n\bigg(\frac{1}{(\rho_1^0)^2} \big|\langle X_i,\overline{\rho\phi}^{(h)}\rangle\big|\bigg)^2 \cdot \bigg(\frac{\|\widehat{\rho\phi}^{(h)}\|_{\infty}}{\rho_1^0}\bigg)^2} \lesssim 1.
\end{eqnarray*}
By condition \eqref{C1} and $m^{-q/(2q+1)} \lesssim 1$, we have
\begin{eqnarray*}
&& \sum_{i=1}^n \bigg| \frac{\|\widehat{\rho\phi}^{(h)}\|^2 - \frac{1}{J_i}\sum_{j=1}^{J_i} \big( \widehat{\rho\phi}^{(h)}(T_{ij}) \big)^2}{(\rho_1^0)^2} \bigg|^2 \cdot \bigg( \frac{1}{(\rho_1^0)^2} \big| \langle X_i, \overline{\rho\phi}^{(h)} \rangle \big| \bigg)^2 \lesssim m^{-2q/(2q+1)}.
\end{eqnarray*}
In addition, since
\begin{eqnarray}
&&\bigg| \frac{V_i(\widehat{\rho\phi}^{(h)})}{W_i(\widehat{\rho\phi}^{(h)})} \bigg| \nonumber\\
&=& \bigg| \frac{ \|\widehat{\rho\phi}^{(h)}\|^2 - \frac{1}{J_i}\sum_{j=1}^{J_i} \big( \widehat{\rho\phi}^{(h)}(T_{ij}) \big)^2 }{ (\rho_1^0)^2 } \cdot \frac{ (\rho_1^0)^2 }{ W_i(\widehat{\rho\phi}^{(h)}) } \bigg| \nonumber \\
&\lesssim& \bigg| \frac{ \|\widehat{\rho\phi}^{(h)}\|^2 - \frac{1}{J_i}\sum_{j=1}^{J_i} \big( \widehat{\rho\phi}^{(h)}(T_{ij}) \big)^2 }{ (\rho_1^0)^2 } \bigg|. \label{eq: bou}
\end{eqnarray}
Combining the above two inequalities,
\begin{eqnarray}\label{bound of first}
(1) = \sqrt{ \sum_{i=1}^n \bigg| \frac{ V_i(\widehat{\rho\phi}^{(h)}) }{ W_i(\widehat{\rho\phi}^{(h)}) } \bigg|^2 \cdot \bigg( \frac{1}{(\rho_1^0)^2} \big| \langle X_i, \overline{\rho\phi}^{(h)} \rangle \big| \bigg)^2 } \lesssim m^{-q/(2q+1)}.
\end{eqnarray}

\textbf{Upper bound of (2)}: Observe that
\begin{eqnarray}\label{bound_aaa}
&&\sum_{i=1}^n\bigg|   \frac{1}{J_i}\sum_{j=1}^{J_i}  Y_{ij}\widehat{\rho\phi}^{(h)}(T_{ij})-\langle X_i,\overline{\rho\phi}^{(h)}\rangle\bigg|^2\nonumber\\
&\lesssim &\sum_{i=1}^n\bigg|   \frac{1}{J_i}\sum_{j=1}^{J_i}  X_{i}(T_{ij})\widehat{\rho\phi}^{(h)}(T_{ij})-\langle X_{i},\widehat{\rho\phi}^{(h)}\rangle\bigg|^2\nonumber\\
 &+&\sum_{i=1}^n\bigg|\frac{1}{J_i}\sum_{j=1}^{J_i}  \varepsilon_{ij}\widehat{\rho\phi}^{(h)}(T_{ij})\bigg|^2\nonumber\\
 &+&\sum_{i=1}^n\bigg|\langle X_{i},\widehat{\rho\phi}^{(h)}\rangle-\langle X_{i},\overline{\rho\phi}^{(h)}\rangle\bigg|^2.
\end{eqnarray}

Notice that 
$
\|\widehat{\rho\phi}^{(h)}\|_{\infty}\lesssim \rho_1^0
$
and
\begin{eqnarray}\label{bound_phi_0}
  \big\|{\phi}^{0}_1\big\|_{\infty} &\lesssim&  \sqrt{\|{\phi}^{0}_1\|^2+\|D^q{\phi}^{0}_1\|^2} \lesssim 1,
\end{eqnarray}
due to Lemma \ref{Le: Bound_function}, and $\|D^q{\phi}^{0}_1\|=\bigg\|\sum_{i=1}^n a_{i1}^0 D^q X_i\bigg\|/\rho_1^0\lesssim 1$ by Assumption \ref{A4}. Besides,
\begin{eqnarray*}
\sum_{i=1}^n{\|(X_i\widehat{\rho\phi}^{(h)})^2\|_{\infty}}
&=&\sum_{i=1}^n{\bigg\| \bigg( \sum_{r=1}^R \rho_r^0 a_{ir}^0 \phi_r^0 \cdot \widehat{\rho\phi}^{(h)} \bigg)^2 \bigg\|_{\infty}}\\
&=& \sum_{i=1}^n \bigg\| \sum_{r=1}^R \rho_r^0 a_{ir}^0 \phi_r^0 \cdot \widehat{\rho\phi}^{(h)} \bigg\|_{\infty}^2 \\
&\leq& \sum_{i=1}^n \left( \sum_{r=1}^R |\rho_r^0 a_{ir}^0| \| \phi_r^0 \|_{\infty} \cdot \| \widehat{\rho\phi}^{(h)} \|_{\infty} \right)^2 \\
&\leq& \sum_{i=1}^n \left( |\rho_1^0 a_{i1}^0| \| \phi_1^0 \|_{\infty} \cdot \| \widehat{\rho\phi}^{(h)} \|_{\infty} + \sum_{r=2}^R |\rho_r^0 a_{ir}^0| \| \phi_r^0 \|_{\infty} \cdot \| \widehat{\rho\phi}^{(h)} \|_{\infty} \right)^2 \\
&\leq& 2 \sum_{i=1}^n \left( \left( \rho_1^0 a_{i1}^0 \| \phi_1^0 \|_{\infty} \cdot \| \widehat{\rho\phi}^{(h)} \|_{\infty} \right)^2 + \left( \sum_{r=2}^R \rho_r^0 a_{ir}^0 \| \phi_r^0 \|_{\infty} \cdot \| \widehat{\rho\phi}^{(h)} \|_{\infty} \right)^2 \right) \\
&=& 2 \| \widehat{\rho\phi}^{(h)} \|_{\infty}^2 \sum_{i=1}^n \left( (\rho_1^0 a_{i1}^0)^2 \| \phi_1^0 \|_{\infty}^2 + \left( \sum_{r=2}^R \rho_r^0 a_{ir}^0 \| \phi_r^0 \|_{\infty} \right)^2 \right) \\
&\leq& 2 \| \widehat{\rho\phi}^{(h)} \|_{\infty}^2 \left( (\rho_1^0)^2 \| \phi_1^0 \|_{\infty}^2 + \sum_{i=1}^n \left( \sum_{r=2}^R \rho_r^0 a_{ir}^0 \| \phi_r^0 \|_{\infty} \right)^2 \right).
\end{eqnarray*}
Now, we bound the second term using \(\left( \sum_{r=2}^R x_r \right)^2 \leq (R-1) \sum_{r=2}^R x_r^2\):
\begin{eqnarray*}
\sum_{i=1}^n \left( \sum_{r=2}^R \rho_r^0 a_{ir}^0 \| \phi_r^0 \|_{\infty} \right)^2 &\leq& (R-1) \sum_{i=1}^n \sum_{r=2}^R \left( \rho_r^0 a_{ir}^0 \| \phi_r^0 \|_{\infty} \right)^2 \\
&\leq& (R-1) \sum_{r=2}^R \left( \rho_r^0 \| \phi_r^0 \|_{\infty} \right)^2 \sum_{i=1}^n (a_{ir}^0)^2 \\
&=& (R-1) \sum_{r=2}^R \left( \rho_r^0 \| \phi_r^0 \|_{\infty} \right)^2 \\
&\leq& (R-1) \sum_{r=2}^R \left( \frac{ \rho_1^0 }{ \kappa } \| \phi_r^0 \|_{\infty} \right)^2 \\
&\lesssim& \frac{ (R-1)^2 (\rho_1^0)^2 }{ \kappa^2 }.
\end{eqnarray*}
Combining the terms, we get
\begin{eqnarray*}
&&2 \| \widehat{\rho\phi}^{(h)} \|_{\infty}^2 \left( (\rho_1^0)^2 \| \phi_1^0 \|_{\infty}^2  + \frac{ (R-1)^2 (\rho_1^0)^2 }{ \kappa^2 } \right) \\
&\lesssim& 2 \| \widehat{\rho\phi}^{(h)} \|_{\infty}^2 (\rho_1^0)^2 \left( 1 + \frac{ (R-1)^2 }{ \kappa^2 } \right).
\end{eqnarray*}
Under Assumption \ref{A7}, which states that $\kappa\gtrsim R$ and hence $(R/\kappa)^2\lesssim 1$, and noting that \(\| \widehat{\rho\phi}^{(h)} \|_{\infty} \lesssim \rho_1^0\), we have
\[
\sum_{i=1}^n{\|(X_i\widehat{\rho\phi}^{(h)})^2\|_{\infty}} \lesssim (\rho_1^0)^4.
\]
Combining with the above inequality, condition \eqref{C1}, and $m^{-q/ (2q+1)}\lesssim 1$, we have
\begin{eqnarray*}
\frac{1}{(\rho_1^0)^4}\sum_{i=1}^n \bigg|   \frac{1}{J_i}\sum_{j=1}^{J_i}  X_{i}(T_{ij})\widehat{\rho\phi}^{(h)}(T_{ij})-\langle X_{i},\widehat{\rho\phi}^{(h)}\rangle\bigg|^2
   \lesssim  m^{-2q/(2q+1)}.
\end{eqnarray*}

Moreover, notice that
\begin{eqnarray*}
\sum_{i=1}^n\bigg|\frac{1}{J_i}\sum_{j=1}^{J_i}  \varepsilon_{ij}\widehat{\rho\phi}^{(h)}(T_{ij})\bigg|^2
&\lesssim& \|\widehat{\rho\phi}^{(h)}\|_{\infty}^2 \sum_{i=1}^n \bigg( \frac{1}{J_i}\sum_{j=1}^{J_i}  \varepsilon_{ij} \bigg)^2 \\
&\leq& \|\widehat{\rho\phi}^{(h)}\|_{\infty}^2 \bigg\{\sum_{i=1}^n \bigg| \frac{1}{J_i}\sum_{j=1}^{J_i}  \varepsilon_{ij} \bigg|\bigg\}^2 \\
&\lesssim& (\rho_1^0)^2 \frac{n \sigma^2 \,x^2}{m},
\end{eqnarray*}
by condition \eqref{C2}, and
\begin{eqnarray*}
\sum_{i=1}^n\bigg|\langle X_{i},\widehat{\rho\phi}^{(h)}\rangle-\langle X_{i},\overline{\rho\phi}^{(h)}\rangle\bigg|^2
=\|\mathcal{X}_n(\widehat{\rho\phi}^{(h)}-\overline{\rho\phi}^{(h)})\|^2
\leq  (\rho_1^0)^2 \big\| \widehat{\rho\phi}^{(h)} - \overline{\rho\phi}^{(h)} \big \|^2
\end{eqnarray*}
due to Lemma \ref{bound_le}.

Combining the above three inequalities with \eqref{bound_aaa}, we have 
\begin{eqnarray}\label{informative_inq}
&&\frac{1}{(\rho_1^0)^4}\sum_{i=1}^n\bigg|   \frac{1}{J_i}\sum_{j=1}^{J_i}  Y_{ij}\widehat{\rho\phi}^{(h)}(T_{ij})-\langle X_i,\overline{\rho\phi}^{(h)}\rangle\bigg|^2\nonumber\\
&\lesssim&   m^{-2q/(2q+1)}+ {\frac{n \sigma^2}{m (\rho_1^0)^2}}\cdot x^2+\frac{1}{(\rho_1^0)^2}\big\| \widehat{\rho\phi}^{(h)} - \overline{\rho\phi}^{(h)} \big \|^2.
\end{eqnarray}

\textbf{Upper bound of (3)}: Notice that 
\begin{eqnarray*}
    \frac{1}{(\rho_1^0)^2}\sqrt{ \sum_{i=1}^n \bigg| \frac{ V_i(\widehat{\rho\phi}^{(h)}) }{ W_i(\widehat{\rho\phi}^{(h)}) } \bigg|^2 \cdot \bigg| \frac{1}{J_i} \sum_{j=1}^{J_i} Y_{ij} \widehat{\rho\phi}^{(h)}(T_{ij}) - \langle X_i, \overline{\rho\phi}^{(h)} \rangle \bigg|^2 } \leq \sup_{i \in [n]} \bigg| \frac{ V_i(\widehat{\rho\phi}^{(h)}) }{ W_i(\widehat{\rho\phi}^{(h)}) } \bigg| \cdot (2).
\end{eqnarray*}
By \eqref{eq: bou} and condition \eqref{C11},
\begin{eqnarray*}
    \sup_{i \in [n]} \bigg| \frac{ V_i(\widehat{\rho\phi}^{(h)}) }{ W_i(\widehat{\rho\phi}^{(h)}) } \bigg| \lesssim \sup_{i \in [n]} \bigg| \frac{ \| \widehat{\rho\phi}^{(h)} \|^2 - \frac{1}{J_i} \sum_{j=1}^{J_i} \big( \widehat{\rho\phi}^{(h)}(T_{ij}) \big)^2 }{ (\rho_1^0)^2 } \bigg| \lesssim 1.
\end{eqnarray*}
Therefore,
\begin{eqnarray}\label{bound_sim}
    (3) \lesssim (2).
\end{eqnarray}

We finally obtain our conclusion by combining \eqref{inequ_kk}, \eqref{bound of first}, \eqref{informative_inq}, and \eqref{bound_sim} in \eqref{Bound_a_fun}.
\end{proof}


\begin{proof}[Proof to Lemma \ref{le_cor}]
Recall
\begin{eqnarray*}
 \overline{\rho\phi}^{(h)} = \sum_{i=1}^n \hat{a}^{(h)}_{i} X_i.
\end{eqnarray*}
This is equivalent to
\begin{eqnarray}\label{le_cor_co}
\overline{\rho\phi}^{(h)} = \arg\min_{\phi\in \mathcal{W}^2_q( \mathcal{T})}
\sum_{i=1}^n\big\|X_i-\hat{a}_{i}^{(h)} \phi\big\|^2.
\end{eqnarray}
Let $\overline{\rho\phi}$ be the minimizer of the expected loss function given $\bm{a}=\hat{\bm{a}}^{(h)}$, i.e.,
\begin{eqnarray*}
\overline{\rho\phi} := \arg\min_{\phi\in \mathcal{W}_q^2(\mathcal{T})}
\mathcal{L}(\infty,\phi,\hat{\bm{a}}^{(h)})
.
\end{eqnarray*}
In the following, we prove that
\begin{eqnarray}
     \| \overline{\rho\phi} - \overline{\rho\phi}^{(h)} \| &\lesssim& \rho_1^0 \sqrt{\nu},\label{est_opt}\\
     \| \overline{\rho\phi} - \widehat{\rho\phi}^{(h)} \|&\lesssim& \rho_1^0 m^{-q/(2q+1)}+  \frac{1}{\nu^{1/(4q)}}\cdot\frac{\sigma}{\sqrt{m}}\cdot x+\sqrt{\frac{n}{m}} \sigma\cdot x+\rho_1^0\operatorname{dist}^2(\hat{\bm{a}}^{(h)},{\bm{a}}_1^0).\label{phi_bb}
\end{eqnarray}
We then prove this lemma by combining the above two inequalities.

\textbf{Proof to \eqref{est_opt}}: Note that $\mathcal{L}(\infty,\overline{\rho\phi},\hat{\bm{a}}^{(h)}) \leq     \mathcal{L}(\infty,\overline{\rho\phi}^{(h)},\hat{\bm{a}}^{(h)})$ by the definition of $\overline{\rho\phi}$, and $\sum_{i=1}^n\big\|X_i-\hat{a}^{(h)}_{i} \overline{\rho\phi}\big\|^2 - \sum_{i=1}^n\big\|X_i-\hat{a}^{(h)}_{i} \overline{\rho\phi}^{(h)}\big\|^2 \geq 0$ by the definition of $\overline{\rho\phi}^{(h)}$,
then
\begin{eqnarray}\label{inq_f}
  0 \leq  \sum_{i=1}^n\big\|X_i-\hat{a}^{(h)}_{i} \overline{\rho\phi}\big\|^2 - \sum_{i=1}^n\big\|X_i-\hat{a}^{(h)}_{i} \overline{\rho\phi}^{(h)}\big\|^2
    \leq \nu \big(\|D^q\overline{\rho\phi}^{(h)}\|^2 - \|D^q\overline{\rho\phi}\|^2\big) \leq
    \nu  \|D^q\overline{\rho\phi}^{(h)}\|^2.
\end{eqnarray}
Since
\begin{eqnarray}\label{bias_bound}
    \|D^q\overline{\rho\phi}^{(h)}\| = \bigg\|\sum_{i=1}^n \hat{a}^{(h)}_{i} D^q{X}_i\bigg\| \lesssim \rho_1^0
\end{eqnarray}
by Assumption \ref{A4}, we have
\begin{eqnarray}\label{inequ}
    \sum_{i=1}^n\big\|X_i - \hat{a}^{(h)}_{i} \overline{\rho\phi}^{(h)}\big\|^2 - \sum_{i=1}^n\big\|X_i - \hat{a}^{(h)}_{i} \overline{\rho\phi}\big\|^2
    \lesssim (\rho_1^0)^2\nu.
\end{eqnarray}
By the Pythagorean theorem, we have
\begin{eqnarray*}
&&\sum_{i=1}^n\big\|X_i - \hat{a}^{(h)}_{i} \overline{\rho\phi}^{(h)}\big\|^2 - \sum_{i=1}^n\big\|X_i - \hat{a}^{(h)}_{i} \overline{\rho\phi}\big\|^2\\
&=&2\sum_{i=1}^n\langle X_i - \hat{a}^{(h)}_{i} \overline{\rho\phi}^{(h)}, \hat{a}^{(h)}_{i} \overline{\rho\phi}^{(h)} - \hat{a}^{(h)}_{i} \overline{\rho\phi}\rangle + \| \overline{\rho\phi}^{(h)} - \overline{\rho\phi} \|^2.
\end{eqnarray*}
We claim that
\begin{eqnarray}\label{proj_phi}
    \sum_{i=1}^n\langle X_i - \hat{a}^{(h)}_{i} \overline{\rho\phi}^{(h)}, \hat{a}^{(h)}_{i} \overline{\rho\phi}^{(h)} - \hat{a}_{i}^{(h)} {\phi}\rangle
    \geq 0,\ \forall \phi\in \mathcal{W}_q^2(\mathcal{T}),
\end{eqnarray}
and therefore,
\begin{eqnarray}\label{eqnnn}
    \sum_{i=1}^n\big\|X_i - \hat{a}^{(h)}_{i} \overline{\rho\phi}^{(h)}\big\|^2 - \sum_{i=1}^n\big\|X_i - \hat{a}^{(h)}_{i} \overline{\rho\phi}\big\|^2 \geq \| \overline{\rho\phi}^{(h)} - \overline{\rho\phi} \|^2.
\end{eqnarray}
By combining \eqref{inequ} and \eqref{eqnnn}, we achieve
\begin{eqnarray*}
  \| \overline{\rho\phi}^{(h)} - \overline{\rho\phi} \|^2 \lesssim (\rho_1^0)^2\nu,
\end{eqnarray*}
then \eqref{est_opt} is proven.

To prove \eqref{proj_phi},
we assume that there exists $\phi\in \mathcal{W}_q^2(\mathcal{T})$ such that
\begin{eqnarray}\label{assup_1}
    \sum_{i=1}^n\langle X_i - \hat{a}^{(h)}_{i} \overline{\rho\phi}^{(h)}, \hat{a}^{(h)}_{i} \overline{\rho\phi}^{(h)} - \hat{a}^{(h)}_{i} {\phi}\rangle
    < 0.
\end{eqnarray}
Let $\phi_v := (1 - v) \overline{\rho\phi}^{(h)} + v \phi,\ v\in [0,1]$,
be a convex combination of $\overline{\rho\phi}^{(h)}$ and $\phi$, and define
\begin{eqnarray*}
    f(v) := \sum_{i=1}^n\|X_i - \hat{a}_{i}^{(h)}\phi_v\|^2.
\end{eqnarray*}
It can be shown that the derivative of $f(v)$ at $v=0$ is negative due to \eqref{assup_1}. Thus, there is a choice of $v\in (0,1]$ such that $f(v) < f(0)$, which is a contradiction to \eqref{le_cor_co}. Therefore, \eqref{proj_phi} holds.

\textbf{Proof to \eqref{phi_bb}:} We first evaluate the Fr\'echet derivatives of the loss functions 
\begin{center}
    $\mathcal{L}(nm,\phi,\hat{\bm{a}}^{(h)})$ and $\mathcal{L}(\infty,\phi,\hat{\bm{a}}^{(h)})$
\end{center}
with respect to $\phi$. Let $\mathcal{B}(\mathcal{H}_1,\mathcal{H}_2)$ contain all bounded operators between two Hilbert spaces $\mathcal{H}_1$ and $\mathcal{H}_2$. Define $\mathcal{D}_{nm}$ and $\mathcal{D}_{\infty}$ as the Fr\'echet derivatives of $\mathcal{L}(nm,\phi,\hat{\bm{a}}^{(h)})$ and $\mathcal{L}(\infty,\phi,\hat{\bm{a}}^{(h)})$ with respect to the function $\phi$, respectively.
For their detailed definitions, refer to Section 3.6 in \citet{hsing2015theoretical}. 
Notice that $\mathcal{D}_{nm}(f),\mathcal{D}_{\infty}(f)\in \mathcal{B}(\mathcal{W}_q^2(\mathcal{T}),\mathbb{R})$, $\forall f\in \mathcal{W}_q^2(\mathcal{T})$.
Furthermore, we show that
\begin{eqnarray}
    \mathcal{D}_{nm}(f)g &=& -\sum_{i=1}^n \frac{2\hat{a}^{(h)}_{i}}{J_i} \sum_{j=1}^{J_i} \big\{ Y_{ij} - \hat{a}^{(h)}_{i} f(T_{ij}) \big\} g(T_{ij}) + 2\nu \langle D^q f, D^q g \rangle, \label{D1} \\
    \mathcal{D}_{\infty}(f)g &=& -2\bigg\langle \sum_{i=1}^n \hat{a}^{(h)}_{i} X_i - f, \, g \bigg\rangle + 2\nu \langle D^q f, D^q g \rangle, \label{D1_inf}
\end{eqnarray}
$\forall f, g \in \mathcal{W}_q^2(\mathcal{T})$. 
The above identities follow directly from the definition of the Fr\'echet derivative.

Similarly, define $\mathcal{D}^2_{\infty}$ as the second Fr\'echet derivative of $\mathcal{L}(\infty,\phi,\hat{\bm{a}}^{(h)})$ with respect to $\phi$. 
By the definition, we can show that $\mathcal{D}^2_{\infty}(f) \in \mathcal{B}(\mathcal{W}_q^2(\mathcal{T}), \mathcal{B}(\mathcal{W}_q^2(\mathcal{T}), \mathbb{R}))$ and
\begin{eqnarray}
   \big\{\mathcal{D}^{2}_{\infty}(\phi)\,f\big\}(g)=2\langle f,g\rangle+2\nu \langle D^q f,D^q g\rangle,\ \forall \phi,f,g \in \mathcal{W}_q^2(\mathcal{T}).\label{D2}
\end{eqnarray}

Based on the Riesz representation theorem in functional analysis, there exists an norm-preserving isometric isomorphism $\mathcal{M}$ from $\mathcal{B}(\mathcal{W}_q^2(\mathcal{T}), \mathbb{R})$ to $\mathcal{W}_q^2(\mathcal{T})$. 
Combining the norm-preserving mapping with \eqref{D2}, Lemma 8.3.4 in \citet{hsing2015theoretical} indicates that 
the operator $\tilde{\mathcal{D}}_{\infty}^2:\mathcal{W}_q^2(\mathcal{T})\to \mathcal{W}_q^2(\mathcal{T})$ defined by $\tilde{\mathcal{D}}_{\infty}^2 f:=\mathcal{M}\big(\mathcal{D}_{\infty}^2(\phi)\,f\big)$ (for any fixed $\phi\in\mathcal{W}_q^2(\mathcal{T})$) is invertible,
and
\begin{eqnarray}\label{sob_second}
   (\tilde{\mathcal{D}}_{\infty}^2)^{-1} f = \frac{1}{2} \sum_{k=1}^\infty \frac{1+\gamma_k}{1+\nu \gamma_k} f_k e_k, \quad \forall f \in \mathcal{W}_q^2(\mathcal{T}),
\end{eqnarray}
where $f = \sum_{k=1}^\infty f_k e_k := \sum_{k=1}^\infty \langle f, e_k \rangle e_k$ with $e_k$ being a set of basis functions of $\mathcal{W}_q^2(\mathcal{T})$. The definition and properties of $e_k$ and $\gamma_k$ are given in Lemma \ref{le:sobo}.

Define $\tilde{\mathcal{D}}_{nm} = \mathcal{M} \mathcal{D}_{nm}$: $\mathcal{W}_q^2(\mathcal{T}) \to \mathcal{W}_q^2(\mathcal{T})$.
With the definition of $\tilde{\mathcal{D}}_{nm}$ and $\tilde{\mathcal{D}}_{\infty}^2$, we 
obtain from a first-order Taylor expansion that
\begin{eqnarray*}
   \tilde{\mathcal{D}}_{nm}(\widehat{\rho\phi}^{(h)}) - \tilde{\mathcal{D}}_{nm}(\overline{\rho\phi}) \approx \tilde{\mathcal{D}}^2_{\infty} (\widehat{\rho\phi}^{(h)} - \overline{\rho\phi})
\end{eqnarray*}
heuristically, ignoring higher-order remainder terms,
where $\tilde{\mathcal{D}}_{nm}(\widehat{\rho\phi}^{(h)})$ is a zero element in $\mathcal{W}_q^2(\mathcal{T})$ by the definition of $\widehat{\rho\phi}^{(h)}$.
As a result, $\widehat{\rho\phi}^{(h)}$ can be approximated by
\begin{eqnarray*}
\widehat{\rho\phi}^{(h)} \approx \overline{\rho\phi} - (\tilde{\mathcal{D}}^2_{\infty})^{-1} \tilde{\mathcal{D}}_{nm}(\overline{\rho\phi}).
\end{eqnarray*}
By this approximation, define
\begin{eqnarray*}
    \widetilde{\rho\phi} := \overline{\rho\phi} - (\tilde{\mathcal{D}}^2_{\infty})^{-1} \tilde{\mathcal{D}}_{nm}(\overline{\rho\phi}).
\end{eqnarray*}
To prove \eqref{phi_bb}, we respectively examine the error bounds $\| \overline{\rho\phi} - \widetilde{\rho\phi} \|^2$ and $\| \widetilde{\rho\phi} - \widehat{\rho\phi}^{(h)} \|^2$.

\textbf{(a) Error bound for $\| \overline{\rho\phi}- \widetilde{\rho\phi}\|^2$}: 

First note that
\begin{eqnarray}\label{eq: tran}
   \langle f,e_k\rangle_{\mathcal{W}_q^2(\mathcal{T})}
    =\langle f,e_k\rangle+\langle D^q f,D^q e_k\rangle
    =f_k+f_k\gamma_k=(1+\gamma_k)f_k,
\end{eqnarray}
by Lemma \ref{le:sobo}. Therefore,
\begin{eqnarray*}
     & &  \| \overline{\rho\phi}- \widetilde{\rho\phi}\big\|^2\\
     &=&\big\|(\tilde{\mathcal{D}}^2_{\infty})^{-1} \tilde{\mathcal{D}}_{nm}(\overline{\rho\phi})\|^2\\
&=& \bigg\|\frac{1}{2}\sum_{k=1}^\infty \frac{1+\gamma_k}{1+\nu \gamma_k} \langle \tilde{\mathcal{D}}_{nm}(\overline{\rho\phi}),e_k\rangle  e_k\bigg\|^2\\
&=& \frac{1}{4}\sum_{k=1}^\infty \frac{(1+\gamma_k)^2}{(1+\nu \gamma_k)^2} \langle \mathcal{M}{\mathcal{D}}_{nm}(\overline{\rho\phi}),e_k\rangle^2\\
&=& \frac{1}{4}\sum_{k=1}^\infty \frac{\langle \mathcal{M}{\mathcal{D}}_{nm}(\overline{\rho\phi}),e_k\rangle^2_{\mathcal{W}_q^2(\mathcal{T})}}{(1+\nu \gamma_k)^2} \\
&=& \frac{1}{4}\sum_{k=1}^\infty \frac{{(\mathcal{D}}_{nm}(\overline{\rho\phi})e_k)^2}{(1+\nu \gamma_k)^2}.
\end{eqnarray*}
The second and fourth ``$=$" are due to \eqref{sob_second} and \eqref{eq: tran}, and the last equality holds due to Riesz representation theorem. 

Recall that
\begin{eqnarray*}
    \mathcal{D}_{nm}(f)g
    =-\sum_{i=1}^n\frac{2\hat{a}^{(h)}_{i}}{J_i}\sum_{j=1}^{J_i}\big\{Y_{ij}-\hat{a}^{(h)}_{i}f(T_{ij})\big\}g(T_{ij})
    +2\nu \langle D^q f,D^q g\rangle.
\end{eqnarray*}
Notice that $\mathcal{D}_{\infty}(\overline{\rho\phi})e_k=0$, $\forall k\geq 1$, by the definition of $\overline{\rho\phi}$, we adopt \eqref{D1_inf} and obtain
\begin{eqnarray}\label{total_bound}
   && \mathcal{D}_{nm}(\overline{\rho\phi})e_k\nonumber\\
   &=&\mathcal{D}_{nm}(\overline{\rho\phi})e_k-\mathcal{D}_{\infty}(\overline{\rho\phi})e_k\nonumber\\
    &=&-2\sum_{i=1}^n\frac{\hat{a}^{(h)}_{i}}{J_i}\sum_{j=1}^{J_i}\{Y_{ij}-\hat{a}^{(h)}_{i}\overline{\rho\phi}(T_{ij})\}e_k(T_{ij})
    +2\bigg\langle \sum_{i=1}^n\hat{a}^{(h)}_{i}X_i-\overline{\rho\phi},e_k\bigg\rangle\nonumber\\
    &=&-2\sum_{i=1}^n\bigg[\frac{1}{J_i}\sum_{j=1}^{J_i}\big\{(\hat{a}^{(h)}_{i})^2\overline{\rho\phi}^{(h)}_1(T_{ij})-(\hat{a}^{(h)}_{i})^2\overline{\rho\phi}(T_{ij})\big\}e_k(T_{ij})-\langle \overline{\rho\phi}^{(h)}-\overline{\rho\phi},e_k\rangle\bigg]\nonumber\\
    &-&2\sum_{i=1}^n\hat{a}^{(h)}_{i}\bigg(\frac{1}{J_i}\sum_{j=1}^{J_i}\varepsilon_{ij}e_k(T_{ij})\bigg)\nonumber\\
    &-&2\sum_{i=1}^n\bigg[\frac{1}{J_i}\sum_{j=1}^{J_i}\big\{\hat{a}^{(h)}_{i}X_{i}(T_{ij})-(\hat{a}^{(h)}_{i})^2\overline{\rho\phi}^{(h)}_1(T_{ij})\big\}e_k(T_{ij})\bigg]\nonumber\\
    &=&(1)+(2)+(3).
\end{eqnarray}
We bound (1), (2), and (3) in the remaining.

\textbf{Upper bound of (1)}: Notice that
\begin{eqnarray*}
&&\sum_{i=1}^n\bigg[\frac{1}{J_i}\sum_{j=1}^{J_i}\big\{ (\hat{a}^{(h)}_{i})^2 \overline{\rho\phi}^{(h)}_1(T_{ij}) - (\hat{a}^{(h)}_{i})^2 \overline{\rho\phi}(T_{ij}) \big\} e_k(T_{ij}) - \langle \overline{\rho\phi}^{(h)} - \overline{\rho\phi}, e_k \rangle \bigg] \\
&=& \sum_{i=1}^n (\hat{a}^{(h)}_{i})^2 \bigg[ \frac{1}{J_i} \sum_{j=1}^{J_i} \big( \overline{\rho\phi}^{(h)}_1(T_{ij}) - \overline{\rho\phi}(T_{ij}) \big) e_k(T_{ij}) - \langle \overline{\rho\phi}^{(h)} - \overline{\rho\phi}, e_k \rangle \bigg] \\
&\leq& \sum_{i=1}^n (\hat{a}^{(h)}_{i})^2 \bigg| \frac{1}{J_i} \sum_{j=1}^{J_i} \big( \overline{\rho\phi}^{(h)}_1(T_{ij}) - \overline{\rho\phi}(T_{ij}) \big) e_k(T_{ij}) - \langle \overline{\rho\phi}^{(h)} - \overline{\rho\phi}, e_k \rangle \bigg| \\
&\leq& \sup_{i\in [n]} \bigg| \frac{1}{J_i} \sum_{j=1}^{J_i} \big( \overline{\rho\phi}^{(h)}_1(T_{ij}) - \overline{\rho\phi}(T_{ij}) \big) e_k(T_{ij}) - \langle \overline{\rho\phi}^{(h)} - \overline{\rho\phi}, e_k \rangle \bigg| 
\end{eqnarray*}
By condition \eqref{C11}, we have 
\begin{eqnarray*}
&& \sup_{i\in [n]} \bigg| \frac{1}{J_i} \sum_{j=1}^{J_i} \big( \overline{\rho\phi}^{(h)}_1(T_{ij}) - \overline{\rho\phi}(T_{ij}) \big) e_k(T_{ij}) - \langle \overline{\rho\phi}^{(h)} - \overline{\rho\phi}, e_k \rangle \bigg| \\
&\lesssim& m^{-q/(2q+1)} \cdot \| (\overline{\rho\phi}^{(h)} - \overline{\rho\phi}) e_k \| + m^{-2q/(2q+1)} \cdot \| (\overline{\rho\phi}^{(h)} - \overline{\rho\phi}) e_k \|_{\infty} \\
&\lesssim& m^{-q/(2q+1)} \cdot \| \overline{\rho\phi}^{(h)} - \overline{\rho\phi} \| + m^{-2q/(2q+1)} \cdot \| \overline{\rho\phi}^{(h)} - \overline{\rho\phi} \|_{\infty}.
\end{eqnarray*}
Notice that $\| \overline{\rho\phi}^{(h)} - \overline{\rho\phi} \| \leq \rho_1^0 \sqrt{\nu}$ due to \eqref{est_opt}, $\| D^q \overline{\rho\phi}^{(h)} \| \lesssim \rho_1^0$ due to \eqref{bias_bound}, and $\| D^q \overline{\rho\phi} \| \leq \| D^q \overline{\rho\phi}^{(h)} \| \lesssim \rho_1^0$ due to \eqref{inq_f}.
Therefore, 
\begin{eqnarray*}
\| \overline{\rho\phi}^{(h)} - \overline{\rho\phi} \|_{\infty} &\lesssim& \| \overline{\rho\phi}^{(h)} - \overline{\rho\phi} \| + \| D^q (\overline{\rho\phi}^{(h)} - \overline{\rho\phi}) \| \\
&\lesssim& \rho_1^0 (\sqrt{\nu} + 1).
\end{eqnarray*}
Since $m^{-q/(2q+1)} \lesssim \nu^{1/(2q)} \lesssim \nu^{1/(4q)}$,
we combine the above results and obtain
\begin{eqnarray}\label{inq_(22)}
&& \bigg| \sum_{i=1}^n \frac{1}{J_i} \sum_{j=1}^{J_i} \big\{ (\hat{a}^{(h)}_{i})^2 \overline{\rho\phi}^{(h)}_1(T_{ij}) - (\hat{a}^{(h)}_{i})^2 \overline{\rho\phi}(T_{ij}) \big\} e_k(T_{ij}) - \langle \overline{\rho\phi}^{(h)} - \overline{\rho\phi}, e_k \rangle \bigg| \nonumber \\
&\lesssim& m^{-q/(2q+1)} \cdot \rho_1^0 \sqrt{\nu} + m^{-2q/(2q+1)} \cdot \rho_1^0 (\sqrt{\nu} + 1) \nonumber \\
&\lesssim& m^{-q/(2q+1)} \cdot \rho_1^0 \cdot \nu^{1/(4q)}.
\end{eqnarray}

\textbf{Upper bound of (2)}: Notice that 
\begin{eqnarray*}
    &&\bigg|\sum_{i=1}^n \hat{a}^{(h)}_{i} \bigg( \frac{1}{J_i} \sum_{j=1}^{J_i} \varepsilon_{ij} e_k(T_{ij}) \bigg) \bigg|\\
    &\leq& \sum_{i=1}^n \big| \hat{a}^{(h)}_{i} \big| \cdot \bigg| \frac{1}{J_i} \sum_{j=1}^{J_i} \varepsilon_{ij} e_k(T_{ij}) \bigg|\\
    &\leq& \sum_{i=1}^n \big| a^{0}_{i1} \big| \cdot \bigg| \frac{1}{J_i} \sum_{j=1}^{J_i} \varepsilon_{ij} e_k(T_{ij}) \bigg| + \sum_{i=1}^n \big| \hat{a}^{(h)}_{i} - a_{i1}^0 \big| \cdot \bigg| \frac{1}{J_i} \sum_{j=1}^{J_i} \varepsilon_{ij} e_k(T_{ij}) \bigg| \\
    &\leq& \sum_{i=1}^n \big| a^{0}_{i1} \big| \cdot \bigg| \frac{1}{J_i} \sum_{j=1}^{J_i} \varepsilon_{ij} \bigg| \cdot \| e_k \|_{\infty} + \| \hat{\bm{a}}^{(h)} - \bm{a}_1^0 \| \cdot \sum_{i=1}^n \bigg| \frac{1}{J_i} \sum_{j=1}^{J_i} \varepsilon_{ij} e_k(T_{ij}) \bigg|  \\
    &\lesssim& \| e_k \|_{\infty}\cdot \bigg( \sum_{i=1}^n \big| a^{0}_{i1} \big| \cdot  \frac{1}{J_i} \sum_{j=1}^{J_i} |\varepsilon_{ij}| + \| \hat{\bm{a}}^{(h)} - \bm{a}_1^0 \| \cdot \sum_{i=1}^n \frac{1}{J_i} \sum_{j=1}^{J_i} | \varepsilon_{ij} |  \bigg).
\end{eqnarray*}
Since $\| e_k \|_{\infty} \lesssim 1$ by Lemma \ref{le:sobo}, we have
\begin{eqnarray*}
    \bigg| \sum_{i=1}^n \hat{a}^{(h)}_{i} \bigg( \frac{1}{J_i} \sum_{j=1}^{J_i} \varepsilon_{ij} e_k(T_{ij}) \bigg) \bigg| \lesssim \sum_{i=1}^n \big| a^{0}_{i1} \big| \cdot  \frac{1}{J_i} \sum_{j=1}^{J_i} |\varepsilon_{ij}| + \| \hat{\bm{a}}^{(h)} - \bm{a}_1^0 \| \cdot \sum_{i=1}^n \frac{1}{J_i} \sum_{j=1}^{J_i} | \varepsilon_{ij} |.
\end{eqnarray*}
By conditions \eqref{C2} and \eqref{C3}, and $\| \hat{\bm{a}}^{(h)} - \bm{a}_1^0 \|\lesssim \operatorname{dist}( \hat{\bm{a}}^{(h)}, {\bm{a}}_1^0 )$, we have
\begin{eqnarray*}
    \bigg| \sum_{i=1}^n \hat{a}^{(h)}_{i} \bigg( \frac{1}{J_i} \sum_{j=1}^{J_i} \varepsilon_{ij} e_k(T_{ij}) \bigg) \bigg| \lesssim x \sqrt{ \frac{n}{m} } \cdot \sigma + \operatorname{dist}( \hat{\bm{a}}^{(h)}, {\bm{a}}_1^0 ) \cdot x \sqrt{ \frac{n}{m} } \cdot \sigma.
\end{eqnarray*}

Using the inequality $ab\leq (a^2+b^2)/2$, we have
\begin{eqnarray*}
   {\operatorname{dist}(\hat{\bm{a}}^{(h)},{\bm{a}}_1^0)} \cdot{\frac{\sqrt{n}x}{\sqrt{m}}}\cdot \sigma
   &\lesssim&
\rho_1^0\operatorname{dist}^2(\hat{\bm{a}}^{(h)},{\bm{a}}_1^0)\cdot \nu^{1/(4q)}
   +\frac{1}{\nu^{1/(4q)}} \cdot{\frac{n x^2}{m\rho_1^0}}\cdot \sigma^2 \\
   &=&
   \rho_1^0\operatorname{dist}^2(\hat{\bm{a}}^{(h)},{\bm{a}}_1^0)\cdot \nu^{1/(4q)}
   +\sqrt{\frac{n}{m}} \sigma\cdot x\cdot \frac{x}{\nu^{1/(4q)}}\cdot \sqrt{\frac{n}{m}} \frac{\sigma}{\rho_1^0}.
\end{eqnarray*}
Notice that $\sqrt{\frac{n}{m}} \frac{\sigma}{\rho_1^0}\cdot x\lesssim \nu^{1/(2q)}$, we combine the above two inequalities and obtain
\begin{eqnarray}\label{bound_(2)}
    &&\bigg|\sum_{i=1}^n\hat{a}^{(h)}_{i1}\bigg(\frac{1}{J_i}\sum_{j=1}^{J_i}\varepsilon_{ij}e_k(T_{ij})\bigg)\bigg|\nonumber\\
    &\lesssim&
    x\sqrt{\frac{n}{m}}\cdot \sigma
    +\rho_1^0\operatorname{dist}^2(\hat{\bm{a}}^{(h)},{\bm{a}}_1^0)\cdot \nu^{1/(4q)}
    +\sqrt{\frac{n}{m}} \sigma\cdot x\cdot \nu^{1/(4q)}.
\end{eqnarray}

We similarly prove the upper bound of (3):
\begin{eqnarray}\label{bound_(1)}
  &&  \sum_{i=1}^n\bigg[\frac{1}{J_i}\sum_{j=1}^{J_i}\big\{\hat{a}^{(h)}_{i}X_{i}(T_{ij})-(\hat{a}^{(h)}_{i})^2\overline{\rho\phi}^{(h)}_1(T_{ij})\big\}e_k(T_{ij})\bigg]\nonumber\\
  & \lesssim &\rho_1^0\operatorname{dist}^2(\hat{\bm{a}}^{(h)},{\bm{a}}_1^0)\cdot \nu^{1/(4q)}+ \rho_1^0m^{-q/(2q+1)}\cdot \nu^{1/(4q)},
\end{eqnarray}
which can be controlled by the terms in \eqref{inq_(22)} and \eqref{bound_(2)}.

\textbf{Combining the upper bounds of (1), (2), and (3)}:
We now examine the upper bound of $\| \overline{\rho\phi}- \widetilde{\rho\phi}\|$. Recall that
\begin{eqnarray*}
   \| \overline{\rho\phi}- \widetilde{\rho\phi}\|^2
   &=& \frac{1}{4}\sum_{k=1}^\infty \frac{\big({ \tilde{\mathcal{D}}_{nm}(\overline{\rho\phi}), e_k } \big )^2}{(1+\nu \gamma_k)^2}\\
  &\lesssim& \sup_{k\geq 1}\big\{ \big({ \tilde{\mathcal{D}}_{nm}(\overline{\rho\phi}), e_k } \big )^2 \big\} \cdot \sum_{k=1}^\infty \frac{1}{(1+\nu \gamma_k)^2}.
\end{eqnarray*}
Note that by Lemma \ref{le:sobo},
\begin{eqnarray*}
    \sum_{k=1}^\infty \frac{1}{(1+\nu \gamma_k)^2}
    &=& q + \sum_{k=1}^\infty \frac{1}{(1+\nu \gamma_{q+k})^2} \\
    &\leq& q + \sum_{k=1}^\infty \frac{1}{(1+C_1\nu k^{2q})^2}
    \leq q + \int_0^\infty \frac{1}{(1+C_1\nu t^{2q})^2}\, \mathrm{d}t \\
    &\lesssim& \frac{1}{\nu^{1/(2q)}}.
\end{eqnarray*}
Combining the above two inequalities with \eqref{total_bound}, \eqref{inq_(22)}, \eqref{bound_(2)}, and \eqref{bound_(1)}, we have
\begin{eqnarray}\label{main_a}
 \| \overline{\rho\phi}- \widetilde{\rho\phi}\|
 &\lesssim&
 \rho_1^0 m^{-q/(2q+1)}
 + \frac{1}{\nu^{1/(4q)}} \cdot \frac{x}{\sqrt{m}} \cdot \sigma
 + \rho_1^0 \operatorname{dist}^2(\hat{\bm{a}}^{(h)}, {\bm{a}}_1^0)
 + \sqrt{\frac{n}{m}}\, \sigma \cdot x.
\end{eqnarray}

\textbf{(b) Error bound for $\|\widehat{\rho\phi}^{(h)}- \widetilde{\rho\phi}\|^2$}: 

Again using \eqref{D2}, we have
\begin{eqnarray}\label{eq_func}
     & &  \|\widehat{\rho\phi}^{(h)}- \widetilde{\rho\phi}\|^2\nonumber\\
     &=&\|\widehat{\rho\phi}^{(h)}-\overline{\rho\phi}+ (\tilde{\mathcal{D}}^2_{\infty})^{-1} \tilde{\mathcal{D}}_{nm}(\overline{\rho\phi})\|^2\nonumber\\
     &=& \|(\tilde{\mathcal{D}}^2_{\infty})^{-1}\big(\tilde{\mathcal{D}}^2_{\infty}(\widehat{\rho\phi}^{(h)}-\overline{\rho\phi})+  \tilde{\mathcal{D}}_{nm}(\overline{\rho\phi})\big)\|^2\nonumber\\
&=& \bigg\|\frac{1}{2}\sum_{k=1}^\infty \frac{1+\gamma_k}{1+\nu \gamma_k} \langle \tilde{\mathcal{D}}^2_{\infty}(\widehat{\rho\phi}^{(h)}-\overline{\rho\phi})+  \tilde{\mathcal{D}}_{nm}(\overline{\rho\phi}),e_k\rangle  e_k\bigg\|^2\nonumber\\
&=& \frac{1}{4}\sum_{k=1}^\infty \frac{(1+\gamma_k)^2}{(1+\nu \gamma_k)^2} \langle \mathcal{M}{\mathcal{D}}^2_{\infty}(\widehat{\rho\phi}^{(h)}-\overline{\rho\phi})+  \mathcal{M}{\mathcal{D}}_{nm}(\overline{\rho\phi}),e_k\rangle^2\nonumber\\
&=& \frac{1}{4}\sum_{k=1}^\infty \frac{\langle \mathcal{M}{\mathcal{D}}^2_{\infty}(\widehat{\rho\phi}^{(h)}-\overline{\rho\phi})+  \mathcal{M}{\mathcal{D}}_{nm}(\overline{\rho\phi}),e_k\rangle^2_{\mathcal{W}_q^2(\mathcal{T})}}{(1+\nu \gamma_k)^2} \nonumber\\
&=& \frac{1}{4}\sum_{k=1}^\infty \frac{\big[\big\{{\mathcal{D}}^2_{\infty}(\widehat{\rho\phi}^{(h)}-\overline{\rho\phi})+  {\mathcal{D}}_{nm}(\overline{\rho\phi})\big\}e_k\big]^2}{(1+\nu \gamma_k)^2}.
\end{eqnarray}
The third and fifth ``$=$" are due to \eqref{sob_second} and \eqref{eq: tran}, and the last equality holds due to Reisz representation theorem. 

Notice that $\mathcal{D}_{nm}(\widehat{\rho\phi}^{(h)})e_k=0$, $\forall k\geq 1$, by the definition of $\widehat{\rho\phi}^{(h)}$. We adopt \eqref{D1} and \eqref{D2} and obtain
\begin{eqnarray}\label{q_func}
&& \big\{{\mathcal{D}}^2_{\infty}(\widehat{\rho\phi}^{(h)}-\overline{\rho\phi})+ {\mathcal{D}}_{nm}(\overline{\rho\phi})\big\}e_k \nonumber\\
&=& \big\{{\mathcal{D}}^2_{\infty}(\widehat{\rho\phi}^{(h)}-\overline{\rho\phi}) + {\mathcal{D}}_{nm}(\overline{\rho\phi}) - {\mathcal{D}}_{nm}(\widehat{\rho\phi}^{(h)})\big\}e_k \nonumber\\
&=& 2\langle \widehat{\rho\phi}^{(h)} - \overline{\rho\phi}, e_k \rangle - \sum_{i=1}^n \frac{2(\hat{a}_{i}^{(h)})^2}{J_i} \sum_{j=1}^{J_i} (\widehat{\rho\phi}^{(h)}(T_{ij}) - \overline{\rho\phi}(T_{ij})) e_k(T_{ij}) \nonumber\\
&=& 2\sum_{i=1}^n (\hat{a}_{i}^{(h)})^2 \bigg[ \langle \widehat{\rho\phi}^{(h)} - \overline{\rho\phi}, e_k \rangle 
      - \frac{1}{J_i} \sum_{j=1}^{J_i} (\widehat{\rho\phi}^{(h)}(T_{ij}) - \overline{\rho\phi}(T_{ij})) e_k(T_{ij}) \bigg]\nonumber\\
&\leq& 2\bigg(\sum_{i=1}^n (\hat{a}_{i}^{(h)})^2\bigg)\cdot 
\sup_{i \in [n]} \bigg| \frac{1}{J_i} \sum_{j=1}^{J_i} (\widehat{\rho\phi}^{(h)}(T_{ij}) - \overline{\rho\phi}(T_{ij})) e_k(T_{ij}) - \langle \widehat{\rho\phi}^{(h)} - \overline{\rho\phi}, e_k \rangle \bigg|\nonumber \\
&=& 2\cdot 
\sup_{i \in [n]} \bigg| \frac{1}{J_i} \sum_{j=1}^{J_i} (\widehat{\rho\phi}^{(h)}(T_{ij}) - \overline{\rho\phi}(T_{ij})) e_k(T_{ij}) - \langle \widehat{\rho\phi}^{(h)} - \overline{\rho\phi}, e_k \rangle \bigg|,
\end{eqnarray}
where we used $\sum_{i=1}^n(\hat{a}^{(h)}_i)^2=\|\hat{\bm{a}}^{(h)}\|_2^2=1$.

By condition \eqref{C11},
\begin{eqnarray*}
&& \sup_{i \in [n]} \bigg| \frac{1}{J_i} \sum_{j=1}^{J_i} (\widehat{\rho\phi}^{(h)}(T_{ij}) - \overline{\rho\phi}(T_{ij})) e_k(T_{ij}) - \langle \widehat{\rho\phi}^{(h)} - \overline{\rho\phi}, e_k \rangle \bigg| \\
&\lesssim& m^{-q/(2q+1)} \cdot \| (\widehat{\rho\phi}^{(h)} - \overline{\rho\phi}) e_k \| + m^{-2q/(2q+1)} \cdot \| (\widehat{\rho\phi}^{(h)} - \overline{\rho\phi}) e_k \|_{\infty} \\
&\lesssim& m^{-q/(2q+1)} \cdot \| \widehat{\rho\phi}^{(h)} - \overline{\rho\phi} \| + m^{-2q/(2q+1)} \cdot \| \widehat{\rho\phi}^{(h)} - \overline{\rho\phi} \|_{\infty}.
\end{eqnarray*}

Note that $\| D^q (\widehat{\rho\phi}^{(h)} - \overline{\rho\phi}) \| \leq \| D^q \widehat{\rho\phi}^{(h)} \| + \| D^q \overline{\rho\phi} \| \leq \| D^q \widehat{\rho\phi}^{(h)} \|+\| D^q \overline{\rho\phi}^{(h)}\|\lesssim \rho_1^0$ due to \eqref{inq_f} and \eqref{bias_bound} and the condition $\| D^q \widehat{\rho\phi}^{(h)} \|\lesssim \rho_1^0$. Therefore,
\begin{eqnarray*}
\| \widehat{\rho\phi}^{(h)} - \overline{\rho\phi} \|_{\infty} &\lesssim& \| \widehat{\rho\phi}^{(h)} - \overline{\rho\phi} \| + \| D^q (\widehat{\rho\phi}^{(h)} - \overline{\rho\phi}) \|\\ 
&\lesssim& \| \widehat{\rho\phi}^{(h)} - \overline{\rho\phi} \| + \rho_1^0.
\end{eqnarray*}
As a result,
\begin{eqnarray*}
&& \sup_{i \in [n]} \bigg| \frac{1}{J_i} \sum_{j=1}^{J_i} (\widehat{\rho\phi}^{(h)}(T_{ij}) - \overline{\rho\phi}(T_{ij})) e_k(T_{ij}) - \langle \widehat{\rho\phi}^{(h)} - \overline{\rho\phi}, e_k \rangle \bigg| \\
&\lesssim& m^{-q/(2q+1)} \cdot \| \widehat{\rho\phi}^{(h)} - \overline{\rho\phi} \| + \rho_1^0 m^{-2q/(2q+1)}.
\end{eqnarray*}
By combining the above inequality with \eqref{eq_func} and \eqref{q_func}, we then obtain
\begin{eqnarray*}
\|\widehat{\rho\phi}^{(h)} - \widetilde{\rho\phi}\| 
\lesssim 
\frac{m^{-q/(2q+1)}}{\nu^{1/(4q)}} \cdot \| \widehat{\rho\phi}^{(h)} - \overline{\rho\phi} \|
+ \frac{m^{-q/(2q+1)}}{\nu^{1/(4q)}} \cdot \rho_1^0 m^{-q/(2q+1)}.
\end{eqnarray*}

Notice that $m^{-q/(2q+1)} \lesssim \nu^{1/(2q)}$. With a suitable $\nu$, we have
\begin{eqnarray*}
\|\widehat{\rho\phi}^{(h)} - \widetilde{\rho\phi}\| \leq \| \widehat{\rho\phi}^{(h)} - \overline{\rho\phi} \| / 2 + \rho_1^0 m^{-q/(2q+1)} / 2.
\end{eqnarray*}
Furthermore,
\begin{eqnarray*}
\| \overline{\rho\phi} - \widetilde{\rho\phi} \| \geq \| \widehat{\rho\phi}^{(h)} - \overline{\rho\phi} \| - \| \widehat{\rho\phi}^{(h)} - \widetilde{\rho\phi} \|,
\end{eqnarray*}
by the triangle inequality.
Combining with the above two inequalities,
\begin{eqnarray*}
\| \overline{\rho\phi} - \widetilde{\rho\phi} \| \geq \| \widehat{\rho\phi}^{(h)} - \widetilde{\rho\phi} \| - \rho_1^0 m^{-q/(2q+1)}.
\end{eqnarray*}
Therefore,
\begin{eqnarray*}
\|\widehat{\rho\phi}^{(h)} - \widetilde{\rho\phi}\|
&\lesssim& 
\rho_1^0 m^{-q/(2q+1)} + \| \overline{\rho\phi} - \widetilde{\rho\phi} \| \\
&\lesssim& 
\rho_1^0 m^{-q/(2q+1)}
+ \frac{1}{\nu^{1/(4q)}} \cdot \frac{x}{\sqrt{m}} \cdot \sigma
+ \rho_1^0 \operatorname{dist}^2(\hat{\bm{a}}^{(h)}, {\bm{a}}_1^0)
+ \sqrt{\frac{n}{m}}\, \sigma \cdot x
\end{eqnarray*}
due to \eqref{main_a}.
\end{proof}

\clearpage
\newpage


\begin{proof}[Proof to Lemma \ref{Le:4}]
In the following proof, we always assume $(\bm{a}_1^0)^\top \bm{\hat{a}}^{(h)} \geq 0$ and $\langle \widehat{\rho\phi}^{(h)},\phi^0_1\rangle \geq 0$ for all $h \geq 0$ as it does not affect the conclusion.

Note that for any positive value $d$,
\begin{eqnarray*}
    \operatorname{dist}(\widehat{\rho\phi}^{(h)},\phi_{1}^0) &\leq& \|d\widehat{\rho\phi}^{(h)} - \phi_{1}^0\| \\
    &\leq& d\|\widehat{\rho\phi}^{(h)} - \overline{\rho\phi}^{(h)}\| +
    \|d\overline{\rho\phi}^{(h)} - \phi_{1}^0\|,
\end{eqnarray*}
due to Lemma \ref{lm:projection}.
We set $d = 1/\rho_1^0$.
By Lemma \ref{le_cor},
\begin{eqnarray*}
&&   \frac{1}{\rho_1^0} \cdot \|\widehat{\rho\phi}^{(h)} - \overline{\rho\phi}^{(h)}\| \\
   &\leq& C\bigg(\frac{1}{\nu^{1/(4q)}} \cdot \frac{\sigma}{\rho_1^0 \sqrt{m}} + \sqrt{\nu}
   + \sqrt{\frac{n}{m}} \cdot \frac{\sigma}{\rho_1^0} \cdot x + m^{-q/(2q+1)}\bigg)
   + \operatorname{dist}^2(\hat{\bm{a}}^{(h)}, {\bm{a}}_1^0) \\
   &\leq& C\bigg(\frac{1}{\nu^{1/(4q)}} \cdot \frac{\sigma}{\rho_1^0 \sqrt{m}} + \sqrt{\nu}
   + \sqrt{\frac{n}{m}} \cdot \frac{\sigma}{\rho_1^0} \cdot x + m^{-q/(2q+1)}\bigg)
   + \operatorname{dist}(\hat{\bm{a}}^{(h)}, {\bm{a}}_1^0).
\end{eqnarray*}
In addition,
\begin{eqnarray*}
   \left\|\frac{\overline{\rho\phi}^{(h)}}{\rho_1^0} - \phi_{1}^0\right\|
   &=& \frac{1}{\rho_1^0} \cdot \left\|\sum_{i=1}^n \left(\hat{a}_{i}^{(h)} - a_{i1}^0\right) X_i\right\| \\
   &\leq& \|\bm{\hat{a}}_{1}^{(h)} - \bm{a}_{1}^0\| \\
   &\leq& \sqrt{2}\operatorname{dist}(\bm{a}_1^0, \bm{\hat{a}}^{(h)})
\end{eqnarray*}
by Lemma \ref{bound_le}. We then obtain \eqref{bound_phi} by combining the above three inequalities.
\end{proof}

\

\subsubsection{Proof of Lemma \ref{le:sobo}}
\begin{Le}\label{le:sobo}
There exists a collection of basis functions $e_k$ in $\mathcal{W}_q^2(\mathcal{T})$ such that
$$
\langle e_{k_1},e_{k_2}\rangle=\mathbb{I}(k_1=k_2)
$$
and
$$
\langle D^qe_{k_1},D^qe_{k_2}\rangle=\mathbb{I}(k_1=k_2)\gamma_{k_1},
$$
where $\gamma_k$s satisfy $\gamma_k=0$, $k\leq q$, and
\begin{eqnarray*}
    C_1k^{2q}\leq \gamma_{k+q}\leq C_2 k^{2q},\ k\geq 1,
\end{eqnarray*}
with $C_1$ and $C_2$ being two constants. In addition,
\begin{eqnarray*}
    \sup_{k\geq 1}\sup_{t\in \mathcal{T}}|e_k(t)|\lesssim 1.
\end{eqnarray*}
\end{Le}

See Section 2.8 in \citet{hsing2015theoretical} for the proof.

\

\subsubsection{Proof of Lemma \ref{bound_le}}
\begin{Le}\label{bound_le}
For any values $c_i$ and $f\in \mathcal{L}^2(\mathcal{T})$,
\begin{eqnarray*}
  \sqrt{\sum_{i=1}^n \big|\langle X_i,f\rangle\big|^2} &\leq& \|\mathcal{X}_n\|_{\infty} \cdot \|f\|, \\
\bigg\|\sum_{i=1}^n c_i X_i \bigg\| &\leq& \|\mathcal{X}_n\|_{\infty} \cdot \sqrt{\sum_{i=1}^n c_i^2},
\end{eqnarray*}
where we abuse notation and denote the operator norm by $\|\cdot\|_{\infty}$.
\end{Le}

\begin{proof}
Since $\sqrt{\sum_{i=1}^n \big|\langle X_i,f\rangle\big|^2} = \|\mathcal{X}_n f\|$, then
\begin{eqnarray*}
    \sqrt{\sum_{i=1}^n \big|\langle X_i,f\rangle\big|^2} \leq \|\mathcal{X}_n\|_{\infty} \cdot \|f\|
\end{eqnarray*}
is obtained by the property of operator norm.

By \eqref{equa_*}, $\mathcal{X}_n^* \bm{c} = \sum_{i=1}^n c_i X_i$. Notice that $\|\mathcal{X}_n^*\|_{\infty} = \|\mathcal{X}_n\|_{\infty}$, which leads to $\|\mathcal{X}_n^* \bm{c}\| \leq \|\mathcal{X}_n\|_{\infty} \cdot \|\bm{c}\|$, $\forall \bm{c}\in \mathbb{R}^n$. The second inequality is proven.
\end{proof}

\

\subsubsection{Proof of Lemma \ref{Le: Bound_function}}
\begin{Le}\label{Le: Bound_function}
For $X\in \mathcal{W}_2^q(\mathcal{T})$,
\begin{eqnarray*}
    \|X\|_{\infty}\lesssim \|X\|_{\mathcal{W}_2^q(\mathcal{T})}.
\end{eqnarray*}
\end{Le}

\begin{proof}
Let $\mathbb{K}$ be the reproducing kernel of $\mathcal{W}_2^q(\mathcal{T})$ with the norm $\|\cdot\|_{\mathcal{W}_2^q(\mathcal{T})}$. By the property of the reproducing kernel,
\begin{eqnarray*}
\sup_{t\in \mathcal{T}}|X(t)|^2
= \sup_{t\in \mathcal{T}} \langle X, \mathbb{K}(\cdot, t) \rangle_{\mathcal{W}_2^q(\mathcal{T})}^2
\leq \|X\|^2_{\mathcal{W}_2^q(\mathcal{T})} \cdot \sup_{t\in \mathcal{T}} |\mathbb{K}(t, t)|.
\end{eqnarray*}
It can be shown that $\sup_{t\in \mathcal{T}} |\mathbb{K}(t, t)|$ is bounded, and
\begin{eqnarray*}
\|X\|^2_{\mathcal{W}_2^q(\mathcal{T})} = \|X\|^2 + \|D^q X\|^2.
\end{eqnarray*}
Combining these results, the conclusion of Lemma \ref{Le: Bound_function} holds.
\end{proof}

\

\subsubsection{Proof of Lemma \ref{lm:projection}}
\begin{Le}\label{lm:projection}
For any $d\in \mathbb{R}$, we have
\begin{eqnarray*}
    \operatorname{dist}(\bm{u},\bm{v}) &\leq& \frac{\|\bm{u} - d \bm{v}\|_2}{\|\bm{u}\|_2}, \quad \forall \bm{u},\bm{v} \in \mathbb{R}^n,\\
 	\operatorname{dist}(f, g) &\leq& \frac{\|f - dg\|_2}{\|f\|_2}, \quad \forall f,g \in \mathcal{L}^2(\mathcal{T}).
\end{eqnarray*}
\end{Le}
\begin{proof}
We only prove the first inequality and the second one can be proven similarly.
\begin{equation*}
\begin{split}
& \operatorname{dist}(\bm{u}, \bm{v})
= \sqrt{1 - \frac{\langle \bm{u}, \bm{v}\rangle^2}{\|\bm{u}\|_2^2 \|\bm{v}\|_2^2}}
\leq \frac{\|\bm{u}-d \bm{v}\|_2}{\|\bm{u}\|_2} \\
\Leftarrow \quad &
1 - \frac{\langle \bm{u}, \bm{v}\rangle^2}{\|\bm{u}\|_2^2 \|\bm{v}\|_2^2}
\leq \frac{\|\bm{u}\|_2^2 - 2 d \langle \bm{u}, \bm{v}\rangle + d^2 \|\bm{v}\|_2^2}{\|\bm{u}\|_2^2} \\
\Leftarrow \quad &
0 \leq \langle \bm{u}, \bm{v}\rangle^2 - 2 d \|\bm{v}\|_2^2 \langle \bm{u}, \bm{v}\rangle + d^2 \|\bm{v}\|_2^4 \\
\Leftarrow \quad &
0 \leq \left( \langle \bm{u}, \bm{v}\rangle - d \|\bm{v}\|_2^2 \right)^2.
\end{split}
\end{equation*}
\end{proof}

\section{Implementation Details of FSVD}

\subsection{Functional Clustering}\label{sec:FC_SM}

We outline the general procedure of functional clustering in Algorithm~\ref{algo: FC}. 
Here, FSVD is utilized for both estimating basis functions and initializing the clustering algorithm. 
For Step~4 in Algorithm~\ref{algo: FC}, we can employ any vector clustering methods to obtain an initial clustering on $\{\hat{\bm{\xi}}_i; i \in [n]\}$. 
Thereafter, the initial estimates for parameters ($\bm{\mu}_h$, $\bm{\Sigma}_h$, $\pi_h$, and $\sigma_h$) can be derived from their empirical estimates based on the initial clustering. 
For the selection of $K$ in Algorithm~\ref{algo: FC}, we can utilize BIC applied to the initial clustering of $\{\hat{\bm{\xi}}_i; i \in [n]\}$ under the Gaussian mixture modeling assumption; see \citet{scrucca2023model} for more details.

\begin{algorithm}[H]
\caption{Functional Clustering by FSVD}\label{algo: FC}
\begin{algorithmic}[1]
\State \textbf{Input:} observed data $\big\{Y_{ij};j\in [J_i],i\in [n]\big\}$, number of clusters $H$, and number of basis functions $K$.
\State Estimate $\{\varphi_k\}_{k\in [K]}$ using the singular functions obtained from Algorithm \ref{algo: FSVD_R}. 
\State Calculate $\hat{\varphi}_k$, $k\leq K$, by Algorithm \ref{algo: FSVD_R}.
\State Calculate $\hat{\xi}_{ik}=\hat{\rho}_k\hat{a}_{ik}$ for $i\in [n]$ and $k\in [K]$, where $\hat{\rho}_k$s and $\hat{a}_{ik}$s are obtained from Algorithm \ref{algo: FSVD_R}.
\State Propose an initial clustering on the vectors $\{\hat{\bm{\xi}}_i := (\hat{\xi}_{i1}, \dots, \hat{\xi}_{iK})^\top; i \in [n]\}$, and calculate initial estimations for $\bm{\mu}_h$, $\bm{\Sigma}_h$, $\pi_h$, $\sigma_h$, $h\in [H]$, based on the clustering result.
\State Given $\hat{\varphi}_k$, $k\leq K$, we implement an EM algorithm on $\big\{Y_{ij};j\in [J_i],i\in [n]\big\}$ to estimate $\mathbb{P}\{Z_i=h\mid \bm{Y}_i\}$, $i\in [n]$ and $h\in [H]$, where the EM algorithm is initialized with the parameters in the last step.
\State \textbf{Output} $\hat{Z}_i=\arg\max_{h\in [H]}\mathbb{P}\{Z_i=h\mid \bm{Y}_i\}$, $i\in [n]$.
\end{algorithmic}
\end{algorithm}

\


\section{Simulation Studies}

\subsection{Initialization of FSVD}\label{sec: init_FSVD}

In this part, we evaluate the effectiveness of the initialization method in Section~\ref{sec:alternative_minimization}. We consider the data generation
\begin{eqnarray}\label{sim: model_init}
     [X_1(t),\dots, X_n(t)]^\top=\sum_{k=1}^K\rho_k\bm{a}_{k}\phi_k(t),\ t\in [0,1].
\end{eqnarray}
Here, $\rho_k=2\exp\big\{(K-k+1) / 2\big\}$, $\{\phi_k;1\leq k\leq K\}$ are the first $K$ non-constant Fourier basis functions, $\bm{a}_k$s are orthonormal random vectors in $\mathbb{R}^n$.
The elements of $\bm{a}_k$ are first sampled from $N(0,1)$, 
and then orthonormalized using the Gram--Schmidt process.

For each $X_i$, we randomly sample the number of time points $J_i$ from $\{4,\dots,8\}$, $\{6,\dots,10\}$ or $\{8,\dots,12\}$; we generate $\{T_{ij};j\in [J_i]\}$ independently from a uniform distribution on $\mathcal{T}=[0,1]$ and generate $Y_{ij}$s according to the measurement model \eqref{mea_model} with $\varepsilon_{ij} \sim N(0, \sigma_i^2)$ and $\sigma^2_i=\mathbb{E}\|X_i\|^2\cdot 5\%$. We set $K=3$ and generated 100 replications for each simulation setting. 

\begin{figure}[h]
\renewcommand{\thefigure}{S1}
    \centering
    \includegraphics[width=1\linewidth]{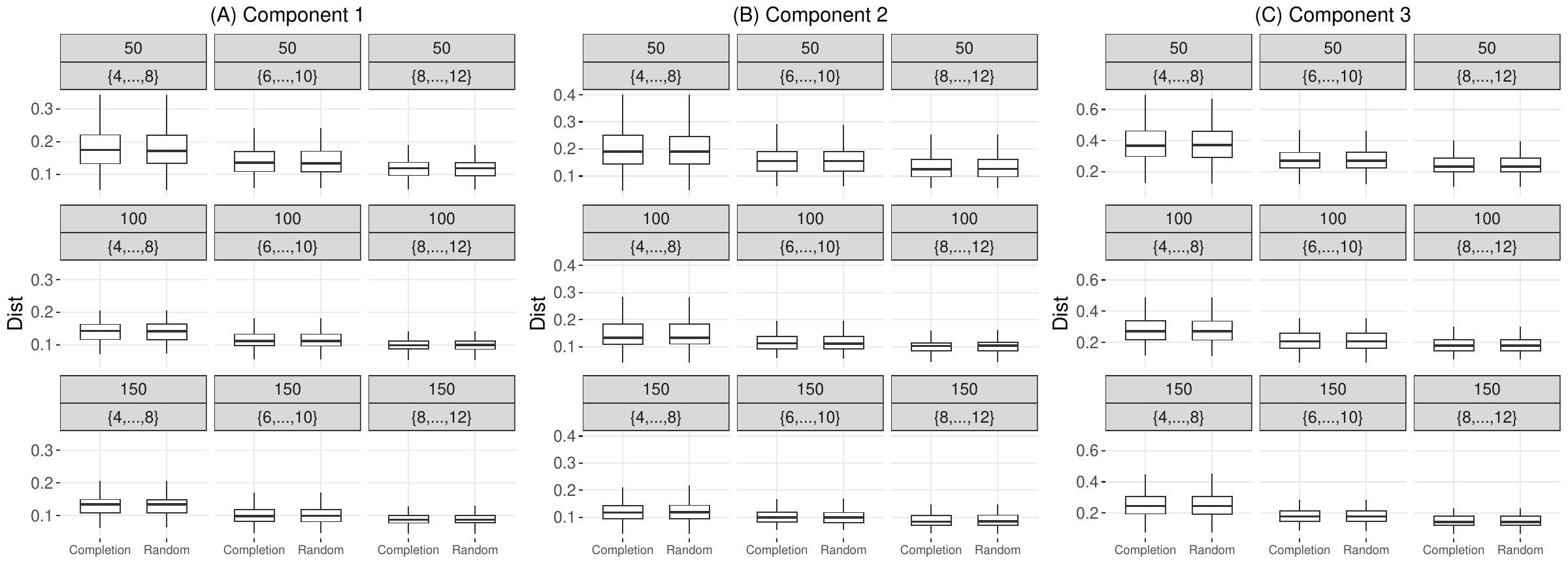}
    \caption{
        Boxplot of the values of $\operatorname{dist}(\cdot, \phi_k)$ 
        under different sample sizes $n$ and numbers of time points $J_i$. 
        Here, ``completion'' and ``random'' indicate matrix-completion 
        initialization and 30-run random initialization, respectively.
    }
    \label{fig:init_FSVD}
\end{figure}

We apply Algorithm~\ref{algo: FSVD_R} with matrix-completion initialization to the generated data. 
For comparison, we also implement FSVD with randomly initialized singular vectors for each component, 
selecting the component that achieves the minimal fitting loss among these initializations. 
The number of random initializations is set to $30$ for each run of FSVD, 
and the elements of the initialized singular vectors are sampled from $N(0,1)$ 
and then normalized. 
This exhaustive search can lead to a better approximation of the global minimum.

We use $\operatorname{dist}(\cdot, \phi_k)$, defined in Part~\ref{sec:theory} 
to measure the distance between the true singular functions and the estimated singular functions obtained 
from different initializations. The results are presented in Figure~\ref{fig:init_FSVD}. 
We observe that the values of $\operatorname{dist}(\cdot, \phi_k)$ under different initializations 
are nearly identical, indicating that the matrix-completion initialization 
achieves comparable performance to random initialization.

\subsection{FSVD for Learning Tasks}\label{sec: sim_SM}

In this part, we compare FSVD with several existing methods in four aspects: functional completion, functional linear regressions, functional clustering, and factor models.
\paragraph*{Simulations on Functional Completion.}
We generate both homogeneous and heterogeneous functional data using the following model:
\begin{eqnarray}\label{sim: model_SM}
     [X_1(t),\dots, X_n(t)]^\top=\sum_{k=1}^K\rho_k(\bm{a}_{k}+\bm{b}_k)\varphi_k(t),\ t\in [0,1].
\end{eqnarray}
Here, $\rho_k=2\exp\big\{(K-k+1) / 2\big\}$, $\{\varphi_k;1\leq k\leq K\}$ are the first $K$ non-constant Fourier basis functions. We construct $\bm{a}_k$s deterministically by setting ${a}_{ik}=\sin\big\{k\pi(i+n/4)/(2n)\big\}$ for $i\in [n], k\in [K]$, letting $\bm{a}_k=(a_{1k},\dots,a_{nk})^\top$, then orthonormalizing $\bm{a}_{k}$s by the Gram--Schmidt process. We draw $b_{ik} \sim N(0, a_{ik}^2)$ independently for each $i, k$ and set $\bm{b}_k=(b_{1k},\dots,b_{nk})^\top$.
Under this setting, $X_i$s are heterogeneous functional data with different mean and covariance functions for each $i$, and $\varphi_k$s are IBFs of $X_i$s, due to Theorem \ref{the: eui_in} c..
We also use \eqref{sim: model_SM} to generate i.i.d.\ functional data by setting $\bm{a}_k$s as zero vectors and generating $b_{ik}\sim N(0,1/n)$ for each $i, k$. As a result, $X_i$s are i.i.d.\ functional data with mean zero with $\varphi_k$s being their eigenfunctions, which corresponds to the setting of FPCA. For each $X_i$, we randomly sample the number of time points $J_i$ from $\{4,\dots,8\}$, $\{6,\dots,10\}$ or $\{8,\dots,12\}$; we generate $\{T_{ij};j\in [J_i]\}$ independently from a uniform distribution on $\mathcal{T}=[0,1]$ and generate $Y_{ij}$s according to the measurement model \eqref{mea_model} with $\varepsilon_{ij} \sim N(0, \sigma_i^2)$ with $\sigma^2_i=\mathbb{E}\|X_i\|^2\cdot 5\%$. We use $K=3$ and generated 100 replications for each simulation setting. 

\begin{figure}[ht!]
\renewcommand{\thefigure}{S2}
\begin{center}
\includegraphics[scale = 0.3, trim=0cm 0cm 0cm 0cm, clip]{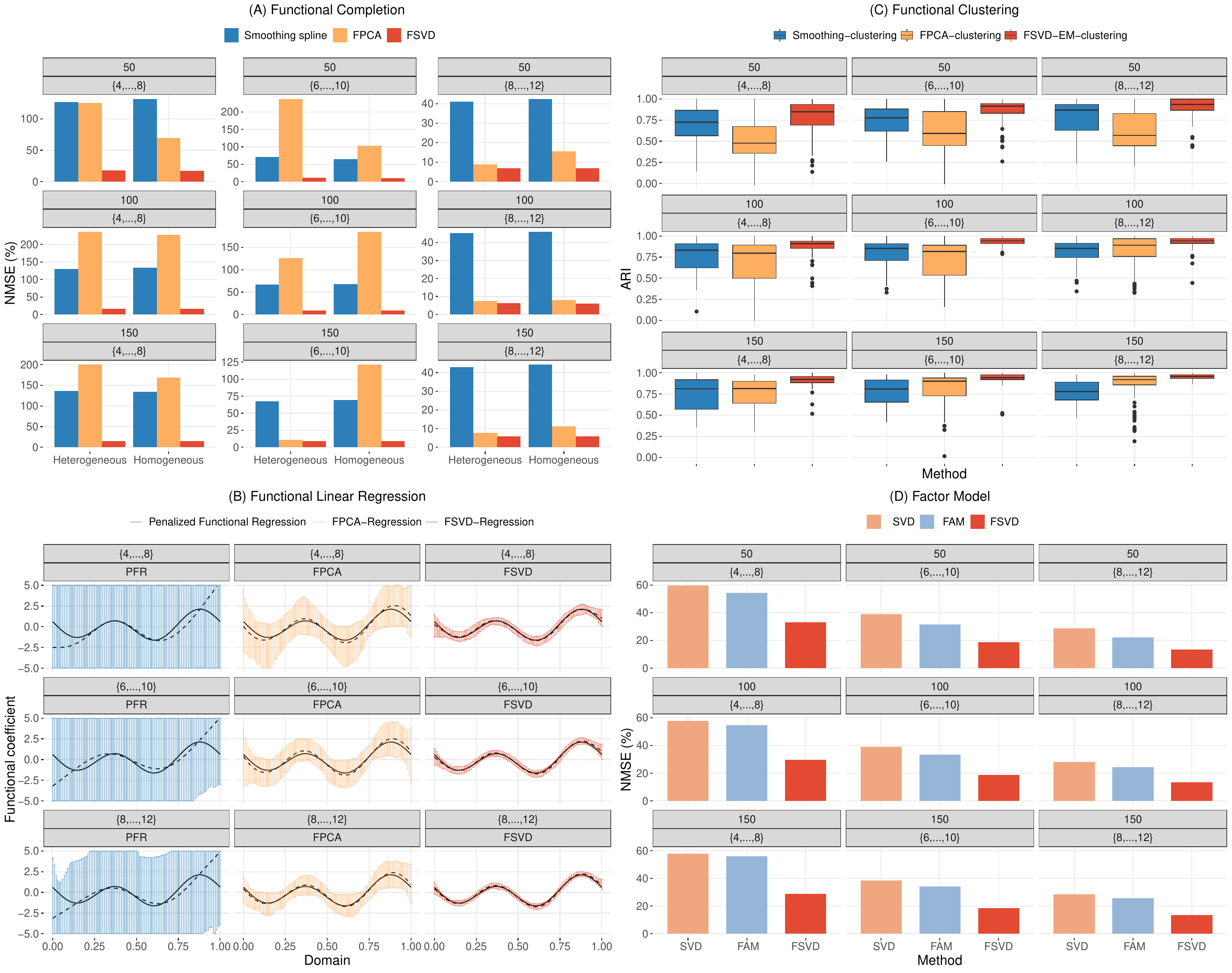}
\end{center}
\caption{\textbf{(A)}: The $\text{NMSE}_{X}$ of \textbf{functional completion} for different methods with sample sizes $n$ (main title) and numbers of time points $J_i$ (subtitle).  \textbf{(B)}: Functional coefficients of \textbf{functional regression} estimated from different methods with different numbers of time points $J_i$. The solid and dotted lines indicate the true functional coefficients and the point-wise means of the estimated functional coefficients from simulation, respectively. The shaded regions represent the 95\% point-wise interval calculated from simulation.
\textbf{(C)}: Box-plots of ARI of \textbf{functional clustering} for different methods with sample sizes $n$ and numbers of time points $J_i$.
\textbf{(D)}:
The $\text{NMSE}_{A}$ of  \textbf{factor model} loadings for different methods with sample sizes $n$ and numbers of time points $J_i$.} \label{fig: combine}
\end{figure}

The average NMSE values over 100 simulations are summarized in Figure~\ref{fig: combine}(A).
We can see that FSVD outperforms both FPCA and the smoothing spline in functional completion under all settings. 
Even when the functional data are i.i.d.\ as assumed by FPCA, FSVD still outperforms FPCA, especially for small $n$ and $J_i$, likely due to the accumulated estimation errors in estimating the covariance structure, which FSVD bypasses. 
The advantage over FPCA on the heterogeneous data is also likely contributed by the violation of the i.i.d.\ assumption that FPCA relies on.

\begin{table}[h]
\renewcommand{\thetable}{S1}
\caption{Estimation accuracy of intrinsic basis functions measured by $\text{dist}(\cdot, \varphi_k)$ under different sample sizes $n$ and the observed number of time points. Under the heterogeneous setting, we only evaluate FSVD since FPCA does not target on intrinsic basis functions.  \label{tab:intrinsic}}
\centering
\renewcommand{\arraystretch}{1.2}
\setlength\tabcolsep{3.5pt}
\footnotesize
\begin{tabular}{c|c|c|ccc|ccc|ccc}
  \hline
\multicolumn{3}{c|}{\multirow{2}{*} {$\text{dist}(\cdot, \varphi_k)$}}
   &  \multicolumn{3}{c|}{$J_i\in \{4,\dots,8\}$} &  \multicolumn{3}{c|}{$J_i\in \{6,\dots,10\}$}& \multicolumn{3}{c}{$J_i\in \{8,\dots,12\}$} \\
    \multicolumn{3}{c|}{}   & $k=1$ & $k=2$ & $k=3$  & $k=1$ & $k=2$ & $k=3$ 
  & $k=1$ & $k=2$ & $k=3$ \\ 
  \hline
  \multirow{6}{*}{\shortstack{Homogeneous\\ case}}
&  \multirow{2}{*}{$n=50$} &
FPCA & 0.29 & 0.37 & 0.74 & 0.25 & 0.32 & 0.61 & 0.23 & 0.31 & 0.58 \\ 
& &  FSVD & 0.25 & 0.26 & 0.36 & 0.21 & 0.22 & 0.25 & 0.20 & 0.21 & 0.23 \\ 
  \cline{2-12}
  &  \multirow{2}{*}{$n=100$} &
FPCA & 0.20 & 0.27 & 0.62 & 0.19 & 0.26 & 0.46 & 0.17 & 0.21 & 0.42 \\ 
&&  FSVD & 0.17 & 0.16 & 0.25 & 0.15 & 0.15 & 0.19 & 0.14 & 0.15 & 0.16 \\ 
       \cline{2-12}
 &        \multirow{2}{*}{$n=150$} &
FPCA & 0.17 & 0.23 & 0.55 & 0.14 & 0.19 & 0.44 & 0.13 & 0.19 & 0.35 \\ 
& &   FSVD & 0.16 & 0.14 & 0.22 & 0.14 & 0.13 & 0.16 & 0.12 & 0.12 & 0.13 \\ 
\hline
  \multirow{3}{*}{\shortstack{Heterogeneous \\ case}}
&  $n=50$ &
 FSVD & 0.22 & 0.25 & 0.41 & 0.20 & 0.22 & 0.30 & 0.18 & 0.20 & 0.22 \\ 
  \cline{2-12}
  & $n=100$ &
FSVD & 0.16 & 0.16 & 0.27 & 0.13 & 0.15 & 0.21 & 0.13 & 0.14 & 0.17 \\ 
       \cline{2-12}
 & $n=150$ &
FSVD & 0.14 & 0.13 & 0.23 & 0.12 & 0.12 & 0.17 & 0.10 & 0.12 & 0.14 \\
 \hline
\end{tabular}
\end{table}

In Table~\ref{tab:intrinsic}, we summarize the estimation accuracy of IBFs using \( \text{dist}(\cdot, \varphi_k) \) defined in Part~\ref{sec:theory}.
Under the homogeneous setting, we adopt the eigenfunctions estimated by FPCA and the singular functions estimated by FSVD to estimate the IBFs. FSVD outperforms FPCA likely because it avoids the need to estimate the covariance structure. Under the heterogeneous setting, we only evaluate FSVD since FPCA does not target on IBFs. In both homogeneous and heterogeneous scenarios, we observe an improvement in FSVD's performance when $J_i$s and $n$ increase, coinciding with Theorem~\ref{Co_bound_sin_intr}.

\paragraph*{Functional Clustering} We generate heterogeneous functional data with $H=3$ clusters using \eqref{sim: model_SM}. Specifically, we set $a_{ik} = a_{hk}$ if $Z_i = h$, where $Z_i$ is randomly drawn from $\{1,\dots,H\}$ to indicate the cluster of $X_i$, and $a_{hk}$ are independently generated from $\text{Uniform}(-1,1)$. We normalize and orthogonalize the vectors $\bm{a}_{k}$ using the Gram--Schmidt algorithm. The $b_{ik}$ are independently generated from $N\left(0, \left(\sum_{i=1}^n a_{ik}^2 / n\right) \times 20\%\right)$. The observation noises $\sigma_i^2$ are set to $\left(\sum_{i=1}^n \mathbb{E}\|X_i\|^2 / n\right) \times 5\%$. The $\rho_k$, $T_{ij}$, and $J_i$ are generated similarly to those in \eqref{sim: model_SM}.

Figure~\ref{fig: combine}(B) shows box plots of ARI values from 100 simulations, where FSVD-based methods achieve superior ARIs over spline-clustering and FPCA-clustering. The lower ARIs of spline-clustering may be due to the inefficiency of B-spline bases in capturing functional patterns, while FPCA-clustering may be affected by the inaccurate estimation of subgroup covariance functions. The FSVD-EM-clustering appropriately captures the functional patterns among heterogeneous functional data, therefore offering the best performance among the three methods.

\paragraph*{Functional Linear Regression}
We generate the functional predictors $X_i$s based on model \eqref{sim: model_SM} under the setting of heterogeneous functional data, and draw $Y_{ij}$s as discrete and noisy measurements of $X_i$s in the same way as the simulations on functional completion. We then construct basis $\{\phi_k; 1\leq k \leq K\}$ as the first $K$ non-constant Fourier basis functions, construct the functional coefficient $\beta = \sum_{k=1}^3 (4-k)^{-1.2} \cdot (-1)^{-k}\varphi_k$,  set $\alpha=0$, draw $\vartheta_i$s independently from $N\!\left( \sqrt{\sum_{i=1}^n\langle X_i,\beta\rangle^2/n} \times 5\%\right)$, and generate $Z_i$s based on \eqref{FLR_model} in the main text.

We present the results of 100 replicated simulations in Figure~\ref{fig: combine}(C). Among the three methods, PFR performs the worst, possibly because the B-spline bases are less efficient for estimating \( \beta \) from data with irregular time points.
Compared to FSVD, $\beta$s estimated using FPCA exhibit larger estimation variances and certain biases. These inaccuracies stem from the heterogeneity of $X_i$s that makes the estimation of eigenfunctions invalid for FPCA. Consequently, FPCA fails to ensure that $\beta$ lies within the space spanned by the eigenfunctions with high probability.

\paragraph*{Factor Model}
Consider the model
\begin{equation*}
    Y_{ij} = \sum_{k=1}^K \rho_k a_{ik} F_k(T_{ij}) + \varepsilon_{ij},\quad i \in [n],\ j \in [J_i],
\end{equation*}
where $K=3$, $\bm{L} = (a_{ik})_{i \in [n], k \in [K]}$ is a fixed loading matrix containing intrinsic basis vectors, $F_1,\dots,F_K$ are random functions, $\varepsilon_{ij}$ are white noises, and $T_{ij}$ are random time points. 
We construct $\bm{a}_k$s deterministically by setting ${a}_{ik}=\sin\big\{k\pi(i+n/4)/(2n)\big\}$ for $i\in [n], k\in [K]$, letting $\bm{a}_k=(a_{1k},\dots,a_{nk})^\top$, and then orthonormalizing $\bm{a}_{k}$s by the Gram--Schmidt process.

The average NMSE values over 100 simulations are presented in Figure~\ref{fig: combine}(D). Among the three methods, SVD performs worst due to errors from data completion and failure to account for temporal smoothness. FAM improves upon SVD by leveraging temporal autocorrelation, but its performance is affected by the non-stationary nature and incomplete observation of the simulated data. Our FSVD method avoids data completion errors and appropriately handles temporal smoothness in non-stationary time series, leading to superior performance. We also observe that the factor loadings estimated by FSVD improve as the number of time points increases for different $n$, aligning with Theorem~\ref{Co_bound_sin_intr_vec}.

\section{Supporting Results}

\subsection{Calculation of Local Kernel-Weighted Cross-Covariance}
For feature $i$ observed as $\{(Y_{ij}, T_{ij})\}_{j=1}^{N_i}$, the local mean is defined as
\[
\hat\mu_i(t_0)
=
\left(\sum_{j=1}^{N_i} K\!\left(\frac{T_{ij}-t_0}{h}\right) Y_{ij}\right)
\Big/
\left(\sum_{j=1}^{N_i} K\!\left(\frac{T_{ij}-t_0}{h}\right)\right),
\]
where $K(u)=\exp(-u^2/2)$. The local cross-covariance between features $i$ and $\ell$ at time $t_0$ is then estimated as
\[
\widehat{\mathrm{Cov}}_{t_0}(i,\ell)
=
\left(\sum_{j=1}^{N_i}\sum_{k=1}^{N_\ell}
w_{ij,\ell k}(t_0)\{Y_{ij}-\hat\mu_i(t_0)\}\{Y_{\ell k}-\hat\mu_\ell(t_0)\}\right)
\Big/
\left(\sum_{j=1}^{N_i}\sum_{k=1}^{N_\ell} w_{ij,\ell k}(t_0)\right),
\]
where $\widehat{\mathrm{Cov}}_{t_0}(i,i)=0$, and the weights are defined as
\[
w_{ij,\ell k}(t_0)
=
K\!\left(\frac{T_{ij}-t_0}{h}\right)\,
K\!\left(\frac{T_{\ell k}-t_0}{h}\right)K\!\Big(\frac{T_{ij}-T_{\ell k}}{b}\Big).
\]
The bandwidth is set to $h=48$ and $b = 12$.
\clearpage

\subsection{Interpretation of Clinical Features}\label{Inter_cli}
\renewcommand{\thetable}{S2}
\begin{table}[h!]
\small
\centering
\caption{Interpretation of Clinical Features}
\begin{tabular}{|m{4cm}|m{10cm}|}
\hline
{Feature} & {Interpretation} \\
\hline
Heart Rate & The number of heartbeats per minute, an important indicator of cardiovascular health. \\
\hline
Respiratory Rate & The number of breaths taken per minute, which can indicate respiratory health and potential distress. \\
\hline
Arterial O2 Saturation & The percentage of oxygen-saturated hemoglobin in the blood, crucial for assessing respiratory function and oxygen delivery. \\
\hline
Blood Pressure & The pressure in arteries during the contraction of the heart muscle, an essential measure of cardiovascular function. \\
\hline
Oxygen Saturation & The overall level of oxygen in the blood, which helps evaluate respiratory efficiency and function. \\
\hline
Base Excess & A measure of excess or deficit of base in the blood, used to assess metabolic acidosis or alkalosis. \\
\hline
Glucose & The level of sugar in the blood, important for diagnosing and managing diabetes. \\
\hline
Creatinine & A waste product from muscle metabolism, used to evaluate kidney function. \\
\hline
INR (PT) & International Normalized Ratio of Prothrombin Time, a measure of blood clotting time, important for patients on anticoagulants. \\
\hline
Lactate & A byproduct of anaerobic metabolism, used to assess tissue hypoxia and sepsis. \\
\hline
Platelet Count & The number of platelets in the blood, crucial for blood clotting and wound healing. \\
\hline
Neutrophils & A type of white blood cell, important for the body's defense against infections. \\
\hline
\end{tabular}
\label{table:features}
\end{table}

\clearpage

\subsection{Illustration of IBFs from Different Clusters}

\begin{figure}[h]
\renewcommand{\thefigure}{S3}
    \centering
    \includegraphics[width=0.9\linewidth]{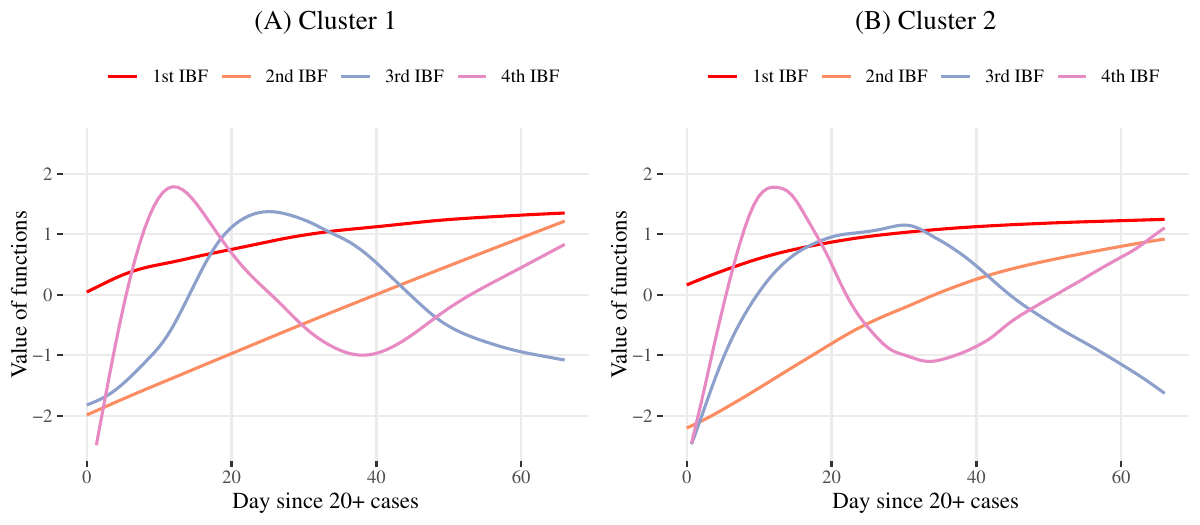}
    \caption{Estimated IBFs of the epidemic functional data from two clusters.}
    \label{fig:IBF_cluters}
\end{figure}

\end{sloppypar}

\end{document}